\RequirePackage{fix-cm}
\documentclass[twocolumn]{svjour3}          
\usepackage{wrapfig}
\usepackage{lipsum}
\usepackage{mwe}

\usepackage{appendix}
\usepackage{amsmath}
\usepackage{graphicx}
\usepackage{lineno}
\usepackage{array}
\usepackage{longtable}
\usepackage{natbib}
\usepackage{siunitx}
\usepackage{multirow}
\usepackage{amssymb}
\usepackage{pifont}
\usepackage{physics}
\usepackage{amsmath} 
\usepackage{array, floatrow, tabularx, makecell, booktabs}%

\DeclareMathOperator*{\argmin}{arg\,min}

\newcommand*\patchAmsMathEnvironmentForLineno[1]{%
	\expandafter\let\csname old#1\expandafter\endcsname\csname #1\endcsname
	\expandafter\let\csname oldend#1\expandafter\endcsname\csname end#1\endcsname
	\renewenvironment{#1}%
	{\linenomath\csname old#1\endcsname}%
	{\csname oldend#1\endcsname\endlinenomath}}%
\newcommand*\patchBothAmsMathEnvironmentsForLineno[1]{%
	\patchAmsMathEnvironmentForLineno{#1}%
	\patchAmsMathEnvironmentForLineno{#1*}}%
\AtBeginDocument{%
	\patchBothAmsMathEnvironmentsForLineno{equation}%
	\patchBothAmsMathEnvironmentsForLineno{align}%
	\patchBothAmsMathEnvironmentsForLineno{flalign}%
	\patchBothAmsMathEnvironmentsForLineno{alignat}%
	\patchBothAmsMathEnvironmentsForLineno{gather}%
	\patchBothAmsMathEnvironmentsForLineno{multline}%
}
\usepackage[table,xcdraw]{xcolor}
\begin{document}
	
	\title{Sparse Algorithms for EEG Source Localization}
	
	\author{Teja Mannepalli, Aurobinda Routray}%
	
	
	\institute{T. Mannepalli \at
		Indian Institute of Technology, Kharagpur, India. PIN: 721 302 \\
		Tel.: +91 96 35 88 5811\\
		\email{mteja134@gmail.com}           
		\and
		A. Routray \at
		Indian Institute of Technology, Kharagpur, India. PIN: 721 302 \\              \email{aurobinda.routray@gmail.com} 
	}
	
	\maketitle
	\sloppy
	\begin{abstract}
		Source localization using EEG is important in diagnosing various physiological and psychiatric diseases related to the brain. The high temporal resolution of EEG helps medical professionals to assess the internal physiology of the brain in a more informative way. The internal sources are obtained from EEG by an inversion process. The number of sources in the brain outnumbers the number of measurements. In this article, a comprehensive review of the state-of-the-art sparse source localization methods in this field is presented. A recently developed method, certainty-based-reduced-sparse-solution (CARSS) is implemented and is examined. A vast comparative study is performed using a sixty-four channel setup involving two source spaces. The first source space has 5,004 sources and the other has 2,004 sources. Four test cases with one, three, five, and seven simulated active source (s) are considered. Two noise levels are also being added to the noiseless data. The CARSS is also evaluated. The results are examined. A real EEG study is also attempted. 
		
		\keywords{Electroencephalograph \and Ill-posed problem \and Source localization \and Sparse signal reconstruction.}  
	\end{abstract}
	
	\section{Introduction}
	\label{intro}
	
	The study of physiological and neurophysiological aspects of the brain is important to diagnose the health of the brain. For this, two techniques are available- radiological and electro-magnetic. Radiological techniques like fMRI, CT that are although non-invasive, involve radiological emissions. It can accurately provide the internal anatomy of the brain. This is used for studying structural anomalies. However, the low temporal resolution does not recommend these methods to analyze the behavioral symptoms of the brain. Therefore, the EEG and MEG have been accepted as the de facto standards for studying behavioral syndromes like epilepsy and sleep disorders. Over the years, effective numerical computing tools have been made available by the researchers for carrying out the inversion processes by way of which surface measurements could be mapped to the internal sources of the cortex.  The high time-resolution of EEG combined with the inversion provides the opportunity to analyze the brain-behavior in more detail (\cite{sanei2013eeg}) (\cite{michel2009electrical}) (\cite{murray2008topographic})(\cite{michel2012towards}). 
	
	\subsection{EEG source localization}
	
	The EEG source localization is the inversion of surface measurements and mapping to the internal sources of the cortex (\cite{grech2008review}). This helps to study and understand brain activity at a better temporal resolution. The sources in the brain outnumber the number of measurements. This results in an ill-posed problem. The source distribution is not sparse at any point in time. The source activity is oscillatory. However, in an experimental setting (like during capture of any event-related potential - ERP) in which many trials can be averaged, the non-task-related activity cancels out. An event-related potential (ERP) is the measured brain response that results in specific sensory, cognitive, or motor event. The problem is treated as a linear model implying a linear combination between EEG measurements and sources. The fact is derived from the linearity of Maxwell's equations of the measurements and sources. The localization of unknown active sources from the known EEG measurements is the \textit{inverse problem}. The linear operator, lead field matrix predicts the electromagnetic fields generated due to an array of sources by the EEG channels. Two main types of approaches are commonly used to solve this problem:
	
	\begin{enumerate}
		\item The parametric or dipole fitting approach. They assume limited dipoles are responsible for explaining the measurements. The location and orientation of the dipoles are unknown.
		\item The non-parametric approach or distributed source model. Several dipoles are distributed uniformly all over the brain. The current article deals with this approach.
	\end{enumerate}
	
	\subsection{Neurological origin of problem}
	
	The neural activations produce electro-magnetic fields in the brain. The human brain contains hundreds of billions of neurons. Some millions of neurons collectively and synchronously fire with a potential in the same orientation resulting in a post-synaptic potential (PSP). The PSP is of the magnitude of about 10 mV with a duration of 10 ms. They will be emerging as a measurable potential on the scalp. EEG measures the collective activity of the neurons. 
	
	Most potentials originate from pyramidal cells. Grey matter is dominated by neuronal cell bodies and synapses and relatively few myelinated axons. They are the 'generators' of the potential. White matter is dominated by the myelinated axons and has relatively few cell bodies. They are the 'transmitters' of the potential.  (\cite{friston2008hierarchical}) (\cite{baillet2001electromagnetic}).  The sources of EEG are the current generators. They are modeled as current dipoles. 
	
	\subsection{Need and importance of the source localization}
	
	The EEG along with source localization is effectively used to analyze a wide range of disorders such as Alzheimer’s disease, epilepsy, seizures, tumors, and brain anomalies (\cite{nolte2009human}) (\cite{sanei2013eeg}). The complex brain pathways under several cognitive as well as emotional tasks can also be studied effectively (\cite{nolte2009human}). Many medical professionals claim EEG as a valuable tool due to its high temporal resolution. This helps to study the brain in detail. 
	
	\subsection{Challenges in Source Localization}
	
	The primary challenge is ill-posedness. Further, the solution has to be unambiguous. The head and brain models need to comply with anatomical and structural constraints of the brain. The data from fMRI can be used for this  (\cite{oostenveld2011fieldtrip}). The process of acquiring pure EEG signals from raw data requires advanced pre-processing techniques and strategies. The generalization of the cortical information across all the subjects is not generally allowed by the medical community. However, such a generalized model can be used to get an overview of internal brain activity.
	
	\subsection{Overview of the current manuscript}
	
	The current manuscript deals with the problem of source localization using EEG. The manuscript is organized as follows. Background deals with the problem definition, characteristics of the problem, and solution strategy. The literature review deals with the current state of art techniques. The next section deals with the challenges posed by this problem and the setbacks of the methods. Five methods are implemented and are compared. The methods are again solved with reduced solution space by a newly developed method by the authors namely certainty-based reduced sparse solution (CARSS) ((\cite{mannepalli2019certainty}). A real data study is evaluated.  
	
	\section{Background}\label{back}
	
	\subsection{Problem definition}
	
	The distributed source models involve inversion of the lead field matrix. The distribution of the current generators in practice is defined over a discrete set of locations. The current dipoles are positioned in the cortex (\cite{dale1993improved}). The source estimates are simply the amplitudes of the dipoles. The distribution is scalar-valued when only their amplitudes are not known. It will be vector-valued if the amplitudes and orientations are unknown. The sources can be either defined over a surface or a volume. The surface-based source models distribute the dipoles over the cortex. The approach involves the 2D surface of the cortex and is first proposed by (\cite{dale1993improved}). The other way is to sample the dipoles on a regular 3D grid in the cortex. 
	
	First, the cortex is sectioned into small 3D volumes called voxels. Each voxel is assumed as a source and is assigned a dipole. The distributed source model has a known dipole location with an unknown dipole moment (\cite{grech2008review}). This has been posed as a problem of sparse signal reconstruction from limited data measurements in the context of an ERP study. The problem has two properties (1) sources outnumbering the measurements. (2) few active sources among many at a particular time instant (\cite{sanei2013eeg}). The distributed source problem is formulated as:
	
	\begin{equation}
	\mathbf{Y = KX + E}\label{esl_prob}
	\end{equation}
	
	$\mathbf{Y}$ is the measurement vector of size $N \times T $ and $\mathbf{K}$ is the forward matrix (or also known as lead field matrix) (\cite{hallez2007review}), (\cite{grech2008review}). $N$ is the number of electrodes on the scalp, and $M$ is the number of voxels. Each voxel is assigned a dipole or three dipoles orthogonal to each other. The degrees of freedom of a source can be considered three (dipoles having 'free' orientations) or one (dipoles having 'fixed' orientations). More on the interpretation of the lead field matrix is clarified in Section - \ref{ss}.
	$\mathbf{X}$ is the magnitude of unknown weights of sources with size $3M \times T $. $\mathbf{E}$ is the noise. The essence of the inverse problem boils down to minimizing the cost function $\mathbf{F}$:
	
	\begin{equation}
	\mathbf{F} = \argmin_{\mathbf{X}} \frac{1}{2} \mathbf{\norm{Y - KX}}_a^{2} + \mathit{f}(\mathbf{X})\label{cost} 
	\end{equation}
	
	The loss function evaluates the sources active from the measurement vector taking care of the noise. Ideally, the minimization of the loss function is equivalent to computing a maximum likelihood estimate under the assumption of Gaussian noise. In reality, a noise covariance matrix needs to be estimated as the noise will not be Gaussian exactly. The regularization parameter $\alpha$ controls the trade-off between the accuracy in the measurements and the noise sensitivity. Also to note, the weights for each of the columns of the lead field matrix need to be incorporated. This is to counterbalance the deep sources. It can be obtained by normalization of each column of $\mathbf{K}$. This is a common way to address this issue.	The $\mathit{f}(\mathbf{X})$ is the penality or regularization term. It introduces an \textit{apriori} knowledge of the solution. Penalization is required to obtain a unique solution. To obtain a globally optimum solution, the $\mathbf{\norm{Y(\tau) - KX(\tau)}}_a^{2}$ and $ \mathit{f}(\mathbf{X(\tau)})$ need to be convex functions.
	
	\subsection{Problem Characteristics}
	
	\begin{enumerate}
		\item \textit{Ill-posedness}: It is a well-known fact that the number of unknown sources outnumbers the number of measurements at any point in time (\cite{sanei2013eeg}). This leads to a case of an under-determined inversion problem having multiple solutions and hence ill-posedness.
		\item \textit{Spatiotemporal Sparsity}: The source distribution is not sparse at any time. However, many trials are averaged in an experimental setting (ERP experiments), so that any non-task-related activity cancels (\cite{grech2008review}). The sources are considered sparse concerning space and time during an ERP. The spatial sparsity is justified only for the experimental setting. 
		
		Also, the evidence from other neuroimaging techniques, such as fMRI and ECoG has revealed the compact nature of cortical activations. The sources are \textit{locally clustered} and \textit{globally sparse}.
		\item \textit{Depth bias}: The shallow sources have a greater effect on the scalp electrodes as compared to the deep sources due to internal attenuation  (\cite{hallez2007review}). The effect needs to be weighed against the depth. The naive inverse problem will favor the shallow sources. 
		\item \textit{Spatial correlation}: The neighboring sources in the cortex are synchronized with each other at any moment (\cite{grech2008review}). This ensures the continuity of the ionic current in the neuron. Also, to note, the uneven cortex surface (having gyrus and sulcus) plays an important role in deciding the continuity of the current field among sources (\cite{baillet2001electromagnetic}). 
		\item \textit{Temporal correlation/ Common Sparsity}: The sources obey the common sparsity (CSP) assumption subject to a temporal neighborhood   (\cite{zhang2011sparse}) (\cite{friston2008hierarchical}) (\cite{michel2004eeg}). A source is estimated to maintain the CSP for 10-20 ms  (\cite{zhang2011sparse}). 
		\item \textit{Anatomical Constraints:} There are primarily two anatomical constraints of the problem widely followed (\cite{baillet1997bayesian} (\cite{baillet2001electromagnetic}.  They are:
		\begin{enumerate}
			\item \textbf{Fixed orientations:} The orientations of the dipoles are believed to be primarily dominated by the ones perpendicular to the cortical surface (\cite{baillet1997bayesian}) (\cite{baillet2001electromagnetic}). This is because the dipoles activating upwards cannot be canceled by other dipoles. This reduces the orientations of the dipole from three to one. Some researchers, however, still consider the orientations of the dipole as three. 
			\item \textbf{Location of the dipoles:} The sources will be present predominantly in the grey matter and limited in the white (\cite{nunez2006electric}) (\cite{dale1993improved}) (\cite{haufe2013critical}). 
		\end{enumerate} 
		\item \textit{Physiological constraints}: Some of the physiological constraints subject to the present problem are:
		
		\begin{enumerate}
			\item \textbf{Smoothness:} The sampling frequency of EEG is in general considered more than adequate for capturing the temporal neuronal activity of the sources. This assures the smoothness of the dipole magnitudes (\cite{dale1993improved}) (\cite{baillet2001electromagnetic}).
			\item \textbf{Functional dependency of cortical patches:}  The cortex is known not to be a smooth surface. The structural complexity of the cortex decides the functional dependency between patches. In some cases, the adjacent patches can be functionally independent (\cite{baillet1997bayesian}). 
			
		\end{enumerate}
		\item \textit{Non-stationarity of the sources}: While the CSP characteristic is valid for some time, the non-stationarity of sources cannot be eluded. The sources are not stationary with time. The instantaneous solvers generally ignore this fact (\cite{gramfort2013time}), (\cite{friston2008multiple}). 
		\item \textit{Characteristics of noise}: During an evoked response study, the 'noise' is the signal from the brain from non-targeted stimuli. This is known as 'baseline noise'. In the case of MEG, the noise co-variances can be estimated using 'empty room' recordings. There are two ways for which noise covariance is estimated- one is the 'resting baseline' and the other is the 'pre-stimulation baseline' (\cite{baillet2001electromagnetic}). The EEG is recorded for the subjects in the resting state. The noise covariance is estimated from these segments of data. The procedure is the 'resting baseline'. The noise-covariance matrix allows the weight of the channels correctly and estimates the noise covariances of each. The noise-covariance matrix provides information about the field and potential patterns representing uninteresting noise sources.
		
	\end{enumerate} 
	
	\subsection{Solution strategy}\label{ss}
	
	\begin{figure}
		\centering
		\includegraphics[width=\linewidth]{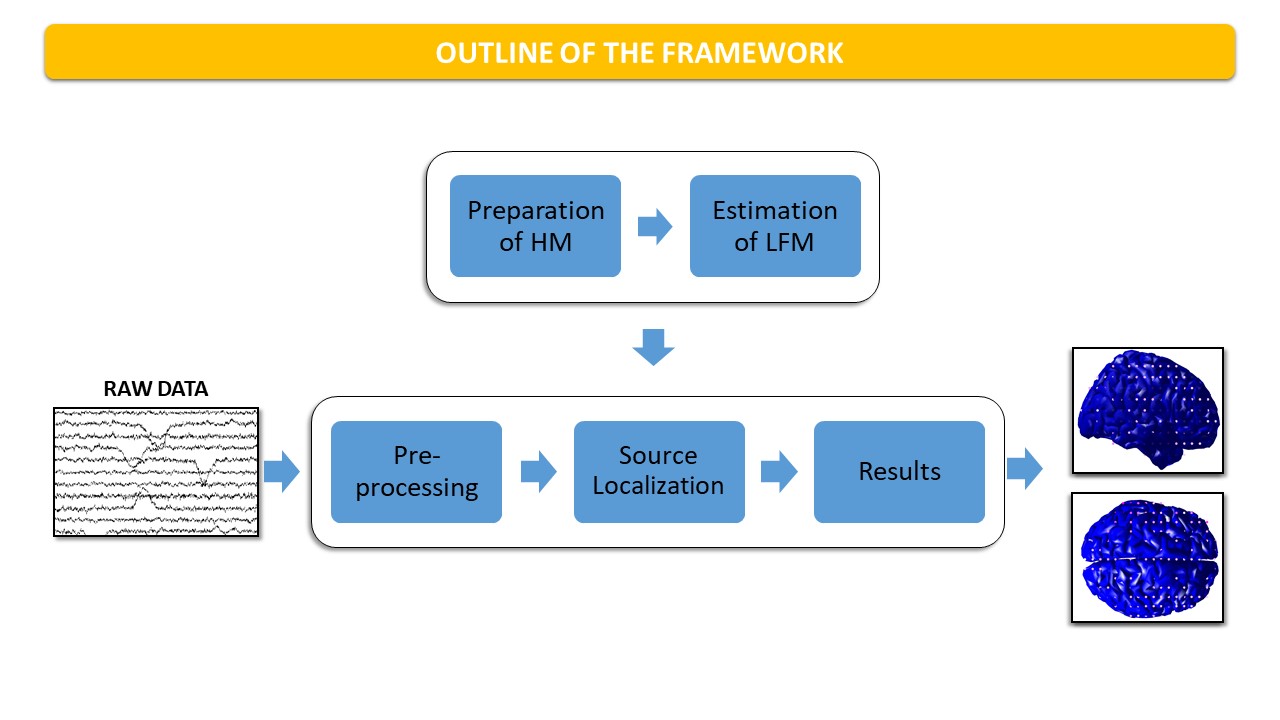}
		\hfill
		\caption{The outline of the approach followed in general. HM: Head model; LFM: lead field matrix}
		\label{meth}
	\end{figure}
	
	Before solving the problem of source localization in a distributed model source scenario, a framework of the data has to be followed. A broad schematic of the methodology followed is presented in Fig. \ref{meth}. 
	
	\subsubsection{Pre-processing} The raw data is pre-processed to remove noise and artifacts.  The artifacts are of two origins- physiological and non-physiological. The artifacts of physiological origin are due to head movements, eye blinks, heartbeat, muscle artifacts, etc. The artifacts of non-physiological origin can be added due to line frequency, cable defects, electronic noises, etc. The artifacts have to be observed and removed carefully. More on the pre-processing of EEG can be found in (\cite{sanei2013eeg}).
	
	\subsubsection{Preparation of the head model} 
	
	The head, skull, and brain models with appropriate measurements are designed. An MRI is converted to the brain model by using a toolbox like Fieldtrip (\cite{oostenveld2011fieldtrip}) or Brainstorm (\cite{baillet2001electromagnetic}) (\cite{mosher2005brainstorm}) or MNE (\cite{gramfort2014mne}) etc. 
	
	\subsubsection{Estimation of the lead field matrix} 
	\hfill\
	
	The dipoles are distributed over the source space. There exist two approaches to prepare a head-model - surface-based approach or volume-based approach.  The source space can be defined as a volume or as a surface. The following procedure adequately fits the volume-based approach. The dipoles are distributed on a regular 2D or 3D grid with needed inter-grid sample distance $m$ within the cortex. A dipole is considered to have either a 'fixed' orientation or 'free' orientation. The orientation of the dipole is fixed perpendicular to the cortical sheet. If then the dipoles have a 'fixed' orientation. If the orientation is left unconstrained, three orthogonal dipoles are placed along the cartesian '\textit{x}', '\textit{y},' and '\textit{z}' directions (\cite{gonccalves2003vivo}) (\cite{hallez2007review}) (\cite{gonccalves2003vivo}) (\cite{hallez2007review}). 
	
	\textit{Lead field matrix}: The potential due to a dipole at $i^{th}$ electrode on the scalp is $\mathbf{k}_j^i = V(\mathbf{r}_{scalp,i},\mathbf{r}_{dipole,j},\mathbf{d});i = 1 \dots N; j = 1\dots M$. $\mathbf{r}_{scalp,i}$ is the position of $i^{th}$ electrode on the scalp. $\mathbf{r}_{dipole,j}$ is the $j^{th}$ source location in the brain. $\mathbf{d} = d_j \mathbf{e}_j$ is the dipole moment with magnitude $d_j = [d_{j,x},d_{j,y},d_{j,z}]$ and orientation $\mathbf{e}_j = \frac{d_j}{\norm{d_j}}$. Each dipole with three orientations results in three column vectors $\mathbf{k}_j = [k_{j,x};k_{j,y};k_{j,z}]$. Each column vector $k_{j,x}$ is of size $N \times 1$. All the column vectors are arranged to frame the lead field matrix 
	
	\begin{equation}
	\mathbf{K} = \begin{bmatrix}
	k^{1}_{1,x} & k^{1}_{1,y}  & k^{1}_{1,z} & k^{1}_{2,x} & k^{1}_{2,y}  & k^{1}_{2,z} & \dots & k^{1}_{M,x} & k^{1}_{M,y} & k^{1}_{M,z} \\
	k^{2}_{1,x} & k^{2}_{1,y}  & k^{2}_{1,z} & k^{2}_{2,x} & k^{2}_{2,y}  & k^{2}_{2,z} & \dots & k^{2}_{M,x} & k^{2}_{M,y} & k^{2}_{M,z} \\
	\vdots & \dots & \vdots  \\
	k^{N}_{1,x} & k^{N}_{1,y}  & k^{N}_{1,z} & k^{N}_{2,x} & k^{N}_{2,y}  & k^{N}_{2,z} & \dots & k^{N}_{M,x} & k^{N}_{M,y} & k^{N}_{M,z} \\
	\end{bmatrix}
	\end{equation}
	
	of size $N \times 3M$. $\mathbf{K}$ contains the information regarding the sources. The lead field matrix $\mathbf{K}$ as in Eq. \ref{esl_prob} is computed using three concentric sphere models or the boundary element method, or finite element method, or finite difference method, etc. according to the choice (\cite{hallez2007review}). 
	
	\section{Literature Review}\label{litre}
	
	The current section is organized into four sub-sections. They are instantaneous solvers, time-block solvers, Bayesian-based learning solvers, tensor-based solvers, and extended source solvers. The instantaneous solvers solve the problem (as in Eq. \ref{esl_prob}) instant by instant independent of each other. They employ vector-based norms to minimize the cost function in Eq. \ref{cost}. 
	
	\subsection{Instantaneous solvers with non-sparse priors}
	
	If the orientations of the sources are free, both the amplitudes and orientations of the current sources need to be estimated. The linear problem to be solved turns out to be:
	
	\begin{equation}
	\mathbf{Y = K X + E}
	\end{equation}
	
	$\mathbf{Y}$ is the measurement vector of size $N \times 1$. $\mathbf{K}$ is the lead field matrix of size $N \times 3M$ and $\mathbf{X}$ has unknown source amplitudes and orientations. $\mathbf{E}$ is noise. The Lagrangian estimate $\mathbf{X^{*}}$ of the true source distribution is given by the solution to the optimization problem:
	
	\begin{equation}
	\mathbf{X^{*}} = \argmin_{\mathbf{X}} \frac{1}{2} \norm{\mathbf{Y - KX}}_{fro}^{2}, \text{ subject to } \norm{\mathbf{X}} \le \epsilon
	\label{eslp}
	\end{equation}
	
	The norm denoted by $fro$ corresponds to the Frobenius norm of the matrix and the parameter $\epsilon$ controls the regularity of the solution. The first part of the equation $\norm{\mathbf{Y - KX}}_{fro}$ is the reconstruction error. The $\norm{\mathbf{X}} \le \epsilon$ means imposing the solution to be small for the Frobenius norm.
	
	Any other norm other than the Frobenius norm can be used in this method. In this regard, there can be many 'minimum norm solvers'. in the EEG community, the minimum-norm solution usually only refers to the minimization of an $\ell_2$ norm (\cite{gramfort2009mapping}). The $\ell_1$ priors become non-differentiable constraints and cannot be cast as this Lagrangian formulation and derive the solution that easily.
	
	On deriving the differentiable and unconstrained problem in Eq. \ref{eslp}, 
	
	\begin{equation}
	F = \argmin_{\mathbf{X}} \frac{1}{2} \mathbf{\norm{Y - KX}}_{fro}^{2} + \alpha \norm{\mathbf{X}}
	\label{mne}
	\end{equation}
	
	Using Woodbury matrix identity, the solution will be:
	
	\begin{equation}
	\mathbf{X = T Y}, \mathbf{T = }\mathbf{K^T[KK^T + \alpha \mathbf{I}]^{\dagger}}
	\label{mnesol}
	\end{equation}
	
	$\mathbf{T}$ is the resolution matrix.
	
	\noindent
	\textit{Regularization constant}: The regularizing parameter $\alpha$ is estimated using standard regularization techniques like Tikhonov regularization. The Brainstorm toolbox developed an efficient technique to estimate a reasonable $\alpha$. By replacing $\mathbf{K}$ with its singular value decomposition (SVD) estimate $\mathbf{USV^{T}}$ in \ref{mnesol}, we get 
	
	\begin{equation}
	\mathbf{X^{*} = K^{T}U(S^{2}} + \alpha \mathbf{I})^{-1} \mathbf{U^{T} Y}
	\end{equation} 
	
	The diagonal entries of $\mathbf{S}$ are the singular values $s_{i,i}$ of $\mathbf{K}$. The singular values are arranged in decreasing order. The matrix $(\mathbf{S^2} + \alpha \mathbf{I})^{-1}$ is a diagonal matrix with diagonal elements $(s_i^2 + \alpha)$. The $\alpha$ obviously should take a value comparable to the $(s_i^2)_i$. The $\alpha$ is analytically first set as 0.01 $s_1^2$. 
	
	A few other commonly known methods for estimating $\alpha$ are L-curve and generalized cross-validation (GCV) (\cite{grech2008review}). The L-curve gives the $\alpha$ at the knee point in the log-plot of the solution $\norm{\mathbf{X}}$ v/s residual $\norm{\mathbf{Y - KX}}$.  
	
	The sLORETA (\cite{pascual2002standardized}) solves the problem using an $\ell_2$ norm. The loss function is minimized and is regularized using the $\ell_2$ norm. The solution is:  
	
	\begin{equation}
	\mathbf{X = T} \text{Y where as } \mathbf{T = K^{T}H [HKK^{T}H + \alpha H]}^{\dagger}\label{slor}
	\end{equation}  
	
	sLORETA assumes that the variability in the estimate is due to the sources. $\mathbf{H}$ is the centering matrix. It is equivalent to an average reference operator. The purpose of $\mathbf{H}$ is to normalize the lead field matrix $\mathbf{K}$. $\mathbf{T}$ is the resolution matrix.  The $\ell_2$ norm cannot provide the sparsity in the solution. The $\ell_2$ norm helps to evenly spread the energy over all the sources. It fails to recover high spatial frequencies. (\cite{gramfort2012mixed}). The $\ell_2$ is less robust to noise compared to the absolute sum $\ell_1$ norm. However, it has found successful and simpler to implement (\cite{pascual2002standardized}) (\cite{pascual1994low}).
	
	The generalized form of MNE to understand the variants in it is:
	
	\begin{equation}
	\mathbf{X = T Y}, \mathbf{T = W}^{-1}\mathbf{K^T[KW^{-1}K^T + \alpha \mathbf{I}]^{\dagger}}
	\label{mnesol1}
	\end{equation}
	
	$\mathbf{W}$ is the weight matrix. The characteristics like 'depth bias' are incorporated in $\mathbf{W}$. $\dagger$ represents pseudo inverse. The minimum norm estimate (MNE) (\cite{dale1993improved}) (\cite{hamalainen1984interpreting}) solves the problem by applying the simple identity operator ($\mathbf{W= I}$). The weighted minimum norm estimate (WMNE) (\cite{grech2008review}) solves the problem by weighting the deep sources ($\mathbf{W} = diag(\norm{\mathbf{K}_i}^2)$, $\mathbf{K}_i$ is the $i^{th}$ column of $\mathbf{K}$). The low resolution impedance tomography (LORETA) (\cite{pascual1994low}) solves by employing a discrete spatial Laplacian operator $(\mathbf{W = B} diag(\norm{\mathbf{\mathbf{K}_i}^2}))$. The method incorporates two factors- the depth of sources and neighborhood correlation of sources.
	
	\subsection{Instantaneous solvers with sparse priors}
	
	The $\ell_2$ norm prior is a non-sparse one although it is differentiable and simple to solve. The main issue with minimum norm solvers is their tendency to smear out the estimated current densities. The solutions delivered will be often quite widespread. 
	
	The instantaneous solvers employing sparse priors need to employ an $\ell_p,p < 2$ norm as the prior. The $\ell_p, p  <  2$ norm has a non-linear combinatorial characteristic unlike $\ell_2$ norm. The $\ell_0$ ideally is supposed to provide the most accurate and sparse solution. However, it is not possible to solve due to the combinatorial explosion. The $\ell_p,p < 2$ norm is used to penalize (or regularize) the solution in (\cite{xu2007lp}) (\cite{ou2009distributed}) to overcome the issues raised by $\ell_2$ norm priors. The $\ell_p,p < 2$ norm methods need to employ non-linear solvers. 
	
	\begin{equation}
	\mathbf{F} = \argmin_X \frac{1}{2} \mathbf{ \norm{Y - KX}}_2^{2} + \alpha (\norm{\mathbf{X}}_p)
	\label{l1} 
	\end{equation}
	
	The $\ell_p,p  <  2$ is employed to regularize the solution. The $\ell_p, p \ge 1$ is convex and $\ell_p, p > 1$ is strictly convex. This makes the respective solvers yield a unique solution.
	
	The conventional minimum $\ell_1$ norm solvers suffer from a poor reconstruction of the smoothness of the source activity over time. The implication is given by the $\ell_1$ norm solvers from a source activity being substantially a "spiky" one or discontinuous. However, it is not "spiky" rather continuous and oscillatory. The $\ell_1$ norm solvers are severely limited by this.
	
	\textit{Iterative Recursive System solvers (ITERS)}:  The basic idea of an ITER, as the name suggests, solves the problem with the same rule iteratively. ITERs aim to produce a sparse solution. A few common ITERs are minimum current estimate (MCE) and focally under-determined system solver (FOCUSS). Both solve the problem instant by instant. MCE employ the $\ell_1$ norm penalization (\cite{matsuura1995selective}). FOCUSS uses an $\ell_2$ norm reweighted scheme to approximate the $\ell_0$ norm regularized method (\cite{gorodnitsky1995neuromagnetic}) (\cite{gorodnitsky1997sparse}). It tries to involve similar to $\ell_0$ norm penalization. The cost function the FOCUSS solver employs is:
	
	\begin{equation}
	\mathbf{x*} = \argmin_x \frac{1}{2} \norm{\mathbf{y - Kx}}_{2}^{2} + \norm{\mathbf{x}}_0
	\end{equation}
	
	$\mathbf{y}$ is $N \times 1$ measurement vector and $\mathbf{x}$ is the $M \times 1$ source vector. The FOCUSS solves the problem in Eq. \ref{esl_prob} employing inversion with its past solution iteratively.
	
	\begin{align}
	\mathbf{C}_i = diag ([\mathbf{X}_{i-1} (1) \dots \mathbf{X}_{i-1} (3M)]) \\
	\mathbf{q}_i = (\mathbf{KC}_i)^{\dagger} Y \\
	\mathbf{X}_i = \mathbf{C}_i \mathbf{q}_i  \text{   unitill converges    }
	\end{align} 
	
	$\mathbf{C}_i$ is the weight factor depending on $(i-1)^{th}$ or past solution. The solver weights its past solution and tries to converge.  \\
	\textit{Issues with ITERS}: ITERS are known to give sparse results. However, the convergence to the local minima instead of the global occurs in many cases (\cite{wipf2006bayesian} ). Adding to this, the ITERS is largely dependent on the initial solution and is highly sensitive to noise. The major issue is the scattering of the solution. 
	
	\subsection{Time block solvers}
	
	The instantaneous solvers employ vector norms to penalize the solution whereas the time block solvers employ mixed norms like the $\ell_{21}$ (\cite{gramfort2012mixed}). While the above solvers deliver a sparse solution, they fail to recover the time profile of the sources. The time-block solvers enable us to collectively analyze the cortex in space and time domains jointly. There are many novel works performed in this domain in recent times that aim to bring out a joint space time-frequency approach (\cite{strohmeier2016iterative}). 
	
	\textit{Two-level mixed norms} : Let $\mathbf{x} \in R^{P}$ be a sequence indexed by a paired indexes $(g, m)$. The two indices are the hierarchy of the coefficients. $g$ means 'group' index and $m$ mean 'membership' index.The idea behind this is that the coefficients within  the same group (indexed by $g$ as $(x_{g,1}, x_{g,2} \dots )$) are \textit{correlated} and coefficients that are distinguished between the groups are \textit{uncorrelated} with each other. Let $p, q$ be the sequence of strictly positive values $\ge 1$. Let $\mathbf{w}$ be weights. The mixed norm of $\mathbf{x}$ is defined as:
	
	\begin{equation}
	\norm{\mathbf{x}}_{p,q} = \sum_{g} ( \sum_{m} ( w_{g,m} |x_{g,m}|^{p} )^{\frac{q}{p}} )^{\frac{1}{q}}
	\end{equation}
	
	By employing a mixed norm, a linkup between $\ell_1$ and $\ell_2$ norms is obtained. Using this the temporal dynamics in the data can be precisely inferred. The cost function for any mixed norm estimate (MxNE) solver is:
	
	\begin{equation}
	\mathbf{X*} = \argmin_{\mathbf{X}} (\frac{1}{2} \norm{\mathbf{Y - KX}}_2^{2} + \alpha f(\mathbf{X}) ) 
	\end{equation}
	
	The mixed norm estimate (MxNE - (\cite{gramfort2012mixed})) utilizes $\ell_{pq}$,($p,q $ be the sequence of strictly positive values $\ge 1$) as a before penalizing the solution (\cite{gramfort2012mixed}) (\cite{gramfort2009mapping}) (\cite{strohmeier2016iterative}). The cost function is:
	
	\begin{equation}
	\mathbf{F} = \argmin_X \frac{1}{2} \norm{\mathbf{Y - KX}}_2^{2} + \alpha \norm{\mathbf{X}}_{21}
	\end{equation}
	
	$\alpha$ is the regularization parameter. The cost function is non-linear and convex. The cost function is solved by proximal gradient methods. They are the extensions of the gradient-based methods used for the optimization of the $\ell_2$ norm functions. 
	
	This $\ell_{21}$ norm can deal with space and time factors simultaneously. The $\ell_{21}$ of $\mathbf{X}$ norm makes a whole group set to zero. If a group is non-zero, all the coefficients in the group will be set to zero. This leads to a row-structured sparsity. It involves sparsity due to the $\ell_1$ norm and gradient in the solution due to the $\ell_2$. The $\ell_{21}$ norm certainly forms a more general class of priors. The work (\cite{gramfort2012mixed}) also reports the use of a three-level $\ell_{212}$ norm with space-time and experimental conditions.  
	
	The work performed in (\cite{strohmeier2016iterative}) reports an efficient approach named iterative reweighted Mixed Norm Estimate (irMxNE). It employs a Frobenius norm to minimize the loss function and a quasi-static $\ell_{0.5}$ norm to penalize the solution. The quasi-static $\ell_{0.5}$ norm makes the cost function a non-convex one. To solve a non-convex function, an approach named block coordinate descent (BCD) is proposed. Also to note, the proposed block coordinate descent (BCD) techniques lead to a faster computation than the fast iterative shrinkage algorithms.
	
	In extension to the mixed norm estimate (MxNE), time-frequency mixed norm estimates (TF-MxNE) proposed in (\cite{gramfort2013time}) utilize $\ell_{21}+\ell_1$ norms. The $\ell_{21}+\ell_1$ norm provides scattered row structures. The method works on the assumption that the time course of each dipole is a linear combination of a few Gabor atoms. 
	
	\begin{equation}
	\mathbf{Y = KX + E = K \xi\rho+ E}
	\end{equation}
	
	The source matrix is decomposed into the Gabor dictionary of atoms as $\mathbf{KJ = K\xi\rho}$. $\mathbf{\rho}$ is the dictionary of $G$ Gabor atoms of size $G \times T$. $\mathbf{\xi}$ has the coefficients of the Gabor decomposition with size $M \times G$. The penalization is accomplished by $\ell_{21}+\ell_1$ norm as $\alpha f(\mathbf{\xi}) = \alpha_{space}\norm{\mathbf{\xi}}_{21} + \alpha_{time}\norm{\mathbf{\xi}}_{1}$, The cost function to be minimized is:
	
	\begin{equation}
	\mathbf{\xi^*} = \argmin_{\mathbf{\xi}} \frac{1}{2} \norm{\mathbf{Y - K\xi \rho} }_{fro}^2 + \alpha \Omega(\xi) 
	\end{equation}
	
	$\alpha \ge 0$ is the penalizing parameter. If $\Omega(\xi) = \norm{\xi}_{1}$, it corresponds to LASSO problem (\cite{tibshirani2005sparsity}). The $\ell_1$ norm does not impose a row structure on source amplitudes. If $\Omega(\xi) = \norm{\xi}_{21}$, it corresponds to MxNE problem (\cite{gramfort2012mixed}). The $\ell_{21}$ prior on $\mathbf{\xi}$ does not produce denoised time series as it does not involve any decomposition. The TF-MxNE employ $\Omega(\xi) = \alpha_{space}\norm{\xi}_{21} + \alpha_{time}\norm{\xi}_{1}$. If $\alpha_{space}$ is large, it will lead to a sparse solution spatially. If $\alpha_{time}$ is large, it will lead to a smooth time series. 
	
	The main aim of the usage of the Gabor dictionary is to retain the time components (or time spectrum) along with frequency. This in turn helps to understand the non-stationarity of the signal. The $\ell_{21}+\ell_1$ provides the row sparse structure and also simultaneously promoting the sparsity of the decomposition.
	
	The $\ell_1-\ell_2$ norm-based Spatio-temporal solver proposed in (\cite{ou2009distributed}) too utilizes the mixed norm $\ell_{12}$ like the MxNE. They are similar estimators but use different algorithms and approaches to solve. The resulting convex SOCP is solved by interior-point methods. This solver uses $\ell_1$ norm in the spatial domain and $\ell_2$ norm in the time domain to solve. The bias is estimated by using SVD of $\mathbf{Y}$. They employ SVD as a temporal basis function.  
	
	\noindent
	\textit{temporal basis function:} The need for temporal basis function is to reconstruct the oscillatory and continuous source activity with good accuracy. It thereby avoids source activity being manifested as discontinuous and "spiky-kind-of" by conventional $\ell_p, p < 1$ norm solvers. To avoid this discontinuity, averaging of the time courses across adjacent sites is performed. However, it reduces the spatial resolution in return. There exists broadly two kinds of approaches to avoid discontinuous source reconstructions. One is building a prior that explicitly incorporates temporal constraint in the cost function as a regularizer (\cite{baillet2001electromagnetic}). They links-up between the present current estimate and the past.
	
	The other different route is by employing temporal basis functions. VESTAL (\cite{huang2006vector}) projects the $\ell_1$ norm solutions into a signal subspace defined by a set of temporal basis functions. The SVD of the measurement vector $\mathbf{Y}$ as $\mathbf{USV^{T}}$ means the linear combinations of vectors in $\mathbf{V^T} = [\mathbf{V}_{(1:\nu)},\mathbf{V}_{(\nu:N)}]$ represents the time activity about the measurements. The subspace having dominant singular vectors $\mathbf{V}_{\nu:N}$ are chosen. The criterion of choosing the 'dominant' vectors are the ones with $\chi^{2}$ values in a certain range after every time point. The $\chi^2$ test reflects that the subspace is not 'over-fitted or 'under fitted. The same subspace $\mathbf{V_{\nu:N}}$ that explains the time activity of the measurements should explain the same for sources. 
	
	The $\ell_1$ norm estimates are employed in (\cite{huang2006vector}). The $\ell_{21}$ norm estimate is employed in (\cite{ou2009distributed}). In \cite{ou2009distributed}, the selection of subspace is such that the $S$ largest SVD components of $\mathbf{Y}$ should "adequately" fits the $\mathbf{Y}$. The variable $S$ is relatively fixed and is chosen appropriately to the best of the knowledge of sources. The purpose of the use of the data-adaptive temporal basis function specifically is to (1) compactly represent the data which reduces the computation, (2) bypass the difficulties related to the experimental setting (They are high variations in the source activity, too many experimental tasks, and subject-subject variations). 
	
	\subsection{Learning based methods}
	
	There exist many novel works using Bayesian strategies utilizing temporal characteristics. The typical maximum aposteriori solvers with Gaussian likelihood lead to convex optimization problems. The minimum norm methods assume a predefined covariance matrix for the sources. The learning-based methods estimate the prior based on the data. The learning based methods involve estimating the posterior distribution $p(\mathbf{X|Y})$ incorporating a prior distribution $p(\mathbf{X})$ on the sources and the likelihood function $p(\mathbf{Y|X})$. The Bayesian representation of Eq. \ref{cost} is:
	
	\begin{equation}
	p(\mathbf{X|Y}) = p(\mathbf{X})p(\mathbf{Y|X})\label{bay}
	\end{equation}
	
	Learning the prior means learning the source covariance matrix $\mathbf{\Sigma_{X}}$ The source matrix $\mathbf{X}$ and error $\mathbf{E}$ are assumed Gaussian with mean and covariance $N(0, \mathbf{\Sigma_X})$ and $N(0, \Sigma_E)$. If $\mathbf{\Sigma_X}$ and $\mathbf{\Sigma_E}$ are known, the $\mathbf{\mathbf{X}}$ can be found out using the cost function:
	
	\begin{gather}\label{r1}
	\mathbf{X^{*} = \argmin_{\mathbf{X}} \norm{\mathbf{Y - KX}}_{\mathbf{\Sigma_E}}^{2} + \norm{\mathbf{X}}^{2}_{\mathbf{\Sigma_X}}} \nonumber \\
	=  \argmin_{\mathbf{X}} ({\mathbf{Y - KX}}^{T}) \mathbf{\Sigma_E}^{-1} (\mathbf{Y - KX}) + \mathbf{X}^{T}\mathbf{\Sigma_X}^{-1}\mathbf{X}   
	\end{gather}
	
	$\norm{\mathbf{X}}_{\mathbf{\Sigma_X}} = trace(\mathbf{X}^{T}\Sigma_X^{-1}\mathbf{X})$  The solution is $\mathbf{X^{*}} = \mathbf{\Sigma_{X}K^{T}(K\mathbf{X}K^{T} + \Sigma_{E})^{-1}Y}$. If $\mathbf{\Sigma_{X} = \mathbf{I}}$ and $\mathbf{\Sigma_{E} = \lambda \mathbf{I}}$, the solution looks similar to minimum-norm. 
	
	The $\ell_2$ norm is the prior here. Learning the $\ell_2$ norm means learning the source covariance matrix $\mathbf{\Sigma_E}$.  If the noise covariance matrix $\mathbf{\Sigma_{E}}$ too along with the source covariance matrix $\Sigma_\mathbf{X}$ are not known, they are \textit{not} fixed apriori. Then the parameters define a \textit{model}. The $p(\mathbf{Y|X})$ is called \textit{likelihood}, $p(\mathbf{X|Y})$ is called \textit{posterior}, $p(\mathbf{X})$ is the prior, and $p(\mathbf{Y})$ is the \textit{model evidence}. In this perspective, the loss function is the likelihood function $p(\mathbf{Y|X})$ and the sparsity constraint is incorporated as the prior distribution $p(\mathbf{X})$. 
	
	The approaches use maximization of the model \textit{evidence}. However, the differences the approaches have are the ways in learning the priors. The concept of model evidence approaches to estimate the source statistics is first performed by (\cite{friston2008hierarchical}), (\cite{friston2008multiple}), and (\cite{friston2002classical}) a.k.a Restricted Maximum Likelihood (ReML). ReML estimates hidden variables (hyperparameters) with an iterative procedure that amounts to Expectation-Maximization update rules. As a procedure in the EM rule, ReML involves maximizing a non-convex likelihood function.
	
	The current problem in Eq. \ref{cost} can be seen in two ways  (1) single measurement vector (SMV), and (2) multiple measurement vector (MMV). The concept of MMV, in general, is that the non-sparse entries in every column in $\mathbf{X}$ will be identical. The Eq. \ref{cost} is re-written as by letting $\mathbf{\iota} = vec(\mathbf{X}) = [X_{11};\dots X_{1M};\dots ;X_{L1};\dots X_{LM}]$ of size $ML \times 1$, $\mathbf{D = K \bigotimes I_L}$ and $\mathbf{y} = vec(Y)$ of size $NL \times 1$. $vec$ represents vectorization of a matrix. The $\mathbf{I}_N$ is the identity matrix:
	
	\begin{align*}
	\mathbf{y} &= \mathbf{K\iota} + \nu  \\
	\mathbf{y} &= \mathbf{\sum^{N}_{i = 1} (\mathbf{k}_i \bigotimes \mathbf{I}_L )\mathbf{x}_i} 
	\end{align*}
	
	$\mathbf{k}_i$ is the $i^{th}$ column of $\mathbf{K}$. $\bigotimes$ represent Kronecker product. This is the 'block sparsity' model.
	
	The sparse Bayesian learning methods are proposed in (\cite{wipf2006bayesian}), (\cite{wipf2007empirical}), (\cite{wipf2009unified}), (\cite{wipf2011latent}), (\cite{zhang2008eeg}), (\cite{zhang2011sparse}), (\cite{zhang2013extension}) and (\cite{cotter2005sparse}). It is abbreviated as bSBL. They are volume-based approaches. They select among a high number of priors that fit the data. The model selection induces the sparsity in the cost function. It is done by maximizing a Gaussian approximation of the evidence. The approach is also known as 'type II maximization'. Let the sources $\mathbf{X}_I, i = 1 \dots M$ be mutually independent with Gaussian probability density function
	
	\begin{equation}
	p(\mathbf{X}_i,\gamma_i,\mathbf{B}_i) \sim N(0,\gamma_i,\mathbf{B}_i))
	\end{equation}
	
	\begin{equation}
	p(\mathbf{Y}|\mathbf{\Sigma_Y}) = \int p(\mathbf{Y|X} p(\mathbf{X|\Sigma_Y})d\mathbf{X} = N(0, \mathbf{\Sigma_Y})
	\end{equation}	
	
	This is equivalent to minimizing the negative log maximum likelihood:
	
	\begin{equation}
	L = -log \text{ } p(\mathbf{Y|\Sigma_{E}}) = -\text{T } log(|\mathbf{\Sigma_{Y}}|) + trace(\mathbf{Y}^{T}\Sigma_{\mathbf{X}}\mathbf{Y})
	\end{equation}
	
	The measurement covariance matrix $\Sigma_{\mathbf{Y}}$ turns out to be:
	
	\begin{equation}
	\Sigma_{\mathbf{Y}} = \mathbf{K\Sigma_{\mathbf{X}}K^{T} + \Sigma_{\mathbf{E}}}  
	\end{equation}
	
	The source covariance matrix $\Sigma_{\mathbf{X}} = \sum_{i = 1}^{M} \gamma_i \mathbf{X_i}$. $\gamma_i$ and $\mathbf{X}_i$ are the hyperparameter and covariance matrices of the $i^{th}$ source. $\mathbf{X}_i$ is the source covariance matrix defined prior. $\mathbf{X}_i$ is zero except at the $i^{th}$ element. This term $\Sigma_{\mathbf{X}} $ imposes sparsity while minimizing the negative log marginal likelihood: The non-negative hyperparameter $\gamma_i$ controls the row sparsity of $\mathbf{X}$. If $\gamma_i$ = 0, the corresponding $i^{th}$ source becomes zero. $\mathbf{B}$ is a positive definite matrix capturing the correlation structure of $\mathbf{X}$. The sparsity is induced via the log term of the measurement covariance matrix in the likelihood. Finally, an optimization procedure that updates the hyperparameters $\gamma_i$ in every iteration is performed. This is done by the evidence maximization update rule or fixed point gradient rule.
	
	If sparsity is imposed on a source, it remains to be sparse. The SBL is known to converge fast. SBL solver assumes that the noise and covariances	are more or less independent of time (\cite{gramfort2009mapping}). In the real case scenario, the changes in $\gamma_i$ occur over time. When the experiment is performed multiple times (trials) and when they are averaged leading to an evoked response, the source covariance matrix is very likely to change from the beginning to the latter (\cite{gramfort2009mapping}).  These are some issues for bSBL solvers.
	
	\subsection{Tensor based methods}
	
	\textit{Overview of Tensors}: The generalization of 2D matrices is often called a tensor of order $N$. The size will be $n_1 \times n_2 \times \dots n_N$. The tensors can be factorized into a sum of a few component rank one tensors (\cite{kolda2009tensor}). A third order tensor $\mathbf{\Gamma}$ of size $M \times N \times O$ can be decomposed as:
	
	\begin{equation}
	\mathbf{\Gamma} = \sum_{k=1}^{R} \mathbf{I}_k \circ \mathbf{J}_k \circ \mathbf{K}_k 
	\end{equation}
	
	$\circ$ represents the vector outer product. $\mathbf{I}_k$, $\mathbf{J}_k$, and $\mathbf{K}_k$ are $k^{th}$ rank one tensors of sizes $M \times 1$, $N \times 1$, and $O \times 1$ respectively. $R$ is the rank of the tensor. The process of factorization is known as CP decomposition (canonical polyadic) (\cite{kolda2009tensor}). The CP decomposition cannot be applied for tensors of order $N  <  3$ or matrices. CPD is similar to singular value decomposition for matrices or specifically, CPD is the generalization of SVD. 
	
	The idea of tensor approaches for EEG source localization is to convert space-time EEG data into space-time-frequency domains which can be done employing a wavelet function. (\cite{becker2014eeg}) (\cite{becker2012multi}). 
	
	\begin{equation}
	\mathbf{\Gamma}(\mathbf{r},\tau,f) = \int_{-\infty}^{\infty} y_i (\mathbf{r},t) \chi (s,t, \tau) dt
	\end{equation} 
	
	$y_i (\mathbf{r},t)$ is the $i^{th}$ channel EEG data. $\chi (s,t,\tau)$ is the wavelet function with scale $s$ and center frequency $f_c$. The resulting $\mathbf{\Gamma}$ is a function of space $\mathbf{r}$, time $\tau$ and frequency $f$. The frequency $f$ can be estimated by $f = \frac{f_c}{sT}$. $T$ is the interval between time samples and $f_c$ is the center frequency. Instead of wavelet, local spatial Fourier transform can also be used (\cite{becker2014eeg}) (\cite{becker2014eeg}). 
	
	\subsection{Extended source methods}
	
	The nature of source activations is locally clustered and globally sparse. The $\ell_2$ norm and sparsity-based localization methods are good at predicting the active source locations. In clinical case studies like epilepsy, some issues arise. They are: (i) they fail to localize sources with a larger spatial extent, and (ii) they fail to discriminate close active sources (\cite{becker2017sissy} ).
	
	The spatial extent of the active source region becomes vital in detecting a focal epileptic zone. The epileptic activity spreads from one brain region to another. This leads to many highly correlated source regions being active simultaneously. To tackle such challenges, the well-known "cortical patch model" is introduced by (\cite{limpiti2006cortical}). A 'patch' is a set of predefined source regions. The beamforming approaches are utilized to identify sources that aim to describe the measurement vector. 
	
	Beamforming methods assume that the measurement data is generated by some cortical sources modeled by an equivalent set of current dipoles. The main issue with beamforming methods is the lack of the number of true sources that are responsible to explain the measurement vector. There are certain strategies to infer the number of true sources before solving the problem. A way of estimating the number is from the eigenvalue decomposition of the measurement covariance matrix. The eigenvalues drop and stay more or less flat. The number of indices $n$ whose eigenvalues are of considerable magnitudes is treated as the number of active sources. In reality, it is quite difficult to see and perceive any kind of drop.  
	
	The 2q-th order Extended Source Multiple Signal Classification (2q-ExSo-MUSIC) algorithm (\cite{birot2011localization} ) and Disk Algorithm	(DA) (\cite{becker2014eeg}) work on the principle of extended source localization. The 'extended source localization methods' aim to localize 'extended sources (or patches)' and works by using a beamforming strategy.
	
	A cortical patch is modeled by many adjacent grid dipoles. Let $p^{th}$ extended source $\Omega_p$ has the grid dipoles from \textit{p} = $1 \dots P$. The source space $\Omega_s$ is divided into extended source $\Omega_p$ and the rest of the sources. The rest of the sources are assumed responsible to emit background activity in the brain.  
	
	\begin{equation}
	\mathbf{X} = \sum_{\mathbf{\Omega_p}} \mathbf{k}_p  \mathbf{x}_{p}^{T} + \sum_{\mathbf{\Omega_b}} \mathbf{k}_b \mathbf{x}_{b}^{T}  = \mathbf{X_p} + \mathbf{X_b}
	\end{equation}
	
	The $\mathbf{X_p}$ explains the EEG data and $\mathbf{X_b}$ indicates the background activity. The objective is to estimate the source signal matrix.  The background activity is considered limited than the active source regions. The dipoles are estimated in the source patch $\Omega_p$ by thresholding the amplitudes of the estimated source signal. Some extended source strategies are reviewed below.
	
	\subsubsection{Variation-Based Sparse Cortical Current Distribution (VB-SCCD)}
	
	The idea of Variation-Based Sparse Cortical Current Distribution (VB-SCCD) (\cite{ding2009reconstructing} ) is to first transform the source domain to another domain that can track variations in current densities better. This is achieved using a linear transform $\mathbf{V}$ to the source space. The source space is triangulated first such that each edge is only shared by two triangles. The operator is defined mathematically as:
	
	\begin{equation}
	\scalebox{0.95}[1]{$
		V = \begin{bmatrix}
		v_{11}& v_{22} & \dots & v_{1N} \\
		\vdots &        & \ddots & \vdots \\
		v_{p1} & v_{2p} & \dots & v_{NP}
		\end{bmatrix}  , 
		\begin{matrix}
		v_{ij} = 1; v_{ik} = -1; & i^{th} \text{ edge with }j, k \\  
		v_{ij} = 0   &  otherwise
		\end{matrix} $}
	\end{equation} 
	
	$V$ is of size $P \times N$. $P$ is the total   umber of edges of triangular elements. They will be non-zero values either 1 or -1. The cost function is:
	
	\begin{equation}
	min \norm{\mathbf{VX}}_{1} \text{ subject to } \norm{\mathbf{Y - KX}}_{2} \le \epsilon
	\label{vbsccd}
	\end{equation}  
	
	The variational map of sources $\norm{V\mathbf{X}}_{1}$ reveals the differences in amplitude between neighboring dipoles. 
	
	\subsubsection{Source Imaging based on Structured Sparsity (SISSY)}
	
	The Variation-Based Sparse Cortical Current Distribution (VB-SCCD) (\cite{ding2009reconstructing}) overcomes the challenge of localizing highly correlated sources. Moreover, the identification of close sources as separate sources remains a challenge. The close sources are identified as a single large one. The issue of neighboring sources is tackled by Source Imaging based on Structured Sparsity (SISSY) (\cite{becker2017sissy}). The cost function of SISSY is:
	
	\begin{equation}
	\min_{\mathbf{X}} \frac{1}{2} \norm{\mathbf{Y-KX}}_F^2 + \lambda (\norm{\mathbf{VX}}_1 + \alpha \norm{\mathbf{X}}_1)
	\end{equation}
	
	(i). SISSY incorporates an additional $\ell_1$ norm to Eq. \ref{vbsccd} to impose sparsity on estimated sources (\cite{gramfort2013time}).
	(ii). SISSY integrates $\ell_{21}$ norm to deal with the temporal structure of sources (\cite{ou2009distributed}). 
	(iii) SISSSY thresholds the background source activity using the 'Automatic thresholding' (AT). The source space is viewed as a graph with each source as a node. The edges between the sources are the edges of the graph. It is based on the watershed algorithm by (\cite{vincent1991watersheds}). The procedure of AT is as follows:
	
	\begin{enumerate}
		\item Applying watershed transforms on the edges of the graph. 
		\item  The identified source regions are merged until only source regions that correspond to a	local amplitude maximum or minimum remain.
		\item The source regions that are responsible for EEG are identified by thresholding (10 \% of maximum value)
	\end{enumerate}
	
	\subsubsection{Truncated Recursively Applied version- Multiple Signal Classification (TRAP-MUSIC)}
	
	Multiple Signal Classification (MUSIC) (\cite{mosher1999source} ) is a well-known strategy. It (i) divides the measurement data into signal and noise subspaces and (ii) checks for each candidate source in the ROI whether it could explain the signal subspace or not (\cite{liu2006efficient} \cite{schmidt1986multiple} ). RAP-MUSIC (\cite{liu2006efficient} \cite{mosher1999source}) is a more sophisticated algorithm that can deal with the issues raised by classical MUSIC.. 
	
	The TRAP-MUSIC (\cite{makela2018truncated} ) methodology is very similar to RAP - MUSIC. The only change is to the reduced signal subspace. The signal subspace is applied a dimensionality reduction at each recursion. The protocol behind the dimensionality reduction is as follows. As it is well known, at each recursion, one source is estimated and is projected out. For the next step, the same source need not be considered in the signal subspace. Let $n$ be the true number of sources, $\tilde{n}$ be the number of sources that are to be estimated. $i$ be the recursion step. After, $i^{th}$ step, the dimension of the remaining subspace is reduced to $\tilde{n} - i$. 
	
	In classical MUSIC, the local peaks of the localizer have to be estimated in one round. The number of peaks gives the number of active dipoles and their source locations. Let $\mu(\mathbf{r})$ be the localizer which is a function of the source location $\mathbf{r}$.
	
	\begin{equation}
	\mu(\mathbf{r}) = \frac{\norm{\mathbf{R_s}\mathbf{k(r)}}}{\norm{\mathbf{k(r)}}}
	\end{equation}
	
	$\mathbf{R_s}$ is the signal space projection. In RAP-MUSIC, one global peak of the localizer has to be found out at every iteration. The localizer function will be at an iteration $i$ is
	
	\begin{equation}
	\mu_i(\mathbf{r}) = \frac{\norm{\mathbf{R_{s,i} Q_i}\mathbf{k(r)}}}{\norm{\mathbf{Q_i k(r)}}}
	\end{equation}
	
	$\mathbf{Q_i}$ is the out projector. TRAP-MUSIC follows very similar procedure as of RAP- MUSIC except the estimation of transformed signal $\mathbf{Q_i}$. The procedure is as follows:
	
	\begin{align}
	\mathbf{B_i} = [\mathbf{k}_1 \dots \mathbf{k}_{i-1}] \nonumber  \\
	\mathbf{Q_i} = \mathbf{I} - \mathbf{B_i}\mathbf{B_i}^{T} \nonumber \\
	svd(\mathbf{Q_i}\mathbf{U_s}) = \mathbf{U}_i\mathbf{\Lambda_i}\mathbf{V}_i^{T}  \nonumber \\
	\mathbf{R_{i}} = \mathbf{U_i}(:,1:\tilde{n})\mathbf{U_i}^{T}(:,1:\tilde{n}) \nonumber \\
	\mu_i(\mathbf{r_i}) = \frac{\norm{\mathbf{R_{s,i} Q_i}\mathbf{k(r)}}}{\norm{\mathbf{Q_i k(r)}}} 
	\end{align}
	
	$\mathbf{B_i}$ has the topographies of previously estimated dipoles. $\mathbf{k_i}$ is the local lead field matrix.  Let $\mathbf{C} = \mathbf{Y}\mathbf{Y}^{T}$ be the covariance matrix of the measurement vector $\mathbf{Y}$. The $\mathbf{C}$ has to be subdivided into signal subspace ($\mathbf{C_o}$) and noise subspace ($\mathbf{C_n}$). Let the measurement vector got added by white noise then, $\mathbf{C_n} = \sigma \mathbf{I}$. The covariance of the noiseless data is $\mathbf{C_o} = \mathbf{UDU}^T$, where $\mathbf{U}$ is an $M \times M$ orthogonal matrix. $\mathbf{U_s} = [\mathbf{U_1} \dots \mathbf{U_n}]$. $n$ is the number of true sources. The eigenvalues $d_j, j = 1 \dots M$ will drop flat after $d_n$ where $n$ is the number of true sources. The need to compute $svd(\mathbf{Q_i}\mathbf{U_s})$ is to find the othogonal projection $\mathbf{R}_i$ on to $span(\mathbf{Q}_i\mathbf{U}_s).$ $i$ is the iteration step, $i = 1 \dots n$. the iteration is continued till a significant drop in the $\mu_i(\mathbf{r_i})$ is observed. After $i>n$ a similar 'plateau' is observed.
	
	\subsection{Certainty based reduced sparse solution(CARSS)}
	
	There is no precise mathematical motivation behind discovering the method. However, it can be well studied through observations or intuitions and is justified by experiments (\cite{mannepalli2019certainty}).
	
	The certainty based reduced sparse solution - CARSS (\cite{mannepalli2019certainty}) reduces the solution space of an inverse problem to only the most certain active sources. This reduces the redundancy of the problem. Let a dipole in a voxel is active. The two key observations from the scalp potential of the dipole on which the method is developed are:
	
	\begin{enumerate}
		\item The 'peak' of the dipole and the 'shape' around the peak, is prominently evident in the measurement vector.
		\item The region of evidence of the peak in the measurement vector is as it is for the dipole. 
	\end{enumerate}
	
	The prominent global extremum of the dipole on the scalp is referred to as the 'peak' of the dipole. Due to the sparse nature of activation, every peak of the active dipoles will be prominently evident in $\mathbf{Y}$. 
	
	The method utilizes the information provided by peaks in $\mathbf{Y}$ to find the most certain sources that have generated the peaks observed in $\mathbf{Y}$. This method has a preliminary stage that has to be done only at the beginning. Stage-I reduces the solution space to the most certain sources. Stage-II solves the inverse problem with reduced solution space.
	
	\section{Challenges}
	
	The methods utilize advanced and robust mathematical tools. They are challenged by some difficulties and setbacks. The solution to the problem is challenged by some key factors. They are:
	
	\begin{enumerate}
		\item \textit{Ill-posedness in the problem}: The problem in the real scenario is greatly ill-posed, making the methods tough to provide an accurate solution. Most of the methods could fail with an increase in the number of sources. 
		
		\item \textit{Uncertainties/noise in the data}: The Bayesian methods and the $\ell_1$ norm-based methods are known to provide accurate solutions to many problems. However, they are also sensitive to noise. The ITER's too are known to mislead the solution under noise. 
		
		The instantaneous solvers solve the problem instant by instant independent of each other. This leads to an implication that the signal, and hence the noise is independent of time. This is reported to be a challenging assumption as the nature of the 'signal' may change with time but 'noise' may not during an evoked response (\cite{gramfort2012mixed}). The noise is usually estimated when the brain is not yet under any stimulus. Besides, there will be measurement noise due to the electrode channels. These all factors may be the issues for instantaneous solvers dealing with noisy data although they prove to be successful. 
		
		\item \textit{Sparsity in the solution}: The source activation is not sparse at any time. The sources show an oscillatory activity but not spiky activity. However, the solution is considered to be sparse during an evoked response study. The averaging over trials (repeated experiments) results in suppression of sources, not of interest. The sparsity promoting $\ell_p, p <  1$ norm is known to provide a sparse solution. The $\ell_p,p \le 1$ is non-convex. The $\ell_1$ norm is convex and can provide sparse solutions. The $\ell_2$ norm provides the root of a squared sum due to which it cannot provide a sparse solution. The characteristic of the $\ell_2$ norm does not allow sparsity. The $\ell_p,p < 1$ solvers produce sparse solutions while $\ell_2$ based solvers cannot. The $\ell_{21}$ norm is at an advantage since it allows link-up between $\ell_2$ and $\ell_1$ norms. 
		
		\item \textit{Comprehensible Results}: The $\ell_2$ norm is known to provide a smooth gradient in its solution. The $\ell_1$ norm can handle noise well than the $\ell_2$ norm. The typical MAP estimators with Gaussian likelihood and $\ell_1$ norm regularization estimators lead to convex optimization problems and will lead to a unique solution. The FOCUSS is known to provide a scattered solution under noise. FOCUSS works to minimize a non-convex $\ell_0$ norm cost function.
		
		\item \textit{Computational time}: In general, many methods provide accuracy by sacrificing computational time. Some methods are too slow to converge. The computation of the inverse of a large matrix, in general, is computationally expensive. 
		
		\item \textit{Convergence issues}: some methods converge to the local minima instead of global. An example is FOCUSS. Some have convergence issues under noise. 
		
		\item \textit{Complying with the ground reality}: Although mathematically well understood and modeled, the solution should be aligned with various anatomical and physiological constraints and realities.  
	\end{enumerate}
	
	\section{Comparative Study}
	\label{comps}
	
	The toolbox utilized for the current study is SEREEGA: Simulating Event-Related EEG Activity developed by (\cite{krol2018sereega} ). All the computations are performed in an x64-bit, 8 GB RAM, Intel(R) core processor Laptop. 
	
	The head model employed is the 'New-York Head model (ICBM-NY)' (\cite{huang2016new}). The lead-field matrices are of size 64 $\times$ 15,012 and 64 $\times$ 6,004. The number of sources is 5,001 and 2,001. The whole computations of generating the EEG signals are performed in the SEREEGA toolbox (\cite{krol2018sereega}). The length of the measurement matrix $\mathbf{Y}$ simulated is of 1000 sample size.
	
	Sparse Bayesian Learning based methods are implemented according to (\cite{zhang2011sparse}) and (\cite{wipf2011latent}). For, SBL according to (\cite{wipf2011latent}), the sparsity data-fit balancing parameter $\lambda$ is considered 0.2. For, (\cite{zhang2011sparse}), the threshold for pruning  small hyperparameters is set to $10^{-3}$. The maximum iterations performed are 50. Mixed norm estimate is implemented according to (\cite{gramfort2012mixed}). sLORETA is implemented according to (\cite{pascual2002standardized}). The regularization parameter is calculated using 'l-curve' method (\cite{hansen1993use}) wherever needed. 
	
	\textit{A' metric}: A' metric estimates the area under the ROC for one HR/FR pair. The Hit rate (HR) is defined as the number of hits. A 'hit' means that the source is localized correctly. False-positive rate (FR) indicates the number of false positives. The higher \textit{A' metric} value indicates a high hit rate compared to the false positive rate. \textit{A' metric} estimates the area under the ROC curve. 
	
	\begin{equation}
	A^{'} = \frac{(HR - FR) + 1 }{2}
	\end{equation}
	
	ROC in general means receiver-operator characteristic (\cite{oikonomou2020novel} \cite{snodgrass1980standardized} \cite{darvas2004mapping}). To estimate the local peak, the procedure as presented in (\cite{wipf2010robust}) is followed.
	
	\textit{Success rate (SR): } A 'hit' for a true source is defined when there is at least one local peak formed within a certain neighborhood of the source's location. (\cite{wipf2010robust} ). The region under consideration is a sphere centered at the true source location and two levels of neighboring sources. SR can be 1 or 0 for a true source. In the aforementioned test cases, the success rate for each source is estimated and the results are tabulated.
	
	\subsection{Monte-carlo simulations}
	
	Four test cases are considered. The random sources with orientation are chosen using in SAREEGA toolbox (\cite{krol2018sereega}). The performance metric is \textit{A metric} (\cite{wipf2010robust}). Each test case has three noise levels - no noise added, pink noise of amplitude one (\cite{krol2018sereega}) (\cite{haufe2019simulation}) and of amplitude four. The number of possibilities considered for each test case will be 360. The detailed breakup is shown below.
	
	\begin{enumerate}
		\item The total number of test-cases is 4. 
		\subitem Test Case-I: one source.
		\subitem Test Case-II: three sources.
		\subitem Test Case-III: five sources.
		\subitem Test Case-IV: seven sources.
		\item Each test-case has three noise levels added.
		\subitem No noise added.
		\subitem pink noise of amplitude one is added.
		\subitem pink noise of amplitude four is added.
		\item Each noise level is evaluated for two source spaces.
		\subitem Five-K source space (has 5,001 X 3 = 15, 001 sources).
		\subitem Two-K source space (has 2,001 X 3 = 6, 001 sources).
		\item Each source space is further estimated with and without CARSS.
		\subitem Not reduced by CARSS.
		\subitem The reduced solution space by CARSS.
		\item The total number of samples considered is 30.
	\end{enumerate} 
	
	The ERP is as shown in Fig. \ref{tc11}. The peak latencies of the ERP are 500, 300, and 200 ms. The peak widths are 200, 300, and 100 ms. The peak amplitudes are 1, 1.2,  and 0.6 $\mu$ V. 
	
\begin{figure}[!htb]
		\centering
		\centering
		\includegraphics[width=0.9\linewidth]{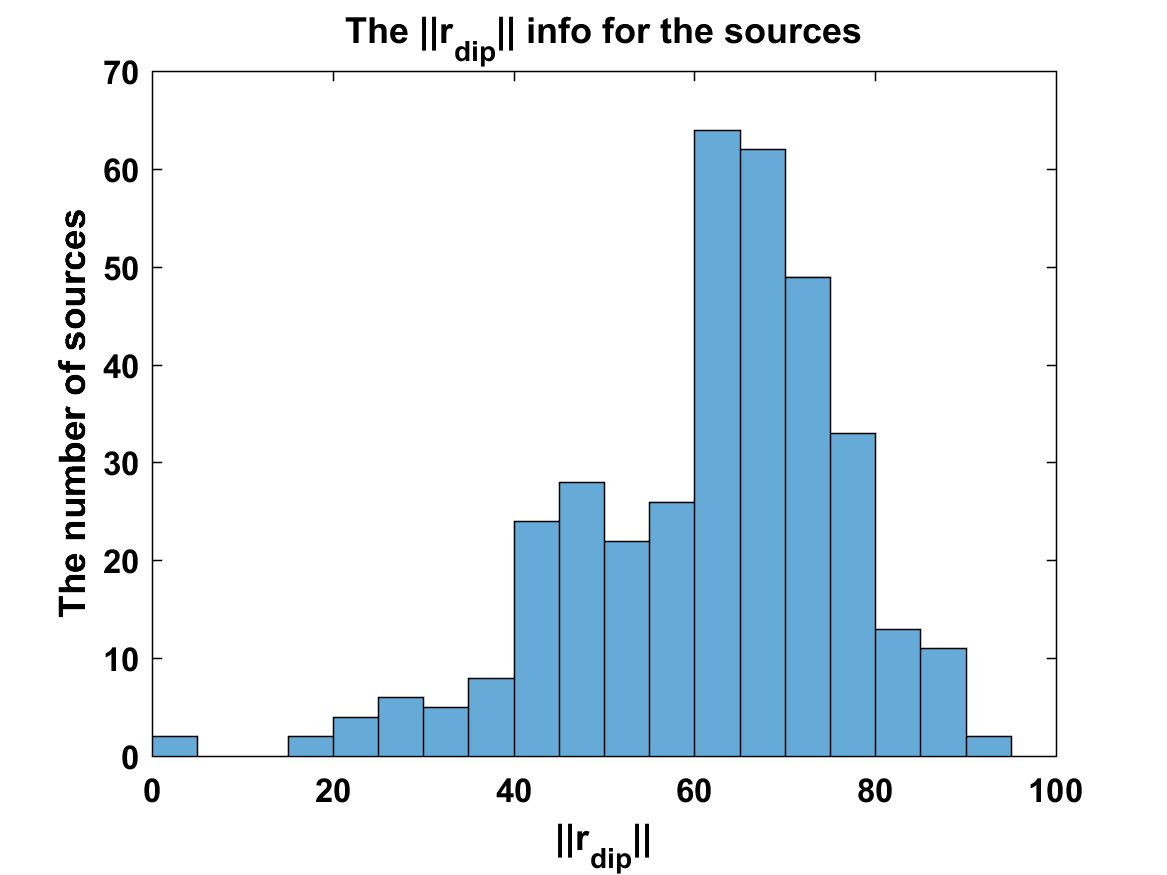}
		\hfill
		\caption{\textbf{TEST CASE-I}: The $\mathbf{r}_{dip}$ (in mm) of all the possibilities (360 in total). }
		\label{mcstc1}
	\end{figure}

	\textit{TEST-CASE I}: Thirty sources with random orientations are chosen. No constraint on the position or orientation of the source is applied. The source can be a deep or shallow one. The results are shown in Table. \ref{mts1}. The source information is given in Fig. \ref{mcstc1}. The results are depicted in Tables \ref{mts1}, \ref{mts1v} and \ref{mts1c}.
	
	\begin{table*}[!h]  
		\caption{\textbf{TEST-CASE I}: The \textit{A' metric} values observed for 5K (left) - and 2K (right) source space. }  
		\label{mts1} 
		\resizebox{.45\textwidth}{!}{
			
			\begin{tabular}{ |l|l|l|l|l|l|l|l| }
				\hline
				\multirow{2}{*}{\textbf{S.No.}} & \multirow{2}{*}{\textbf{Method}} & \multicolumn{3}{c|}{\textbf{without CARSS}}                                                                              & \multicolumn{3}{c|}{\textbf{with CARSS}}                                                                                 \\ \cline{3-8} 
				&                                  & \multicolumn{1}{c|}{\textit{Non}} & \multicolumn{1}{c|}{\textit{p - 1}} & \multicolumn{1}{c|}{\textit{p - 4}} & \multicolumn{1}{c|}{\textit{Non}} & \multicolumn{1}{c|}{\textit{p - 1}} & \multicolumn{1}{c|}{\textit{p - 4}} \\ \hline
				1                               & FOCUSS                           & 0.92                                   & 0.60                                   & 0.44                                   & 0.99                                   & 0.78                                   & 0.55                                   \\ \hline
				2                               & MxNE                             & 0.62                                   & 0.58                                   & 0.46                                   & 0.86                                   & 0.82                                   & 0.76                                   \\ \hline
				3.                              & SBL WIPF                         & 0.60                                   & 0.56                                   & 0.47                                   & 0.90                                   & 0.78                                   & 0.65                                   \\ \hline
				4                               & SBL- Zhang                       & 0.61                                   & 0.57                                   & 0.47                                   & 0.85                                   & 0.79                                   & 0.76                                   \\ \hline
				5                               & sLORETA                          & 0.60                                   & 0.56                                   & 0.41                                   & 0.82                                   & 0.75                                   & 0.55                                   \\ \hline
		\end{tabular}      }
		\resizebox{.45\textwidth}{!}{
			
			\begin{tabular}{ |l|l|l|l|l|l|l|l|}
				\hline
				\multicolumn{1}{|c|}{\multirow{2}{*}{\textbf{S.No.}}} & \multicolumn{1}{c|}{\multirow{2}{*}{\textbf{Method}}} & \multicolumn{3}{c|}{\textbf{without CARSS}}                                                   & \multicolumn{3}{c|}{\textbf{with CARSS}}                                                      \\ \cline{3-8} 
				\multicolumn{1}{|c|}{}                                & \multicolumn{1}{c|}{}                                 & \multicolumn{1}{c|}{Non} & \multicolumn{1}{c|}{p - 1} & \multicolumn{1}{c|}{p - 4} & \multicolumn{1}{c|}{Non} & \multicolumn{1}{c|}{p - 1} & \multicolumn{1}{c|}{p - 4} \\ \hline
				1                                                     & FOCUSS                                                & 0.90                          & 0.69                          & 0.50                          & 0.98                          & 0.88                          & 0.71                          \\ \hline
				2                                                     & MxNE                                                  & 0.87                          & 0.76                          & 0.65                          & 0.97                          & 0.95                          & 0.89                          \\ \hline
				3.                                                    & SBL WIPF                                              & 0.85                          & 0.69                          & 0.55                          & 0.98                          & 0.90                          & 0.78                          \\ \hline
				4                                                     & SBL- Zhang                                            & 0.84                          & 0.70                          & 0.62                          & 0.98                          & 0.94                          & 0.90                          \\ \hline
				5                                                     & sLORETA                                               & 0.75                          & 0.65                          & 0.55                          & 0.90                          & 0.85                          & 0.77                          \\ \hline
		\end{tabular}        } 
		
	\end{table*}

	\begin{table*}[!h]
		\caption{\textbf{TEST-CASE I}: The 95\% Conf. Int. [upper limit, lower limit] of the \textit{A' metric} values observed for 5K (up) - and 2K (down) source space.}
		\label{mts1v} 
		\resizebox{.45\textwidth}{!}{
			\begin{tabular}{|l|l|l|l|l|l|l|l|}
				\hline
				\multicolumn{1}{|c|}{\multirow{2}{*}{\textbf{S.No.}}} & \multicolumn{1}{c|}{\multirow{2}{*}{\textbf{Method}}} & \multicolumn{3}{c|}{\textbf{without CARSS}}                                                                     & \multicolumn{3}{c|}{\textbf{with CARSS}}                                                                        \\ \cline{3-8} 
				\multicolumn{1}{|c|}{}                                & \multicolumn{1}{c|}{}                                 & \multicolumn{1}{c|}{\textit{Nonon}} & \multicolumn{1}{c|}{\textit{p - 1}} & \multicolumn{1}{c|}{\textit{p - 4}} & \multicolumn{1}{c|}{\textit{Nonon}} & \multicolumn{1}{c|}{\textit{p - 1}} & \multicolumn{1}{c|}{\textit{p - 4}} \\ \hline
				1.                                                    & FOCUSS                                                & +4 to -6                            & +8 to -7.5                          & +9 to -11                           & +2 to -3                            & +3.5 to -6                          & +7.3 to -2.6                        \\ \hline
				2.                                                    & MxNE                                                  & +3 to -5                            & + 4.8 to -5                         & +6.6 to -7                          & +1.8 to -2.5                        & +2.5 to -4.1                        & +3.35 to -2.73                      \\ \hline
				3.                                                    & SBL - Wipf                                            & +2 to -6                            & +3.6 to -4.52                       & + 5.9 to - 8.5                      & +1.7 to -1.6                        & +3.2 to -1.2                        & +4.2 to - 5.7                       \\ \hline
				4.                                                    & SBL -Zhang                                            & +2.5 to -3.5                        & + 3.5 to -3.9                       & +4.6 to -5.85                       & +1.8 to -2.0                        & + 3.3 to -2.95                      & +4.8 to -4.2                        \\ \hline
				5.                                                    & sLORETA                                               & +4.5 to -3.6                        & +5.2 to -3.2                        & +6.6 to -4.5                        & + 2.9 to -3.5                       & +4.4 to -4.2                        & +5.7 to -5.1                        \\ \hline
		\end{tabular}		} 
		\resizebox{.47\textwidth}{!}{
			\begin{tabular}{|l|l|l|l|l|l|l|l|}
				\hline
				\multicolumn{1}{|c|}{\multirow{2}{*}{\textbf{S.No.}}} & \multicolumn{1}{c|}{\multirow{2}{*}{\textbf{Method}}} & \multicolumn{3}{c|}{\textbf{without CARSS}}                                                                     & \multicolumn{3}{c|}{\textbf{with CARSS}}                                                                        \\ \cline{3-8} 
				\multicolumn{1}{|c|}{}                                & \multicolumn{1}{c|}{}                                 & \multicolumn{1}{c|}{\textit{Nonon}} & \multicolumn{1}{c|}{\textit{p - 1}} & \multicolumn{1}{c|}{\textit{p - 4}} & \multicolumn{1}{c|}{\textit{Nonon}} & \multicolumn{1}{c|}{\textit{p - 1}} & \multicolumn{1}{c|}{\textit{p - 4}} \\ \hline
				1.                                                    & FOCUSS                                                & +3.6 to -4                          & +7.5 to -8                          & +9.5 to -8.9                        & +2.05 to -2.3                       & +3 to -5.65                         & +8.9 to -7.7                        \\ \hline
				2.                                                    & MxNE                                                  & +2.5 to -3.4                        & + 4.05 to -3.65                     & +6.85 to -6.55                      & +1.1 to -1.75                       & +1.3 to -2.6                        & +2.85 to -2.25                      \\ \hline
				3.                                                    & SBL - Wipf                                            & +2.25 to -3.75                      & +3.20 to -4.20                      & + 5.5 to - 6.65                     & +1.8 to -1.45                       & +2.85 to -1.60                      & +3.95 to - 4.25                     \\ \hline
				4.                                                    & SBL -Zhang                                            & +2.85 to -3.15                      & + 3.95 to -3.50                     & +4.90 to -4.65                      & +1.15 to -2.0                       & +2.8 to -2.05                       & +4.25 to -4.85                      \\ \hline
				5.                                                    & sLORETA                                               & +3.5 to -3.9                        & +6.2 to -1.85                       & +6.15 to -5.5                       & + 2.15 to -2.35                     & +3.65 to -3.05                      & +4.55 to -4.35                      \\ \hline
			\end{tabular}
		}
	\end{table*}
	
	\begin{table}[]
		\resizebox{.95\textwidth}{!}{
			\caption{\textit{TEST CASE- I}: The no. of sources CARSS reduced to is presented}
			\label{mts1c}
			\begin{tabular}{|l|l|l|l|l|l|l|l|}
				\hline
				\multicolumn{1}{|c|}{\multirow{2}{*}{\textbf{S.No.}}} & \multicolumn{1}{c|}{\multirow{2}{*}{\textbf{Noise level}}} & \multicolumn{3}{c|}{\textbf{Five - K (15,000)}}                                                              & \multicolumn{3}{c|}{\textbf{Two - K (6, 000)}}                                                               \\ \cline{3-8} 
				\multicolumn{1}{|c|}{}                                & \multicolumn{1}{c|}{}                                      & \multicolumn{1}{c|}{\textit{Avg.}} & \multicolumn{1}{c|}{\textit{Max.}} & \multicolumn{1}{c|}{\textit{Min.}} & \multicolumn{1}{c|}{\textit{Avg.}} & \multicolumn{1}{c|}{\textit{Max.}} & \multicolumn{1}{c|}{\textit{Min.}} \\ \hline
				1.                                                    & No noise                                                   & 2531                               & 3725                               & 732                                & 1367                               & 1852                               & 990                                \\ \hline
				2.                                                    & pink - 1                                                   & 2916                               & 3678                               & 870                                & 1351                               & 1670                               & 576                                \\ \hline
				3.                                                    & pink - 4                                                   & 3373                               & 4415                               & 2191                               & 1335                               & 1852                               & 413                                \\ \hline
		\end{tabular}  }
	\end{table}

	\textit{TEST-CASE II}: Three sources with random orientation has been chosen . Thy are chosen  such that (i) $Ds(\mathbf{r}_i,\mathbf{r}_j) \ge 80, i, j = 1 \dots 30$, and (ii) $\mathbf{r}_i \ge 90, i = 1 \dots 30$. The information about the sources chosen is given in Table. \ref{ds2}. The total number of possibilities is 360. The results are depicted in Tables \ref{mts2}, \ref{mts2v} and \ref{mts2c}.
	
	\begin{table}[]
		\caption{The distance between each source $Ds(\mathbf{r}_i,\mathbf{r}_j)$ and their radius of curvature $\norm{r}_{dip}$}
		\label{ds2}
		\begin{tabular}{|l|l|l|l|l|}
			\hline
			\textit{\textbf{Sources}} & \textbf{$\norm{r}_{dip}$} & \textbf{S-I} & \textbf{S-II} & \textbf{S-III} \\ \hline
			\textit{S-I}              & 94.8                      & -            & 110.5         & 100.9          \\ \hline
			\textit{S-II}             & 116.7                     & 110.0        & -             & 120.5          \\ \hline
			\textit{S-III}            & 81.2                      & 100.9        & 110.5         & -              \\ \hline
		\end{tabular}
	\end{table}
	
	\begin{table*}
		\caption{{ \textbf{TEST-CASE II}: The \textit{A' metric} values observed for 5K (left) - and 2K (right) source space. }}   \label{mts2} 
		\resizebox{.45\textwidth}{!}{
			\begin{tabular}{|l|l|l|l|l|l|l|l|}
				\hline
				\multicolumn{1}{|c|}{\multirow{2}{*}{\textbf{S.No.}}} & \multicolumn{1}{c|}{\multirow{2}{*}{\textbf{Method}}} & \multicolumn{3}{c|}{\textbf{without CARSS}}                                                                              & \multicolumn{3}{c|}{\textbf{with CARSS}}                                                                                 \\ \cline{3-8} 
				\multicolumn{1}{|c|}{}                                & \multicolumn{1}{c|}{}                                 & \multicolumn{1}{c|}{\textit{Non}} & \multicolumn{1}{c|}{\textit{p - 1}} & \multicolumn{1}{c|}{\textit{p - 4}} & \multicolumn{1}{c|}{\textit{Non}} & \multicolumn{1}{c|}{\textit{p - 1}} & \multicolumn{1}{c|}{\textit{p - 4}} \\ \hline
				1                                                     & FOCUSS                                                & 0.55                                   & 0.10                                   & 0.05                                   & 0.82                                   & 0.26                                   & 0.10                                   \\ \hline
				2                                                     & MxNE                                                  & 0.42                                   & 0.30                                   & 0.21                                   & 0.61                                   & 0.50                                   & 0.35                                   \\ \hline
				3.                                                    & SBL WIPF                                              & 0.39                                   & 0.20                                   & 0.06                                   & 0.66                                   & 0.45                                   & 0.12                                   \\ \hline
				4                                                     & SBL- Zhang                                            & 0.31                                   & 0.30                                   & 0.15                                   & 0.55                                   & 0.45                                   & 0.24                                   \\ \hline
				5                                                     & sLORETA                                               & 0.26                                   & 0.22                                   & 0.12                                   & 0.45                                   & 0.27                                   & 0.14                                   \\ \hline
		\end{tabular} }
		\resizebox{.45\textwidth}{!}{
			\begin{tabular}{|l|l|l|l|l|l|l|l|}
				\hline
				\multicolumn{1}{|c|}{\multirow{2}{*}{\textbf{S.No.}}} & \multicolumn{1}{c|}{\multirow{2}{*}{\textbf{Method}}} & \multicolumn{3}{c|}{\textbf{without CARSS}}                                                   & \multicolumn{3}{c|}{\textbf{with CARSS}}                                                      \\ \cline{3-8} 
				\multicolumn{1}{|c|}{}                                & \multicolumn{1}{c|}{}                                 & \multicolumn{1}{c|}{Non} & \multicolumn{1}{c|}{p - 1} & \multicolumn{1}{c|}{p - 4} & \multicolumn{1}{c|}{Non} & \multicolumn{1}{c|}{p - 1} & \multicolumn{1}{c|}{p - 4} \\ \hline
				1                                                     & FOCUSS                                                & 0.65                          & 0.22                          & 0.10                          & 0.82                          & 0.29                          & 0.20                          \\ \hline
				2                                                     & MxNE                                                  & 0.51                          & 0.44                          & 0.38                          & 0.76                          & 0.72                          & 0.60                          \\ \hline
				3.                                                    & SBL WIPF                                              & 0.50                          & 0.29                          & 0.16                          & 0.80                          & 0.33                          & 0.21                          \\ \hline
				4                                                     & SBL- Zhang                                            & 0.45                          & 0.39                          & 0.20                          & 0.70                          & 0.59                          & 0.48                          \\ \hline
				5                                                     & sLORETA                                               & 0.39                          & 0.26                          & 0.16                          & 0.45                          & 0.21                          & 0.10                          \\ \hline
		\end{tabular} } \\
	\end{table*}
	
	\begin{table*}[h]
		\caption{\textbf{TEST-CASE II}: The 95\% Conf. Int. [upper limit, lower limit] of the \textit{A' metric} values observed for 5K (left) - and 2K (right) source space. The \% values are given in the table.}
		\label{mts2v} 
		\resizebox{.48\textwidth}{!}{
			\begin{tabular}{|l|l|l|l|l|l|l|l|}
				\hline
				\multicolumn{1}{|c|}{\multirow{2}{*}{\textbf{S.No.}}} & \multicolumn{1}{c|}{\multirow{2}{*}{\textbf{Method}}} & \multicolumn{3}{c|}{\textbf{without CARSS}}                                                                     & \multicolumn{3}{c|}{\textbf{with CARSS}}                                                                        \\ \cline{3-8} 
				\multicolumn{1}{|c|}{}                                & \multicolumn{1}{c|}{}                                 & \multicolumn{1}{c|}{\textit{Nonon}} & \multicolumn{1}{c|}{\textit{p - 1}} & \multicolumn{1}{c|}{\textit{p - 4}} & \multicolumn{1}{c|}{\textit{Nonon}} & \multicolumn{1}{c|}{\textit{p - 1}} & \multicolumn{1}{c|}{\textit{p - 4}} \\ \hline
				1.                                                    & FOCUSS                                                & 4 to -3                             & +14 to -16                          & +25 to -19                          & +3 to -3                            & + 7 to -11                          & + 15 to -14                         \\ \hline
				2.                                                    & MxNE                                                  & +3 to -4                            & + 5 to -3                           & + 7 to -5                           & + 2 to -5                           & +6 to -4                            & +5 to -6                            \\ \hline
				3.                                                    & SBL - Wipf                                            & +4 to -6                            & +6 to -5                            & +7 to - 5                           & +3 to -3                            & +4 to -2                            & +4 to -4                            \\ \hline
				4.                                                    & SBL -Zhang                                            & +4 to -6                            & + 5 to -3                           & +6 to -7                            & +3 to -3                            & +5 to -4                            & +6 to -5                            \\ \hline
				5.                                                    & sLORETA                                               & +5 to -5                            & +6 to -5                            & +7 to -7                            & +4 to -3                            & +5 to -4                            & +7 to -3                            \\ \hline
		\end{tabular}	}  
		\resizebox{.48\textwidth}{!}{
			\begin{tabular}{|l|l|l|l|l|l|l|l|}
				\hline
				\multicolumn{1}{|c|}{\multirow{2}{*}{\textbf{S.No.}}} & \multicolumn{1}{c|}{\multirow{2}{*}{\textbf{Method}}} & \multicolumn{3}{c|}{\textbf{without CARSS}}                                                                     & \multicolumn{3}{c|}{\textbf{with CARSS}}                                                                        \\ \cline{3-8} 
				\multicolumn{1}{|c|}{}                                & \multicolumn{1}{c|}{}                                 & \multicolumn{1}{c|}{\textit{Nonon}} & \multicolumn{1}{c|}{\textit{p - 1}} & \multicolumn{1}{c|}{\textit{p - 4}} & \multicolumn{1}{c|}{\textit{Nonon}} & \multicolumn{1}{c|}{\textit{p - 1}} & \multicolumn{1}{c|}{\textit{p - 4}} \\ \hline
				1.                                                    & FOCUSS                                                & 5 to -2                             & +9 to -11                           & +25 to -19                          & +4 to -3                            & + 8 to -10                          & + 17 to -12                         \\ \hline
				2.                                                    & MxNE                                                  & +3 to -3                            & + 4 to -3                           & + 6 to -5                           & + 2 to -5                           & +3 to -3                            & +4 to -6                            \\ \hline
				3.                                                    & SBL - Wipf                                            & +3 to -4                            & +5 to -5                            & +8 to - 6                           & +3 to -3                            & +6 to -4                            & +6 to -6                            \\ \hline
				4.                                                    & SBL -Zhang                                            & +3 to -4                            & + 4 to -4                           & +5 to -4                            & +3 to -3                            & +4 to -4                            & +5 to -5                            \\ \hline
				5.                                                    & sLORETA                                               & +4 to -4                            & +5 to -4                            & +6 to -5                            & +3 to -3                            & +4 to -4                            & +7 to -5                            \\ \hline
			\end{tabular}
			
		}
	\end{table*} 
	
	\begin{table}[]
		\caption{\textit{TEST CASE- II}: The table shows the no. of sources CARSS reduced the solution space is presented}
		\label{mts2c}
		\resizebox{.95\textwidth}{!}{
			\begin{tabular}{|l|l|l|l|l|l|l|l|}
				\hline
				\multicolumn{1}{|c|}{\multirow{2}{*}{\textbf{S.No.}}} & \multicolumn{1}{c|}{\multirow{2}{*}{\textbf{Noise level}}} & \multicolumn{3}{c|}{\textbf{Five - K (15,000)}}                                                              & \multicolumn{3}{c|}{\textbf{Two - K (6, 000)}}                                                               \\ \cline{3-8} 
				\multicolumn{1}{|c|}{}                                & \multicolumn{1}{c|}{}                                      & \multicolumn{1}{c|}{\textit{Avg.}} & \multicolumn{1}{c|}{\textit{Max.}} & \multicolumn{1}{c|}{\textit{Min.}} & \multicolumn{1}{c|}{\textit{Avg.}} & \multicolumn{1}{c|}{\textit{Max.}} & \multicolumn{1}{c|}{\textit{Min.}} \\ \hline
				1.                                                    & No noise                                                   & 3646                               & 4849                               & 1503                               & 1767                               & 2372                               & 958                                \\ \hline
				2.                                                    & pink - 1                                                   & 4248                               & 5870                               & 1735                               & 1955                               & 2925                               & 1281                               \\ \hline
				3.                                                    & pink - 4                                                   & 4478                               & 6280                               & 2051                               & 2255                               & 1590                               & 2925                               \\ \hline
		\end{tabular} }
	\end{table}
	
	\textit{TEST-CASE - III}: Five sources are chosen . Their conditions for selection are: (i) Three deep sources $\mathbf{r_{dip}} \le 60$ (ii) An intermittent source $80 \le \mathbf{r_{dip}} \le 60$ (iii) Distance between the sources $Ds(\mathbf{r}_i,\mathbf{r}_j) \ge 70$, and a (iv) shallow source $\mathbf{r_{dip}} \ge 70$.  The information about the sources chosen is given in \ref{ds3}. The results are depicted in Tables \ref{mts3}, \ref{mts3v} and \ref{mts3c}.
	
	\begin{table}[]
		\caption{The distance between each source $Ds(\mathbf{r}_i,\mathbf{r}_j)$ and their radius of curvature $\norm{r}_{dip}$}
		\label{ds3}
		\begin{tabular}{|l|l|l|l|l|l|l|}
			\hline
			S.No. & $\norm{r_{dip}}$ & S-I & S-II & S-III & S-IV & S-V   \\ \hline
			S-I   & 39.7             & -   & 72.2 & 100.8 & 81.2 & 110.0 \\ \hline
			S-II  & 44.8             &     & -    & 69.8  & 92.6 & 102.5 \\ \hline
			S-III & 62.6             &     &      & -     & 96.6 & 76.2  \\ \hline
			S-IV  & 74.9             &     &      &       & -    & 122.1 \\ \hline
			S-V   & 90.6             &     &      &       &      & -     \\ \hline
		\end{tabular}
	\end{table}

	\begin{table*}
		\caption{{ \textbf{TEST-CASE III}: The \textit{A' metric} values observed for 5K (left) - and 2K (right) source space. } }  \label{mts3} 
		
		\resizebox{.45\textwidth}{!}{
			\begin{tabular}{|l|l|l|l|l|l|l|l|}
				\hline
				\multicolumn{1}{|c|}{\multirow{2}{*}{\textbf{S.No.}}} & \multicolumn{1}{c|}{\multirow{2}{*}{\textbf{Method}}} & \multicolumn{3}{c|}{\textbf{without CARSS}}                                                                              & \multicolumn{3}{c|}{\textbf{with CARSS}}                                                                                 \\ \cline{3-8} 
				\multicolumn{1}{|c|}{}                                & \multicolumn{1}{c|}{}                                 & \multicolumn{1}{c|}{\textit{Non}} & \multicolumn{1}{c|}{\textit{p - 1}} & \multicolumn{1}{c|}{\textit{p - 4}} & \multicolumn{1}{c|}{\textit{Non}} & \multicolumn{1}{c|}{\textit{p - 1}} & \multicolumn{1}{c|}{\textit{p - 4}} \\ \hline
				1                                                     & FOCUSS                                                & 0.20                                   & 0.01                                   & 0.01                                   & 0.32                                   & 0.05                                   & 0.01                                   \\ \hline
				2                                                     & MxNE                                                  & 0.14                                   & 0.10                                   & 0.05                                   & 0.22                                   & 0.14                                   & 0.09                                   \\ \hline
				3.                                                    & SBL WIPF                                              & 0.10                                   & 0.02                                   & 0.01                                   & 0.25                                   & 0.09                                   & 0.05                                   \\ \hline
				4                                                     & SBL- Zhang                                            & 0.19                                   & 0.11                                   & 0.05                                   & 0.18                                   & 0.13                                   & 0.10                                   \\ \hline
				5                                                     & sLORETA                                               & 0.15                                   & 0.08                                   & 0.04                                   & 0.17                                   & 0.10                                   & 0.08                                   \\ \hline
		\end{tabular}	}
		\resizebox{.45\textwidth}{!}{
			\begin{tabular}{|l|l|l|l|l|l|l|l|}
				\hline
				\multicolumn{1}{|c|}{\multirow{2}{*}{\textbf{S.No.}}} & \multicolumn{1}{c|}{\multirow{2}{*}{\textbf{Method}}} & \multicolumn{3}{c|}{\textbf{without CARSS}}                                                   & \multicolumn{3}{c|}{\textbf{with CARSS}}                                                      \\ \cline{3-8} 
				\multicolumn{1}{|c|}{}                                & \multicolumn{1}{c|}{}                                 & \multicolumn{1}{c|}{Non} & \multicolumn{1}{c|}{p - 1} & \multicolumn{1}{c|}{p - 4} & \multicolumn{1}{c|}{Non} & \multicolumn{1}{c|}{p - 1} & \multicolumn{1}{c|}{p - 4} \\ \hline
				1                                                     & FOCUSS                                                & 0.37                          & 0.01                          & 0.01                          & 0.39                          & 0.05                          & 0.01                          \\ \hline
				2                                                     & MxNE                                                  & 0.26                          & 0.22                          & 0.16                          & 0.33                          & 0.26                          & 0.17                          \\ \hline
				3.                                                    & SBL WIPF                                              & 0.22                          & 0.12                          & 0.01                          & 0.26                          & 0.15                          & 0.02                          \\ \hline
				4                                                     & SBL- Zhang                                            & 0.23                          & 0.18                          & 0.15                          & 0.37                          & 0.30                          & 0.21                          \\ \hline
				5                                                     & sLORETA                                               & 0.17                          & 0.11                          & 0.06                          & 0.21                          & 0.16                          & 0.11                          \\ \hline
		\end{tabular} } \\
	\end{table*}
	
	\begin{table*}[h]
		\caption{\textbf{TEST-CASE III}: The 95\% Conf. Int. [upper limit, lower limit] of the \textit{A' metric} values observed for 5K (left) - and 2K (right) source space. The \% values are given in the table.}
		\label{mts3v} 
		\resizebox{.48\textwidth}{!}{
			\begin{tabular}{|l|l|l|l|l|l|l|l|}
				\hline
				\multicolumn{1}{|c|}{\multirow{2}{*}{\textbf{S.No.}}} & \multicolumn{1}{c|}{\multirow{2}{*}{\textbf{Method}}} & \multicolumn{3}{c|}{\textbf{without CARSS}}                                                                     & \multicolumn{3}{c|}{\textbf{with CARSS}}                                                                        \\ \cline{3-8} 
				\multicolumn{1}{|c|}{}                                & \multicolumn{1}{c|}{}                                 & \multicolumn{1}{c|}{\textit{Nonon}} & \multicolumn{1}{c|}{\textit{p - 1}} & \multicolumn{1}{c|}{\textit{p - 4}} & \multicolumn{1}{c|}{\textit{Nonon}} & \multicolumn{1}{c|}{\textit{p - 1}} & \multicolumn{1}{c|}{\textit{p - 4}} \\ \hline
				1.                                                    & FOCUSS                                                & 7 to -5                             & +22 to -20                          & +35 to -49                          & +6 to -3                            & + 17 to -14                         & + 30 to -15                         \\ \hline
				2.                                                    & MxNE                                                  & +5 to -4                            & + 8 to -7                           & + 9 to -9                           & + 4 to -6                           & +6 to -7                            & +6 to -8                            \\ \hline
				3.                                                    & SBL - Wipf                                            & +7 to -4                            & +9 to -7                            & + 10 to - 9                         & +4 to -5                            & +6 to -7                            & +8 to -8                            \\ \hline
				4.                                                    & SBL -Zhang                                            & +7 to -5                            & + 8 to -6                           & +7 to -8                            & +6 to -5                            & +6 to -6                            & +7 to -6                            \\ \hline
				5.                                                    & sLORETA                                               & +9 to -5                            & +8 to -8                            & +7 to -10                           & +4 to -4                            & +6 to -7                            & +7 to -7                            \\ \hline
			\end{tabular}
		}
		\resizebox{.48\textwidth}{!}{
			\begin{tabular}{|l|l|l|l|l|l|l|l|}
				\hline
				\multicolumn{1}{|c|}{\multirow{2}{*}{\textbf{S.No.}}} & \multicolumn{1}{c|}{\multirow{2}{*}{\textbf{Method}}} & \multicolumn{3}{c|}{\textbf{without CARSS}}                                                                     & \multicolumn{3}{c|}{\textbf{with CARSS}}                                                                        \\ \cline{3-8} 
				\multicolumn{1}{|c|}{}                                & \multicolumn{1}{c|}{}                                 & \multicolumn{1}{c|}{\textit{Nonon}} & \multicolumn{1}{c|}{\textit{p - 1}} & \multicolumn{1}{c|}{\textit{p - 4}} & \multicolumn{1}{c|}{\textit{Nonon}} & \multicolumn{1}{c|}{\textit{p - 1}} & \multicolumn{1}{c|}{\textit{p - 4}} \\ \hline
				1.                                                    & FOCUSS                                                & 6 to 5                              & +20 to -16                          & +30 to -36                          & +5 to -5                            & + 16 to -11                         & + 25 to -16                         \\ \hline
				2.                                                    & MxNE                                                  & +4 to -3                            & + 7 to -5                           & + 6 to -7                           & + 4 to -2                           & +6 to -8                            & +5 to -9                            \\ \hline
				3.                                                    & SBL - Wipf                                            & +6 to -3                            & +9 to -6                            & + 10 to - 10                        & +3 to -4                            & +6 to -5                            & +8 to -9                            \\ \hline
				4.                                                    & SBL -Zhang                                            & +6 to -4                            & + 8 to -5                           & +8 to -8                            & +6 to -6                            & +5 to -6                            & +8 to -6                            \\ \hline
				5.                                                    & sLORETA                                               & +7 to -7                            & +8 to -8                            & +10 to -10                          & +5 to -5                            & +6 to -7                            & +7 to -7                            \\ \hline
			\end{tabular}

		}
	\end{table*}

	\begin{table}[]
		\caption{\textit{TEST CASE- III}: The no. of sources CARSS reduced to is presented}
		\label{mts3c}
		\resizebox{.95\textwidth}{!}{
			\begin{tabular}{|l|l|l|l|l|l|l|l|}
				\hline
				\multicolumn{1}{|c|}{\multirow{2}{*}{\textbf{S.No.}}} & \multicolumn{1}{c|}{\multirow{2}{*}{\textbf{Noise level}}} & \multicolumn{3}{c|}{\textbf{Five - K (15,000)}}                                                              & \multicolumn{3}{c|}{\textbf{Two - K (6, 000)}}                                                               \\ \cline{3-8} 
				\multicolumn{1}{|c|}{}                                & \multicolumn{1}{c|}{}                                      & \multicolumn{1}{c|}{\textit{Avg.}} & \multicolumn{1}{c|}{\textit{Max.}} & \multicolumn{1}{c|}{\textit{Min.}} & \multicolumn{1}{c|}{\textit{Avg.}} & \multicolumn{1}{c|}{\textit{Max.}} & \multicolumn{1}{c|}{\textit{Min.}} \\ \hline
				1.                                                    & No noise                                                   & 4745                               & 6961                               & 2598                               & 2231                               & 2781                               & 1713                               \\ \hline
				2.                                                    & pink - 1                                                   & 5284                               & 7656                               & 2691                               & 2381                               & 3508                               & 1685                               \\ \hline
				3.                                                    & pink - 4                                                   & 5417                               & 7331                               & 3379                               & 2603                               & 3601                               & 1145                               \\ \hline
		\end{tabular}  }
	\end{table}
	
	\textit{TEST-CASE IV}: Seven sources are chosen . Their conditions for selection are: (i) Four deep sources distance between them is $Ds(\mathbf{r}_i,\mathbf{r}_j) \le 55$ (ii) An intermittent source $60 \le \mathbf{r_{dip}} \le 70$ , and (iii) two shallow source $\mathbf{r_{dip}} \ge 70$. The information about the sources is given in Table. \ref{dts4}. The results are depicted in Tables \ref{mts4}, \ref{mts4v} and \ref{mts4c}.
	
	\begin{table}[]
		\caption{The distance between S-I and each source $Ds(\mathbf{r}_i,\mathbf{r}_j)$ and their radius of curvature $\norm{r}_{dip}$}
		\label{dts4}
		\resizebox{.95\textwidth}{!}{
			\begin{tabular}{|l|l|l|l|l|l|l|l|}
				\hline
				\textbf{S.No.}   & \textbf{S-I} & \textbf{S-II} & \textbf{S-III} & \textbf{S-IV} & \textbf{S-V} & \textbf{S-VI} & \textbf{S-VII} \\ \hline
				S-I              &              & 49.2          & 50.8           & 53.8          & 69.7         & 76.9          & 75.9           \\ \hline
				$\norm{r_{dip}}$ & 36.2         & 46.7          & 59.5           & 66.4          & 83.7         & 92.8          & 96.8           \\ \hline
		\end{tabular}  }
	\end{table}
	
	\begin{table*}
		\caption{{ \textbf{TEST-CASE IV}: The \textit{A' metric} values observed for 5K (left) - and 2K (right) source space. }}  \label{mts4} 
		\resizebox{.45\textwidth}{!}{
			\begin{tabular}{|l|l|l|l|l|l|l|l|}
				\hline
				\multicolumn{1}{|c|}{\multirow{2}{*}{\textbf{S.No.}}} & \multicolumn{1}{c|}{\multirow{2}{*}{\textbf{Method}}} & \multicolumn{3}{c|}{\textbf{without CARSS}}                                                                              & \multicolumn{3}{c|}{\textbf{with CARSS}}                                                                                 \\ \cline{3-8} 
				\multicolumn{1}{|c|}{}                                & \multicolumn{1}{c|}{}                                 & \multicolumn{1}{c|}{\textit{Non}} & \multicolumn{1}{c|}{\textit{p - 1}} & \multicolumn{1}{c|}{\textit{p - 4}} & \multicolumn{1}{c|}{\textit{Non}} & \multicolumn{1}{c|}{\textit{p - 1}} & \multicolumn{1}{c|}{\textit{p - 4}} \\ \hline
				1                                                     & FOCUSS                                                & 0.0500                                 & 0.0001                                 & 0.0001                                 & 0.3200                                 & 0.0006                                 & 0.0001                                 \\ \hline
				2                                                     & MxNE                                                  & 0.0900                                 & 0.0100                                 & 0.0040                                 & 0.1100                                 & 0.0400                                 & 0.0160                                 \\ \hline
				3.                                                    & SBL WIPF                                              & 0.0700                                 & 0.0001                                 & 0.0001                                 & 0.1200                                 & 0.0002                                 & 0.0001                                 \\ \hline
				4                                                     & SBL- Zhang                                            & 0.0600                                 & 0.0080                                 & 0.0010                                 & 0.1000                                 & 0.0500                                 & 0.0080                                 \\ \hline
				5                                                     & sLORETA                                               & 0.0020                                 & 0.0005                                 & 0.0001                                 & 0.0800                                 & 0.0100                                 & 0.0005                                 \\ \hline
		\end{tabular}				}
		\resizebox{.45\textwidth}{!}{ 
			\begin{tabular}{|l|l|l|l|l|l|l|l|}
				\hline
				\multicolumn{1}{|c|}{\multirow{2}{*}{\textbf{S.No.}}} & \multicolumn{1}{c|}{\multirow{2}{*}{\textbf{Method}}} & \multicolumn{3}{c|}{\textbf{without CARSS}}                                                   & \multicolumn{3}{c|}{\textbf{with CARSS}}                                                      \\ \cline{3-8} 
				\multicolumn{1}{|c|}{}                                & \multicolumn{1}{c|}{}                                 & \multicolumn{1}{c|}{Non} & \multicolumn{1}{c|}{p - 1} & \multicolumn{1}{c|}{p - 4} & \multicolumn{1}{c|}{Non} & \multicolumn{1}{c|}{p - 1} & \multicolumn{1}{c|}{p - 4} \\ \hline
				1                                                     & FOCUSS                                                & 0.1000                        & 0.0001                        & 0.0001                        & 0.3200                        & 0.0008                        & 0.0001                        \\ \hline
				2                                                     & MxNE                                                  & 0.1800                        & 0.0600                        & 0.0200                        & 0.1900                        & 0.0800                        & 0.0250                        \\ \hline
				3.                                                    & SBL WIPF                                              & 0.1100                        & 0.0001                        & 0.0001                        & 0.1300                        & 0.0003                        & 0.0001                        \\ \hline
				4                                                     & SBL- Zhang                                            & 0.1000                        & 0.0600                        & 0.0300                        & 0.1000                        & 0.0900                        & 0.0060                        \\ \hline
				5                                                     & sLORETA                                               & 0.0020                        & 0.0010                        & 0.0001                        & 0.0030                        & 0.0100                        & 0.0005                        \\ \hline
		\end{tabular} } \\
	\end{table*}
	
	\begin{table*}[h]
		\caption{\textbf{TEST-CASE IV}: The 95\% Conf. Int. [upper limit, lower limit] of the \textit{A' metric} values observed around mean for 5K (left) - and 2K (right) source space. The \% values are given in the table.}
		\label{mts4v} 
		\resizebox{.48\textwidth}{!}{
			\begin{tabular}{|l|l|l|l|l|l|l|l|}
				\hline
				\multicolumn{1}{|c|}{\multirow{2}{*}{\textbf{S.No.}}} & \multicolumn{1}{c|}{\multirow{2}{*}{\textbf{Method}}} & \multicolumn{3}{c|}{\textbf{without CARSS}}                                                                     & \multicolumn{3}{c|}{\textbf{with CARSS}}                                                                        \\ \cline{3-8} 
				\multicolumn{1}{|c|}{}                                & \multicolumn{1}{c|}{}                                 & \multicolumn{1}{c|}{\textit{Nonon}} & \multicolumn{1}{c|}{\textit{p - 1}} & \multicolumn{1}{c|}{\textit{p - 4}} & \multicolumn{1}{c|}{\textit{Nonon}} & \multicolumn{1}{c|}{\textit{p - 1}} & \multicolumn{1}{c|}{\textit{p - 4}} \\ \hline
				1.                                                    & FOCUSS                                                & 8 to 10                             & +45 to -40                          & +46 to -46                          & +7 to -9                            & + 40 to -40                         & + 45 to -38                         \\ \hline
				2.                                                    & MxNE                                                  & +8 to -9                            & + 10 to -11                         & + 10 to -14                         & + 8 to -8                           & +10 to -12                          & +9 to -14                           \\ \hline
				3.                                                    & SBL - Wipf                                            & +8 to -6                            & +11 to -12                          & + 14 to - 15                        & +8 to -7                            & +10 to -11                          & + 15 to - 15                        \\ \hline
				4.                                                    & SBL -Zhang                                            & +8 to -8                            & + 10 to -12                         & +10 to -15                          & +8 to -8                            & + 10 to -10                         & +13 to -12                          \\ \hline
				5.                                                    & sLORETA                                               & +7 to -10                           & +12 to -12                          & +12 to -14                          & +7 to -11                           & +10 to -12                          & +10 to -14                          \\ \hline
		\end{tabular}   }
		\resizebox{.48\textwidth}{!}{
			\begin{tabular}{|l|l|l|l|l|l|l|l|}
				\hline
				\multicolumn{1}{|c|}{\multirow{2}{*}{\textbf{S.No.}}} & \multicolumn{1}{c|}{\multirow{2}{*}{\textbf{Method}}} & \multicolumn{3}{c|}{\textbf{without CARSS}}                                                                     & \multicolumn{3}{c|}{\textbf{with CARSS}}                                                                        \\ \cline{3-8} 
				\multicolumn{1}{|c|}{}                                & \multicolumn{1}{c|}{}                                 & \multicolumn{1}{c|}{\textit{Nonon}} & \multicolumn{1}{c|}{\textit{p - 1}} & \multicolumn{1}{c|}{\textit{p - 4}} & \multicolumn{1}{c|}{\textit{Nonon}} & \multicolumn{1}{c|}{\textit{p - 1}} & \multicolumn{1}{c|}{\textit{p - 4}} \\ \hline
				1.                                                    & FOCUSS                                                & +9 to -9                            & +45 to -35                          & +45 to -45                          & +7 to -7                            & + 40 to -40                         & + 45 to -45                         \\ \hline
				2.                                                    & MxNE                                                  & +7 to -9                            & + 9 to -10                          & + 10 to -11                         & + 7 to -7                           & +9 to -9                            & +10 to -11                          \\ \hline
				3.                                                    & SBL - Wipf                                            & +6 to -7                            & +10 to -10                          & + 15 to - 15                        & +8 to -8                            & +10 to -10                          & + 14 to - 14                        \\ \hline
				4.                                                    & SBL -Zhang                                            & +8 to -8                            & + 9 to -9                           & +12 to -12                          & +7 to -7                            & + 8 to -8                           & +11 to -11                          \\ \hline
				5.                                                    & sLORETA                                               & +10 to -10                          & +12 to -12                          & +16 to -16                          & + 8 to -8                           & +10 to -10                          & +15 to -15                          \\ \hline
			\end{tabular}
		}
	\end{table*}

	\begin{table}[]
		\caption{\textit{TEST CASE- IV}: The no. of sources CARSS reduced to is presented}
		\label{mts4c}
		\resizebox{0.95\textwidth}{!}{
			\begin{tabular}{|l|l|l|l|l|l|l|l|}
				\hline
				\multicolumn{1}{|c|}{\multirow{2}{*}{\textbf{S.No.}}} & \multicolumn{1}{c|}{\multirow{2}{*}{\textbf{Noise level}}} & \multicolumn{3}{c|}{\textbf{Five - K (15,000)}}                                                              & \multicolumn{3}{c|}{\textbf{Two - K (6, 000)}}                                                               \\ \cline{3-8} 
				\multicolumn{1}{|c|}{}                                & \multicolumn{1}{c|}{}                                      & \multicolumn{1}{c|}{\textit{Avg.}} & \multicolumn{1}{c|}{\textit{Max.}} & \multicolumn{1}{c|}{\textit{Min.}} & \multicolumn{1}{c|}{\textit{Avg.}} & \multicolumn{1}{c|}{\textit{Max.}} & \multicolumn{1}{c|}{\textit{Min.}} \\ \hline
				1.                                                    & No noise                                                   & 5929                               & 2952                               & 6327                               & 1826                               & 319                                & 2260                               \\ \hline
				2.                                                    & pink - 1                                                   & 6117                               & 2997                               & 7571                               & 2125                               & 1080                               & 2284                               \\ \hline
				3.                                                    & pink - 4                                                   & .7963                              & 3377                               & 9813                               & 4117                               & 2952                               & 5335                               \\ \hline
		\end{tabular}  }
	\end{table}
	
	\begin{table*}[]
		\caption{The success rate (SR) for the sources in all the test cases in the 5K-SS. The three values in each cell designate the case when no-noise, pink noise of amplitude one and four are considered. 'deep' indicates deep sources, 'int.' indicates intermittent sources and 'shall.' indicates shallow sources}
		\label{sr1}
		\resizebox{.98\textwidth}{!}{
			\begin{tabular}{|l|l|l|l|l|l|l|l|l|l|l|l|l|l|l|l|l|}
				\hline
				\multicolumn{1}{|c|}{\textbf{\begin{tabular}[c]{@{}c@{}}TEST\\ CASE\end{tabular}}}          & \multicolumn{2}{c|}{\textbf{TC-1}}                                                                                                                                            & \multicolumn{2}{c|}{\textbf{TC-2}}                                                                                                                                                          & \multicolumn{6}{c|}{\textbf{TC-3}}                                                                                                                                                                                                                                                                                                                                                                                                                                                                                                                            & \multicolumn{6}{c|}{\textbf{TC-4}}                                                                                                                                                                                                                                                                                                                                                                                                                                                                                                                          \\ \hline
				\multicolumn{1}{|c|}{\textit{\begin{tabular}[c]{@{}c@{}}ACTIVTED\\ SOURCE(S)\end{tabular}}} & \multicolumn{1}{c|}{\textit{\begin{tabular}[c]{@{}c@{}}S-I \\ {[}30{]}\end{tabular}}} & \multicolumn{1}{c|}{\textit{\begin{tabular}[c]{@{}c@{}}S-I \\ {[}30{]}\end{tabular}}} & \multicolumn{1}{c|}{\textit{\begin{tabular}[c]{@{}c@{}}S-I,II,III \\ {[}90{]}\end{tabular}}} & \multicolumn{1}{c|}{\textit{\begin{tabular}[c]{@{}c@{}}S-I,II,III \\ {[}90{]}\end{tabular}}} & \multicolumn{1}{c|}{\textit{\begin{tabular}[c]{@{}c@{}}3 deep\\ {[}30{]}\end{tabular}}} & \multicolumn{1}{c|}{\textit{\begin{tabular}[c]{@{}c@{}}1 int.\\ {[}10{]}\end{tabular}}} & \multicolumn{1}{c|}{\textit{\begin{tabular}[c]{@{}c@{}}1 shall.\\ {[}10{]}\end{tabular}}} & \multicolumn{1}{c|}{\textit{\begin{tabular}[c]{@{}c@{}}3 deep\\ {[}30{]}\end{tabular}}} & \multicolumn{1}{c|}{\textit{\begin{tabular}[c]{@{}c@{}}1 int.\\ {[}10{]}\end{tabular}}} & \multicolumn{1}{c|}{\textit{\begin{tabular}[c]{@{}c@{}}1 shall.\\ {[}10{]}\end{tabular}}} & \multicolumn{1}{c|}{\textit{\begin{tabular}[c]{@{}c@{}}4 deep\\ {[}40{]}\end{tabular}}} & \multicolumn{1}{c|}{\textit{\begin{tabular}[c]{@{}c@{}}1 int\\ {[}10{]}\end{tabular}}} & \multicolumn{1}{c|}{\textit{\begin{tabular}[c]{@{}c@{}}2 shall.\\ {[}20{]}\end{tabular}}} & \multicolumn{1}{c|}{\textit{\begin{tabular}[c]{@{}c@{}}4 deep\\ {[}40{]}\end{tabular}}} & \multicolumn{1}{c|}{\textit{\begin{tabular}[c]{@{}c@{}}1 int\\ {[}10{]}\end{tabular}}} & \multicolumn{1}{c|}{\textit{\begin{tabular}[c]{@{}c@{}}2 shall.\\ {[}20{]}\end{tabular}}} \\ \hline
				\multicolumn{1}{|c|}{}                                                                      & \multicolumn{1}{c|}{\begin{tabular}[c]{@{}c@{}}without\\ CARSS\end{tabular}}          & \multicolumn{1}{c|}{\begin{tabular}[c]{@{}c@{}}with\\ CARSS\end{tabular}}             & \multicolumn{1}{c|}{\begin{tabular}[c]{@{}c@{}}without\\ CARSS\end{tabular}}                 & \multicolumn{1}{c|}{\begin{tabular}[c]{@{}c@{}}with\\ CARSS\end{tabular}}                    & \multicolumn{3}{c|}{\begin{tabular}[c]{@{}c@{}}without\\ CARSS\end{tabular}}                                                                                                                                                                                                  & \multicolumn{3}{c|}{\begin{tabular}[c]{@{}c@{}}with\\ CARSS\end{tabular}}                                                                                                                                                                                                     & \multicolumn{3}{c|}{\begin{tabular}[c]{@{}c@{}}without\\ CARSS\end{tabular}}                                                                                                                                                                                                 & \multicolumn{3}{c|}{\begin{tabular}[c]{@{}c@{}}with\\ CARSS\end{tabular}}                                                                                                                                                                                                    \\ \hline
				FOCUSS                                                                                      & 30,29,29                                                                              & 30,30,30                                                                              & 85,78,72                                                                                     & 90,86,80                                                                                     & 19,6,1                                                                                  & 8,2,2                                                                                   & 9,6,4                                                                                     & 21,7,2                                                                                  & 9,1,2                                                                                   & 10,7,6                                                                                    & 5,0,0                                                                                   & 6,1,1                                                                                  & 15,6,1                                                                                    & 6,0,0                                                                                   & 7,1,1                                                                                  & 17,8,7                                                                                    \\ \hline
				MxNE                                                                                        & 30,30,30                                                                              & 30,30,30                                                                              & 84,85,78                                                                                     & 90,90,89                                                                                     & 20,16,12                                                                                & 7,6,5                                                                                   & 9,8,8                                                                                     & 22,18,16                                                                                & 10, 8,7                                                                                 & 10,10,9                                                                                   & 8,5,2                                                                                   & 7,6,4                                                                                  & 14,11,9                                                                                   & 9,8,6                                                                                   & 8,7,6                                                                                  & 18, 15,14                                                                                 \\ \hline
				SBL - Wipf                                                                                  & 30,30,30                                                                              & 30,30,30                                                                              & 89,82,76                                                                                     & 90,82,81                                                                                     & 21,13,9                                                                                 & 9,6,4                                                                                   & 9,8,7                                                                                     & 25,16,11                                                                                & 10,6,4                                                                                  & 10,8,8                                                                                    & 6,2,0                                                                                   & 6,4,2                                                                                  & 13,7,5                                                                                    & 5,3,1                                                                                   & 7,6,4                                                                                  & 14,12,10                                                                                  \\ \hline
				SBL - Zhang                                                                                 & 30,30,30                                                                              & 30,30,30                                                                              & 88,82,78                                                                                     & 90,88,86                                                                                     & 19,15,12                                                                                & 8,7,6                                                                                   & 9,9,7                                                                                     & 23,19,16                                                                                & 10,9,8                                                                                  & 10,10,9                                                                                   & 7,6,6                                                                                   & 7,7,5                                                                                  & 14,11,8                                                                                   & 9,7,7                                                                                   & 8,6,6                                                                                  & 16,15,13                                                                                  \\ \hline
				sLORETA                                                                                     & 30,29,29                                                                              & 30,30,30                                                                              & 85,80,76                                                                                     & 90,86,78                                                                                     & 15,11,8                                                                                 & 7,6,5                                                                                   & 8,7,6                                                                                     & 18,14,10                                                                                & 8,7,6                                                                                   & 10,9,8                                                                                    & 5,4,2                                                                                   & 6,4,4                                                                                  & 11,9,8                                                                                    & 5,5,5                                                                                   & 8,5,5                                                                                  & 12,10,8                                                                                   \\ \hline
		\end{tabular} }
	\end{table*}
	
	\begin{table*}[]
		\caption{The success rate (SR) for the sources in all the test cases in the 2K-SS. The three values in each cell designate the case when no-noise, pink noise of amplitude one and four are considered.  'deep' indicates deep sources, 'int.' indicates intermittent sources and 'shall.' indicates shallow sources}
		\label{sr2}
		\resizebox{.98\textwidth}{!}{
			\begin{tabular}{|l|l|l|l|l|l|l|l|l|l|l|l|l|l|l|l|l|}
				\hline
				\multicolumn{1}{|c|}{\textbf{\begin{tabular}[c]{@{}c@{}}TEST\\ CASE\end{tabular}}}          & \multicolumn{2}{c|}{\textbf{TC-1}}                                                                                                                                            & \multicolumn{2}{c|}{\textbf{TC-2}}                                                                                                                                                          & \multicolumn{6}{c|}{\textbf{TC-3}}                                                                                                                                                                                                                                                                                                                                                                                                                                                                                                                            & \multicolumn{6}{c|}{\textbf{TC-4}}                                                                                                                                                                                                                                                                                                                                                                                                                                                                                                                          \\ \hline
				\multicolumn{1}{|c|}{\textit{\begin{tabular}[c]{@{}c@{}}ACTIVTED\\ SOURCE(S)\end{tabular}}} & \multicolumn{1}{c|}{\textit{\begin{tabular}[c]{@{}c@{}}S-I \\ {[}30{]}\end{tabular}}} & \multicolumn{1}{c|}{\textit{\begin{tabular}[c]{@{}c@{}}S-I \\ {[}30{]}\end{tabular}}} & \multicolumn{1}{c|}{\textit{\begin{tabular}[c]{@{}c@{}}S-I,II,III \\ {[}90{]}\end{tabular}}} & \multicolumn{1}{c|}{\textit{\begin{tabular}[c]{@{}c@{}}S-I,II,III \\ {[}90{]}\end{tabular}}} & \multicolumn{1}{c|}{\textit{\begin{tabular}[c]{@{}c@{}}3 deep\\ {[}30{]}\end{tabular}}} & \multicolumn{1}{c|}{\textit{\begin{tabular}[c]{@{}c@{}}1 int.\\ {[}10{]}\end{tabular}}} & \multicolumn{1}{c|}{\textit{\begin{tabular}[c]{@{}c@{}}1 shall.\\ {[}10{]}\end{tabular}}} & \multicolumn{1}{c|}{\textit{\begin{tabular}[c]{@{}c@{}}3 deep\\ {[}30{]}\end{tabular}}} & \multicolumn{1}{c|}{\textit{\begin{tabular}[c]{@{}c@{}}1 int.\\ {[}10{]}\end{tabular}}} & \multicolumn{1}{c|}{\textit{\begin{tabular}[c]{@{}c@{}}1 shall.\\ {[}10{]}\end{tabular}}} & \multicolumn{1}{c|}{\textit{\begin{tabular}[c]{@{}c@{}}4 deep\\ {[}40{]}\end{tabular}}} & \multicolumn{1}{c|}{\textit{\begin{tabular}[c]{@{}c@{}}1 int\\ {[}10{]}\end{tabular}}} & \multicolumn{1}{c|}{\textit{\begin{tabular}[c]{@{}c@{}}2 shall.\\ {[}20{]}\end{tabular}}} & \multicolumn{1}{c|}{\textit{\begin{tabular}[c]{@{}c@{}}4 deep\\ {[}40{]}\end{tabular}}} & \multicolumn{1}{c|}{\textit{\begin{tabular}[c]{@{}c@{}}1 int\\ {[}10{]}\end{tabular}}} & \multicolumn{1}{c|}{\textit{\begin{tabular}[c]{@{}c@{}}2 shall.\\ {[}20{]}\end{tabular}}} \\ \hline
				\multicolumn{1}{|c|}{}                                                                      & \multicolumn{1}{c|}{\begin{tabular}[c]{@{}c@{}}without\\ CARSS\end{tabular}}          & \multicolumn{1}{c|}{\begin{tabular}[c]{@{}c@{}}with\\ CARSS\end{tabular}}             & \multicolumn{1}{c|}{\begin{tabular}[c]{@{}c@{}}without\\ CARSS\end{tabular}}                 & \multicolumn{1}{c|}{\begin{tabular}[c]{@{}c@{}}with\\ CARSS\end{tabular}}                    & \multicolumn{3}{c|}{\begin{tabular}[c]{@{}c@{}}without\\ CARSS\end{tabular}}                                                                                                                                                                                                  & \multicolumn{3}{c|}{\begin{tabular}[c]{@{}c@{}}with\\ CARSS\end{tabular}}                                                                                                                                                                                                     & \multicolumn{3}{c|}{\begin{tabular}[c]{@{}c@{}}without\\ CARSS\end{tabular}}                                                                                                                                                                                                 & \multicolumn{3}{c|}{\begin{tabular}[c]{@{}c@{}}with\\ CARSS\end{tabular}}                                                                                                                                                                                                    \\ \hline
				FOCUSS                                                                                      & 30,30,29                                                                              & 30,30,30                                                                              & 90,88,82                                                                                     & 90,89,88                                                                                     & 20,6,5                                                                                  & 9,6,5                                                                                   & 10,5,2                                                                                    & 21,10, 6                                                                                & 10,7,5                                                                                  & 10,8,6                                                                                    & 8,2,1                                                                                   & 7,2,3                                                                                  & 16,5,2                                                                                    & 8,3,3                                                                                   & 7.4,3                                                                                  & 16,9,2                                                                                    \\ \hline
				MxNE                                                                                        & 30,30,30                                                                              & 30,30,30                                                                              & 90,86,82                                                                                     & 90,90,90                                                                                     & 22,18,16                                                                                & 8,7,6                                                                                   & 10,8,8                                                                                    & 25,19,18                                                                                & 10, 8,7                                                                                 & 10,10,9                                                                                   & 9,8,7                                                                                   & 8,6,5                                                                                  & 15,12,10                                                                                  & 10,8,8                                                                                  & 8,8,6                                                                                  & 19,17,15                                                                                  \\ \hline
				SBL - Wipf                                                                                  & 30,30,30                                                                              & 30,30,30                                                                              & 90,88,86                                                                                     & 90,89,88                                                                                     & 23,15,10                                                                                & 10,6,4                                                                                  & 10,7,7                                                                                    & 25,17,12                                                                                & 10,7,5                                                                                  & 10,8,8                                                                                    & 9,7,2                                                                                   & 8,5,4                                                                                  & 17,10,5                                                                                   & 9,8,2                                                                                   & 8,6,3                                                                                  & 16,11,9                                                                                   \\ \hline
				SBL - Zhang                                                                                 & 30,30,30                                                                              & 30,30,30                                                                              & 90,90,88                                                                                     & 90,90,90                                                                                     & 22,18,15                                                                                & 10,8,6                                                                                  & 10,9,7                                                                                    & 25,20,18                                                                                & 10,9,8                                                                                  & 10,10,9                                                                                   & 8,6,5                                                                                   & 8,7,6                                                                                  & 16,11,11                                                                                  & 9,6,5                                                                                   & 8,7,7                                                                                  & 16,12,13                                                                                  \\ \hline
				sLORETA                                                                                     & 30,30,29                                                                              & 30,30,30                                                                              & 88,86,86                                                                                     & 90,89,88                                                                                     & 20,16,14                                                                                & 8,7,5                                                                                   & 9,7,7                                                                                     & 19,16,13                                                                                & 8,7,6                                                                                   & 10,9,8                                                                                    & 7,6,4                                                                                   & 7,5,2                                                                                  & 13,10,9                                                                                   & 7,6,5                                                                                   & 8,6,3                                                                                  & 14,11,8                                                                                   \\ \hline
		\end{tabular}  }
	\end{table*}
	
	\subsection{CARSS evaluation}
	
	The detailed analysis of CARSS is evaluated in (\cite{mannepalli2019certainty}). However, two challenging possibilities arise for any source localization method in general. They are (i) deep sources and (ii) near sources. In this work, the aforementioned aspects of CARSS are analyzed.
	
	\textit{DEEP SOURCES:} Estimation of deep sources is a challenge for any source localization method. The evidence of the peak is critical for CARSS to detect. Thirty deep sources whose $\norm{r}_{dip} \le 40$ are chosen. The  $\mu(\norm{r}_{dip})$ is 27.5. CARSS detected all the deep sources even under the presence of noise comfortably. The only issue if at all arise is the number of certain sources. The results are presented in Table. \ref{tcc1}. Only 5K-source space is considered.
	
	\begin{table}[]
		\caption{\textbf{CARSS EVALUATION - 1: }The detection of deep sources by CARSS}
		\label{tcc1}
		\resizebox{.95\textwidth}{!}{
			\begin{tabular}{|l|l|l|l|l|l|}
				\hline
				\multicolumn{1}{|c|}{\textit{\textbf{}}} & \multicolumn{1}{c|}{\textbf{Mean}} & \multicolumn{1}{c|}{\textbf{Min.}} & \multicolumn{1}{c|}{\textbf{Max.}} & \multicolumn{1}{c|}{\textbf{\begin{tabular}[c]{@{}c@{}}Detection\\ status\end{tabular}}} & \multicolumn{1}{c|}{\textbf{\begin{tabular}[c]{@{}c@{}}Certainty\\ (0-1)\end{tabular}}} \\ \hline
				\textit{No noise}                        & 3105                               & 4051                               & 1904                               & 30/30                                                                                    & 0.9951                                                                                  \\ \hline
				\textit{pink-1}                          & 3857                               & 4998                               & 1846                               & 30/30                                                                                    & 0.9613                                                                                  \\ \hline
				\textit{pink-4}                          & 4198                               & 5510                               & 1896                               & 30/30                                                                                    & 0.9454                                                                                  \\ \hline
		\end{tabular}}
	\end{table}
	
	\textit{DEEP SOURCE AND ITS NEAR SOURCE}: A deep source whose $\norm{r}_{dip} \le 40$ and its near-source whose $Ds({r}_{dip,i}, {r}_{dip,j}) \le 40$ are chosen . The results are presented in Table. \ref{tcc2}. Only 5K-source space is considered. The results show inconsistency. Although CARSS succeeded half of the possibilities, it could not guarantee all. 
	
	\begin{table}[]
		\caption{\textbf{CRASS EVALUATION - 2: }The detection of deep and near sources by CARSS}
		\label{tcc2}
		\resizebox{.95\textwidth}{!}{
			\begin{tabular}{|l|l|l|l|l|l|l|l|}
				\hline
				\multicolumn{1}{|c|}{\multirow{2}{*}{\textit{\textbf{}}}} & \multicolumn{1}{c|}{\multirow{2}{*}{\textbf{Mean}}} & \multicolumn{1}{c|}{\multirow{2}{*}{\textbf{Min.}}} & \multicolumn{1}{c|}{\multirow{2}{*}{\textbf{Max.}}} & \multicolumn{3}{c|}{\textbf{\begin{tabular}[c]{@{}c@{}}Detection\\ status\end{tabular}}} & \multicolumn{1}{c|}{\multirow{2}{*}{\textbf{\begin{tabular}[c]{@{}c@{}}Certainty\\ (0-1)\end{tabular}}}} \\ \cline{5-7}
				\multicolumn{1}{|c|}{}                                    & \multicolumn{1}{c|}{}                               & \multicolumn{1}{c|}{}                               & \multicolumn{1}{c|}{}                               & S-I,S-II                      & S-I/ S-II                     & None                     & \multicolumn{1}{c|}{}                                                                                    \\ \hline
				\textit{No noise}                                         & 2213                                                & 6318                                                & 556                                                 & 18/30                         & 10/30                         & 2/30                     & 0.8585;0.6325                                                                                            \\ \hline
				\textit{pink-1}                                           & 2915                                                & 7523                                                & 1010                                                & 16/30                         & 11/30                         & 3/30                     & 0.6644;0.5247                                                                                            \\ \hline
				\textit{pink-4}                                           & 3135                                                & 7959                                                & 254                                                 & 14/30                         & 12/30                         & 4/30                     & 0.5251; 0.3995                                                                                           \\ \hline
		\end{tabular} }
	\end{table}
	
	\subsection{Extended source algorithms}
	
	Two test cases are considered. Thirty simulations are performed for each test case with randomly chosen sources. The random sources with orientation are chosen using in SAREEGA toolbox (\cite{krol2018sereega}). The methods aimed to be studied are Truncated RAP-MUSIC (\cite{makela2018truncated} ) and Source Imaging based on Structured Sparsity (SISSY) (\cite{becker2017sissy} ). The motivation behind the study is not to compare the methods but is how well the integration with CARSS will go. The initial estimate of the number of sources considered is $\tilde{n} = n + 2$. 
	
	The performance metric utilized is Dipole Localization Error (DLe) (\cite{yao2005evaluation} ) and Spatial Dispersion (\cite{zhu2014reconstructing} \cite{molins2008quantification} ). The DLe investigates the similarity between true source configuration and the estimated. Let $n$ and $\tilde{n}$ be the number of true and estimated dipoles and let $I$ and $\tilde{I}$ be the true and estimated set of dipole indices of the active patch. The DLe is defined as:
	
	\begin{equation}
	\scalebox{0.95}[1]{$
		DLE = \frac{1}{2n} \sum_{k \in I} \min_{\forall \mathbf{l}} \norm{\mathbf{r_{k} - \mathbf{r_{l}}}} - 
		\frac{1}{2\tilde{n}} \sum_{l \in \tilde{I}} \min_{\forall \mathbf{k}} \norm{\mathbf{r_{l} - \mathbf{r_{k}}}}     $} 
	\end{equation} 
	
	The SD estimates the spatial blur of estimated sources as compared with the same of true sources. 
	
	\begin{equation}
	SD = \sqrt{\frac{\sum_{p = 1}^{N_{s}}\sum_{q = d_{n}}d_{pq}^{2}\left \| \tilde{s}_{q} \right \|^{2}_{2}}{\sum_{q=1}^{N_{e}} \left \| \tilde{s}_{q} \right \|^{2}_{2}}}
	\end{equation}
	
	Understanding the equations in simpler terms, they can be described as (i) DLe is the average distance of the maxima of the estimated source to the true sources, and (ii) SD is the spatial extent of the estimated source as compared to the same of the true source.  
	
	\textit{TEST-CASE (Ext.)-I}: 
	
	One source with random orientation is considered. The performance of the methods with CARSS is better than without. CARSS reduced the source space giving fewer sources to search in. Both the methods located the dipoles correctly. The 95 \% confidence limit over the mean is mentioned. The number of sources is estimated as one by the methods ($\tilde{n} = n + 2;$). In TRAP-MUSIC, a significant drop in the scanning value of the localizer ($\mu_i$) shown a significant drop just as (\cite{makela2018truncated} ) after $n=1$. 
	
	The results are presented in Table. \ref{ext1} and Table. \ref{ext2}.
	
	\begin{table*}[]
		\caption{\textit{TEST-CASE (Ext.)-I}:The performance metrics of TRAP-MUSIC when a source is activated.}
		\label{ext1}
		\resizebox{0.95\textwidth}{!}{
			\begin{tabular}{|l|l|l|l|l|l|l|l|}
				\hline
				\multicolumn{1}{|c|}{\multirow{2}{*}{\textbf{S.No.}}} & \multicolumn{1}{c|}{\multirow{2}{*}{\textbf{Noise levels}}} & \multicolumn{3}{c|}{\textbf{without CARSS}}                          & \multicolumn{3}{c|}{\textbf{with CARSS}}                                               \\ \cline{3-8} 
				\multicolumn{1}{|c|}{}                                & \multicolumn{1}{c|}{}                                       & \textit{Success Rate} & \textit{DLE (mm)} & \textit{SD (mm)}         & \textit{Success Rate} & \textit{DLE (\textbackslash{}mu)} & \textit{SD (mm)}           \\ \hline
				1.                                                    & 0 dB                                                        & 30/30                 & 0                 & 11.00 (+12 \% to - 2 \%) & 30/30                 & 0                                 & 4.95.00 (+10 \% to - 2 \%) \\ \hline
				2.                                                    & pink - 1                                                    & 30/30                 & 0                 & 29.25 (+21 \% to - 8 \%) & 30/30                 & 0                                 & 29.25 (+21 \% to - 8 \%)   \\ \hline
				3.                                                    & pink - 4                                                    & 30/30                 & 2.11              & 40.01(+20 \% to - 5 \%)  & 30/30                 & 0                                 & 40.01(+20 \% to - 5 \%)    \\ \hline
		\end{tabular}  }
	\end{table*}

	\begin{table*}[]
		\caption{\textit{TEST-CASE (Ext.)-I}:The performance metrics of SISSY  when a source is activated.}
		\label{ext2}
		\resizebox{0.95\textwidth}{!}{
			\begin{tabular}{|l|l|l|l|l|l|l|l|}
				\hline
				\multicolumn{1}{|c|}{\multirow{2}{*}{\textbf{S.No.}}} & \multicolumn{1}{c|}{\multirow{2}{*}{\textbf{Noise levels}}} & \multicolumn{3}{c|}{\textbf{without CARSS}}                          & \multicolumn{3}{c|}{\textbf{with CARSS}}                                               \\ \cline{3-8} 
				\multicolumn{1}{|c|}{}                                & \multicolumn{1}{c|}{}                                       & \textit{Success Rate} & \textit{DLE (mm)} & \textit{SD (mm)}         & \textit{Success Rate} & \textit{DLE (\textbackslash{}mu)} & \textit{SD (mm)}           \\ \hline
				1.                                                    & 0 dB                                                        & 30/30                 & 0                 & 9.00 (+15 \% to - 3 \%)  & 30/30                 & 0                                 & 4.66.00 (+11 \% to - 2 \%) \\ \hline
				2.                                                    & pink - 1                                                    & 30/30                 & 0.26              & 19.95 (+25 \% to - 7 \%) & 30/30                 & 0                                 & 12.55 (+32 \% to - 9 \%)   \\ \hline
				3.                                                    & pink - 4                                                    & 30/30                 & 4.05              & 66.51(+10 \% to - 15 \%) & 30/30                 & 0                                 & 45.01(+19 \% to - 11 \%)   \\ \hline
		\end{tabular}  }
	\end{table*}
	
	\textit{TEST-CASE (Ext.) II}
	
	Three deep sources with random orientations are considered. The performance of the methods with CARSS is better than without. CARSS reduced the source space giving fewer sources to search in. The 95 \% confidence limit over the mean is mentioned. The results are presented in Table. \ref{ext21} and Table. \ref{ext22}. the number of sources is estimated as three by the methods ($\tilde{n} = n + 2;$). In TRAP-MUSIC, a significant drop in the scanning value of the localizer ($\mu_i, i = 1 \dots \tilde{n}$) shown a significant drop just as (\cite{makela2018truncated} ) after $n=3$. 
	
	\begin{table*}[]
		\caption{\textit{TEST-CASE (Ext.)-II}:The performance metrics of SISSY when three deep sources are activated.}
		\label{ext21}
		\resizebox{0.95\textwidth}{!}{
			\begin{tabular}{|l|l|l|l|l|l|l|l|}
				\hline
				\multicolumn{1}{|c|}{\multirow{2}{*}{\textbf{S.No.}}} & \multicolumn{1}{c|}{\multirow{2}{*}{\textbf{Noise levels}}} & \multicolumn{3}{c|}{\textbf{without CARSS}}                          & \multicolumn{3}{c|}{\textbf{with CARSS}}                                              \\ \cline{3-8} 
				\multicolumn{1}{|c|}{}                                & \multicolumn{1}{c|}{}                                       & \textit{Success Rate} & \textit{DLE (mm)} & \textit{SD (mm)}         & \textit{Success Rate} & \textit{DLE (\textbackslash{}mu)} & \textit{SD (mm)}          \\ \hline
				1.                                                    & 0 dB                                                        & 22/30                 & 20.56             & 12.00 (+10 \% to - 6 \%) & 29/30                 & 12.25                             & 6.00 (+10 \% to - 10 \%)  \\ \hline
				2.                                                    & pink - 1                                                    & 18/30                 & 45.40             & 39.50 (+16\% to - 10 \%) & 24/30                 & 22.95                             & 21.24 (+17\% to - 12 \%)  \\ \hline
				3.                                                    & pink - 4                                                    & 10/30                 & 52.35             & 55.95(+29 \% to - 20 \%) & 19/30                 & 39.95                             & 27.56 (+35 \% to - 22 \%) \\ \hline
		\end{tabular}  }
	\end{table*}

	\begin{table*}[]
		\caption{\textit{TEST-CASE (Ext.)-II}:The performance metrics of TRAP- MUSIC when three deep sources are activated.}
		\label{ext22}
		\resizebox{0.95\textwidth}{!}{
			\begin{tabular}{|l|l|l|l|l|l|l|l|}
				\hline
				\multicolumn{1}{|c|}{\multirow{2}{*}{\textbf{S.No.}}} & \multicolumn{1}{c|}{\multirow{2}{*}{\textbf{Noise levels}}} & \multicolumn{3}{c|}{\textbf{without CARSS}}                           & \multicolumn{3}{c|}{\textbf{with CARSS}}                                              \\ \cline{3-8} 
				\multicolumn{1}{|c|}{}                                & \multicolumn{1}{c|}{}                                       & \textit{Success Rate} & \textit{DLE (mm)} & \textit{SD (mm)}          & \textit{Success Rate} & \textit{DLE (\textbackslash{}mu)} & \textit{SD (mm)}          \\ \hline
				1.                                                    & 0 dB                                                        & 23/30                 & 23.65             & 8.00 (+14 \% to - 4 \%)   & 28/30                 & 9.95                              & 4.70  (+11 \% to - 1 \%)  \\ \hline
				2.                                                    & pink - 1                                                    & 19/30                 & 40.40             & 21.00 (+20\% to - 8 \%)   & 25/30                 & 22.50                             & 20.50 (+15\% to - 8 \%)   \\ \hline
				3.                                                    & pink - 4                                                    & 10/30                 & 49.55             & 42.25 (+39 \% to - 30 \%) & 17/30                 & 39.95                             & 35.75 (+26 \% to - 13 \%) \\ \hline
		\end{tabular}   }
	\end{table*}
	
	\subsection{Real data experiment}
	
	The data utilized for the current study is downloaded from http://kdd.ics.uci.edu/databases/eeg/eeg.html. The aim of the study to examine EEG correlates of genetic predisposition to alcoholism (\cite{ingber1997statistical}) (\cite{ingber1998statistical} ) (\cite{snodgrass1980standardized} ). It contains measurements from 64 electrodes placed on the scalp sampled at 256 Hz (3.9-msec epoch) for 1 second. 
	
	There were two sets of subjects: the alcoholic group and the control group. Each subject was exposed to either a single stimulus (S1) or to two stimuli (S1 and S2) which were pictures of objects chosen from the 1980 Snodgrass and Vanderwart picture set. When two stimuli were shown, they were presented in either a matched condition where S1 was identical to S2 or in a non-matched condition where S1 differed from S2.
	
	The three versions of the EEG data set are provided. The small dataset version has two subjects - one from alcoholics and the other from the control group. For each subject, three matching paradigms are provided. They are $c_1$ (one presentation only), $c_m$ (match to the previous presentation) and $c_n$ (no-match to the previous presentation). Ten trial runs are given. For the current study, a small dataset is utilized. The large dataset has 120 trials for 120 subjects. The objective of the current study does not fit considering large data hence the small dataset is chosen. 
	
	\begin{figure*}
		\centering
		\begin{minipage}{0.75\textwidth}
			\centering
			\includegraphics[width=0.9\textwidth]{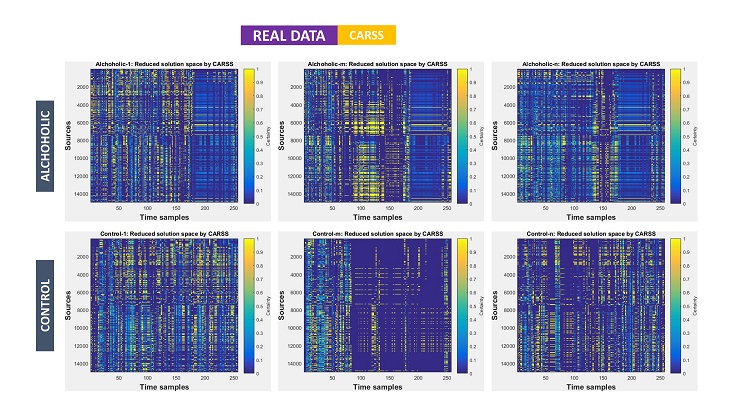} 
			\caption{\textit{REAL DATA}: The reduced solution space by CARSS is presented. The x-axis represent time samples and y-axis represent sources indices. CARSS reduced between 8 \% to 30 \% of the total solution space. }
			\label{rdc}
		\end{minipage}
	\end{figure*}
	
	\subsection{Simulated data tests}
	
	\noindent
	\textit{Test Case-I}: \\
	
	The dipole is located at [17.08, -4.21, 75.74] mm oriented along [1,0,0] is activated. The distance from the origin of Source-I $\norm{\mathbf{r}}$ (S-I) is 77 mm. 
	
\begin{figure}[!h]
		\centering
		\centering
		\includegraphics[width=0.9\linewidth]{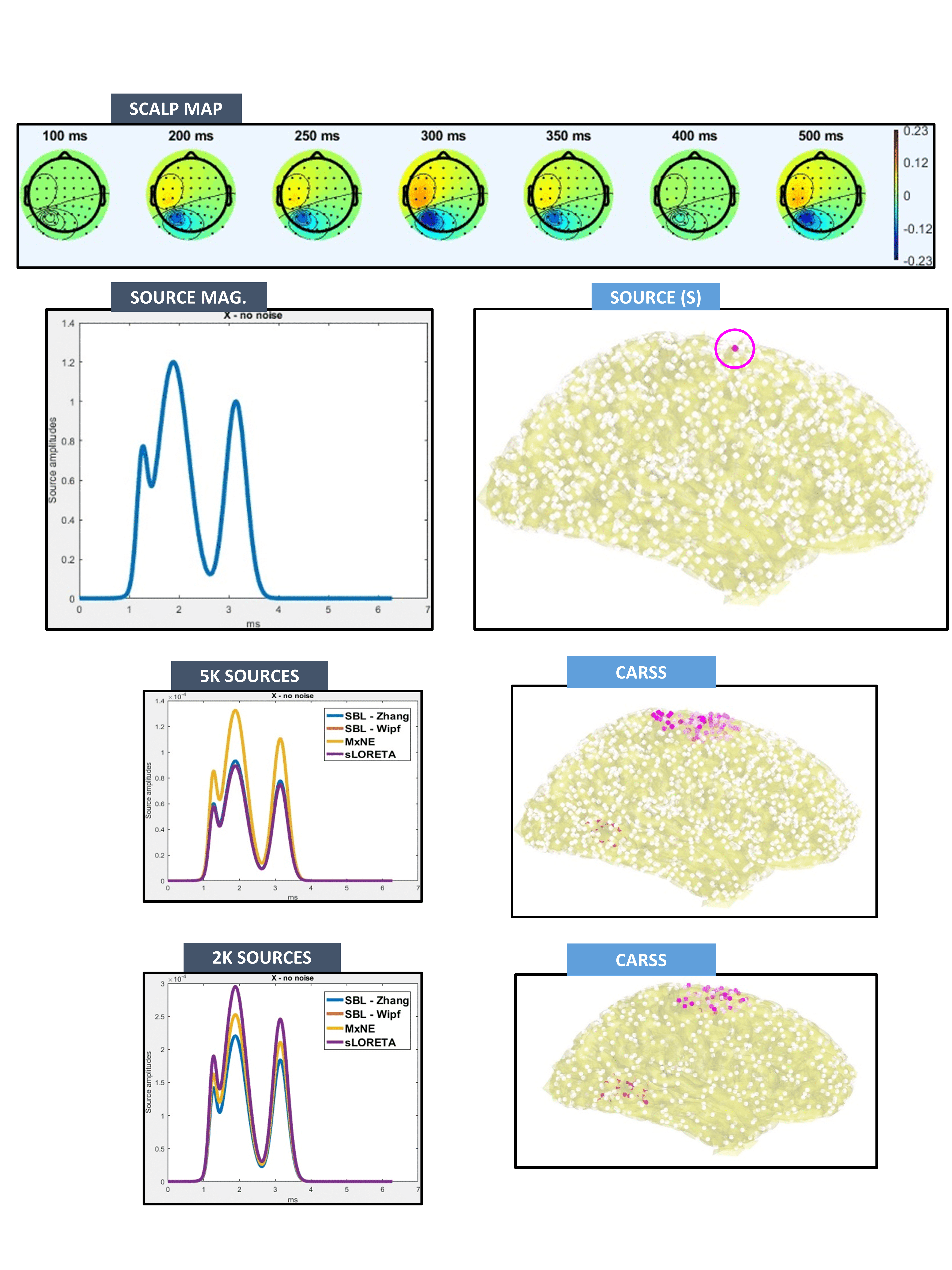}
		\hfill
		\caption{\textbf{SAMPLE TEST CASE-I}: Scalp maps of the simulated ERP. The peak latencies are 500, 300, and 200 ms. The peak widths are 200, 300 and 100 ms. The peak amplitudes are 1, 1.2,  and 0.6 $\mu$ V. The x-axis represent $t*2\pi/T,t=1 \dots 1000, T=1000$. The ground-truth source activation pattern $\mathbf{X}$ is shown. The reconstructed source magnitude $\mathbf{X}^{*}$ by the methods in the 2K-SS and 5K-SS are shown. The total number of sources in the 5K-source-space (5K-ss) are 15,012 and in the 2K-source-space (2K-ss) 6,012. CARSS reduced to 875 in 2K-SS and 1256 in 5K-SS when no noise is added. }
		\label{tc11}
	\end{figure}

	The ERP is as shown in Fig. \ref{tc11}. The peak latencies of the ERP are 500, 300, and 200 ms. The peak widths are 200, 300, and 100 ms. The peak amplitudes are 1, 1.2, and 0.6 $\mu$ V. 
	
	The sensor noise of amplitude 5\% and 10\% are added to the noiseless measurement data. This noise has no dependencies across channels or samples.
	
	The CARSS identified all the sources and reduced the SS to 875 out of 15012 in the 5K-SS. The CARSS identified all the sources and reduced the SS to 1256 out of 6012 in the 2K-SS. The sources are identified even under the presence of noise.
	
	The details can be observed from Fig. \ref{tc11}.
	
	\noindent
	\textit{Test Case-II}: \\
	
\begin{figure*}[!h]
		\centering
		\centering
		\includegraphics[width=0.9\linewidth]{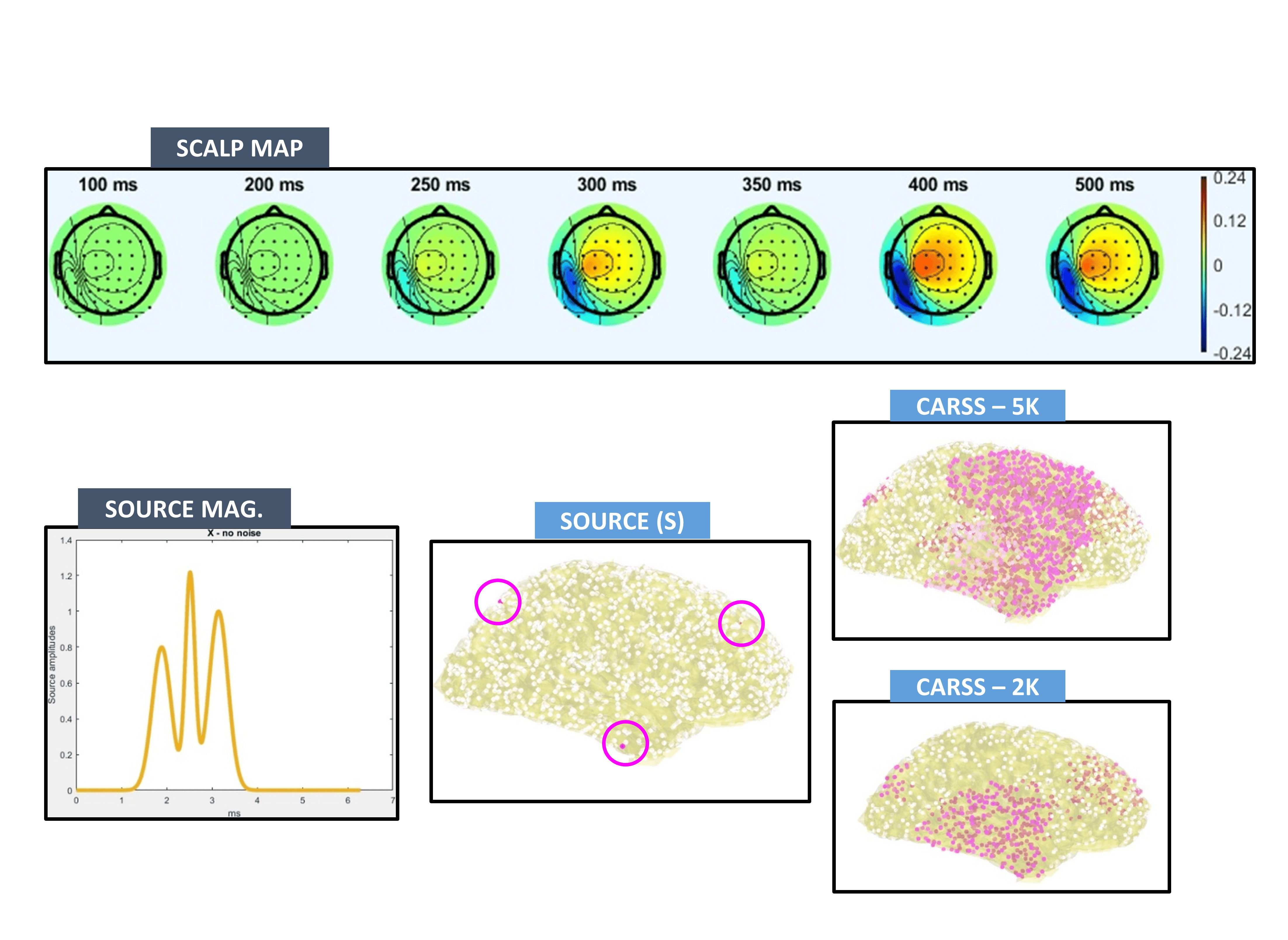}
		\hfill
		\caption{\textbf{SAMPLE TEST CASE-II:} Scalp maps of the simulated ERP. The peak latencies are 500, 400, and 300 ms. The peak widths are 200, 100 and 200 ms. The peak amplitudes are 1, 1.2,  and 0.8 $\mu$ V. The ground-truth source activation pattern for all the sources $\mathbf{X}$ is shown. The source is located at [-39.2764, 42.20, 35.88] mm,	[28.77, -74.93, 53.13] mm, and [55.69, -7.10, -41.42] mm oriented along [1,0,0], [0,1,0], and [0,0,1]. The total number of sources in the 5K-source-space (5K-ss) are 15,012 and in the 2K-source-space (2K-ss) 6,012. CARSS reduced to 4285 in 2K-SS and 1313 in 5K-SS under no noise. The length of the signal is 1000 ms ($T = 1000$). The x-axis represent each sample $t*2\pi/T,t=1 \dots 1000, T=1000$.}
		\label{tc21}
	\end{figure*}

	The dipoles are located at [-39.2764, 42.20, 35.88] mm,	[28.77, -74.93, 53.13] mm, and [55.69, -7.10, -41.42] mm oriented along [1,0,0], [0,1,0], and [0,0,1] are activated. All the three ERPs are the same as shown in Fig. \ref{tc21}. The peak latencies of the ERP are 500, 400, and 300 ms. The peak widths are 200, 100 and 200 ms. The peak amplitudes are 1, 1.2,  and 0.8 $\mu$ V. 
	
	\begin{enumerate}
		\item Distance from the origin $\norm{\mathbf{r_{i}}},i=1$ of S-I is 67.9 mm. Distance between S-I and others $\norm{\mathbf{r}_{i,j}},j=2 \dots 5$ is 136.6 mm.
		\item Distance from the origin $\norm{\mathbf{r_{i}}},i=2$ of S-II is 96.3 mm. Distance between S-II and others $\norm{\mathbf{r}_{i,j}},j=2 ,3$ is 119.4 mm.
		\item Distance from the origin $\norm{\mathbf{r_{i}}},i=3$ of S-III is 69.8 mm.
	\end{enumerate}
	
	The brown noise of amplitude one and four are added to the noiseless measurement data. There is no such specific aim in adding brown noise.  
	
	The CARSS identified all the sources and reduced the SS to 4258 out of 15012 in the 5K-SS. The CARSS identified all the sources and reduced the SS to 1313 out of 6012 in the 2K-SS. The sources are identified even under the presence of noise.
	
	The details can be observed from Fig. \ref{tc21}.
	
	\noindent
	\textit{Test case-III}
	
\begin{figure*}[!h]
		\centering
		\centering
		\includegraphics[width=0.85\linewidth]{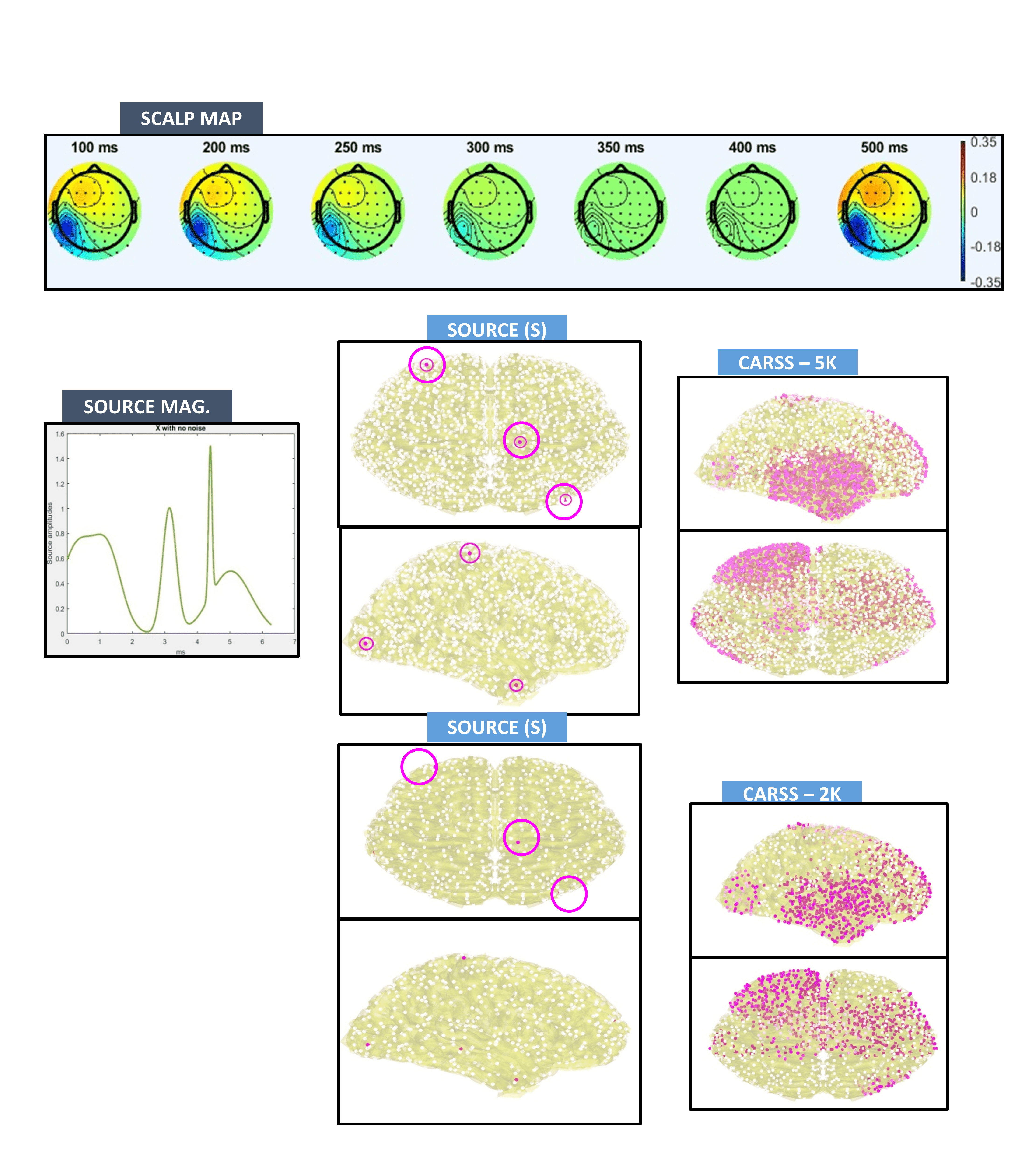}
		\hfill
		\caption{\textbf{SAMPLE TEST CASE-III:} Scalp maps of the simulated ERP. The peak latencies are 500, 400, and 300 ms. The peak widths are 200, 100 and 200 ms. The peak amplitudes are 1, 1.2,  and 0.8 $\mu$ V. The ground-truth source activation pattern for all the sources $\mathbf{X}$ is shown. The source is located at [-35.65, 5.63, 38.84] mm,	[-65.19, -39.30, -14.21] mm, [15.45, -27.58, 73.34] mm, [39.05, -73.27, -3.30] mm, and [-32.99, 60.39, 13.64] mm oriented along [1,0,0], [0,1,0], [0,0,1], [0,1,0] and [1,0,0]. The total number of sources in the 5K-source-space (5K-ss) are 15,012 and in the 2K-source-space (2K-ss) 6,012. CARSS reduced to 4285 in 2K-SS and 1313 in 5K-SS under no noise. The length of the signal is 1000 ms ($T = 1000$). The x-axis represent each sample $t*2\pi/T,t=1 \dots 1000, T=1000$.}
		\label{tc31}
	\end{figure*}

	The dipoles are located at [-35.65, 5.63, 38.84] mm,	[-65.19, -39.30, -14.21] mm, [15.45, -27.58, 73.34] mm, [39.05, -73.27, -3.30] mm, and [-32.99, 60.39, 13.64] mm oriented along [1,0,0], [0,1,0], [0,0,1], [0,1,0] and [1,0,0] are activated. All the three ERPs are the same as shown in Fig. \ref{tc31}.  The peak latencies are 500, 400, and 300 ms. The peak widths are 200, 100 and 200 ms. The peak amplitudes are 1, 1.2,  and 0.8 $\mu$ V. 
	
	\begin{enumerate}
		\item Distance from the origin $\norm{\mathbf{r_{i}}},i=1$ of S-I is 53.0 mm.
		Distance between S-I and others $\norm{\mathbf{r}_{i,j}},j=2 \dots 5$ is 113.1 mm.
		
		\item Distance from the origin $\norm{\mathbf{r_{i}}},i=2$ of S-II is 77.4 mm.
		Distance between S-II and others $\norm{\mathbf{r}_{i,j}},j=3 \dots 5$ is 119.6 mm.
		
		\item Distance from the origin $\norm{\mathbf{r_{i}}},i=3$ of S-III is 79.9 mm.
		Distance between S-I and others $\norm{\mathbf{r}_{i,j}},j=4, 5$ is 103.7 mm.
		
		\item Distance from the origin $\norm{\mathbf{r_{i}}},i=4$ of S-VI is 101.2 mm.
		Distance between S-IV and others $\norm{\mathbf{r}_{i,j}},j = 5$ is 170 mm.
		
		\item Distance from the origin $\norm{\mathbf{r_{i}}},i=5$ of S-V is 70.4 mm.
	\end{enumerate}

	The white noise of amplitude 10 \% and 40 \% of the maximum measurement amplitude is added to the same.
	
	The CARSS identified all the sources and reduced the SS to 6930 out of 15012 in the 5K-SS. The CARSS identified all the sources and reduced the SS to 2843 out of 6012 in the 2K-SS. The sources are identified even under the presence of noise.
	
	The details can be observed from Fig. \ref{tc31}.
	
\begin{figure*}[!h]
		\centering
		\centering
		\includegraphics[width=0.85\linewidth]{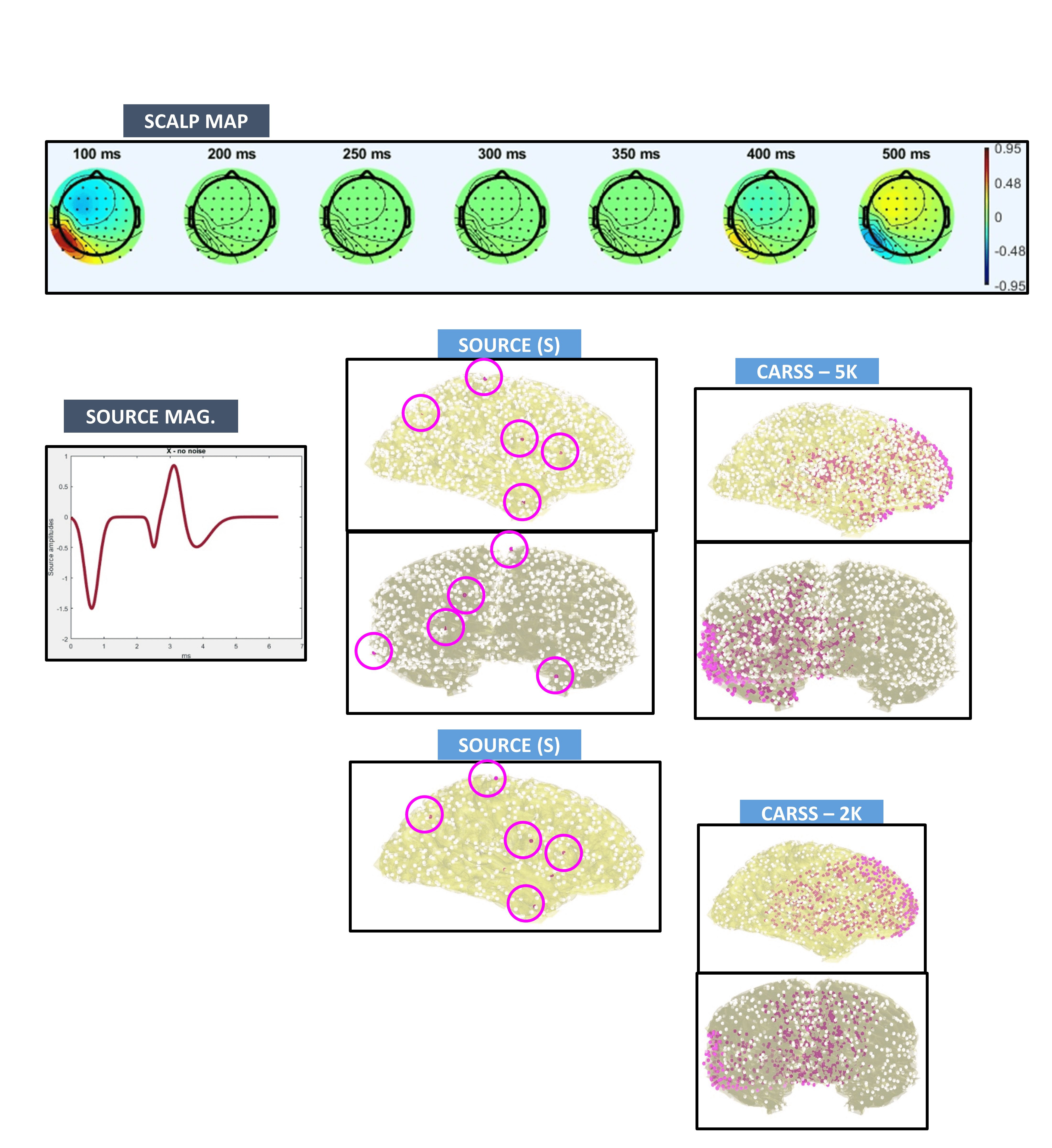}
		\hfill
		\caption{\textbf{SAMPLE TEST CASE-IV:} Scalp maps of the simulated ERP. The peak latencies are 100, 400, 500, and 600 ms. The peak widths are 200,100,200, and 400 ms. The peak amplitudes are -1.5, -0.5, 1, and -0.5 $\mu$ V. The ground-truth source activation pattern for all the sources $\mathbf{X}$ is shown. The source is located at [-69.24, -30.62, -16.12] mm, [-34.91, 18.87, 2.29] mm, [-33.65, 6.60, -17.48], [-35.97, 7.97, -15.03], [-16.69, -76.23, 39.63] mm, [25.91, 0.68, -39.21] mm and [4.38, -28.65, 77.3] mm oriented along [1,0,0], [0,1,0], [0,0,1], [0,1,0], [1,0,0], [0,1,0] and [0,0,1]. The total number of sources in the 5K-source-space (5K-ss) are 15,012 and in the 2K-source-space (2K-ss) 6,012. CARSS reduced to 3980 in 2K-SS and 1819 in 5K-SS under no noise. The length of the signal is 1000 ms ($T = 1000$). The x-axis represent each sample $t*2\pi/T,t=1 \dots 1000, T=1000$.}
		\label{tc41}
	\end{figure*}

	\noindent
	\textit{Test case-IV}:
	
	The dipoles is located at [-69.24, -30.62, -16.12] mm, [-34.91, 18.87, 2.29] mm, [-33.65, 6.60, -17.48], [-35.97, 7.97, -15.03], [-16.69, -76.23, 39.63] mm, [25.91, 0.68, -39.21] mm and [4.38, -28.65, 77.3] mm oriented along [1,0,0], [0,1,0], [0,0,1], [0,1,0], [1,0,0], [0,1,0] and [0,0,1] are activated. All the seven ERPs are shown in Fig. \ref{tc41}.  The peak latencies are 100, 400, 500, and 600 ms. The peak widths are 200,100,200, and 400 ms. The peak amplitudes are -1.5, -0.5, 1, and -0.5 $\mu$ V.  
	
	\begin{enumerate}
		\item Distance from the origin $\norm{\mathbf{r_{i}}},i=1$ of S-I is 77.4.
		Distance between S-I and others $\norm{\mathbf{r}_{i,j}},j=2 \dots 7$ is 63.0 mm.
		
		\item Distance from the origin $\norm{\mathbf{r_{i}}},i=2$ of S-II is 39.8 mm.
		Distance between S-II and others $\norm{\mathbf{r}_{i,j}},j=3 \dots 7$ is 23.3 mm.
		
		\item Distance from the origin $\norm{\mathbf{r_{i}}},i=3$ of S-III is 38.5 mm.
		Distance between S-III and others $\norm{\mathbf{r}_{i,j}},j=4 \dots 7$ is 35.7 mm.
		
		\item Distance from the origin $\norm{\mathbf{r_{i}}},i=4$ of S-VI is 39.6 mm.
		Distance between S-IV and others $\norm{\mathbf{r}_{i,j}},j=5 \dots 7$ is 75.1 mm.
		
		\item Distance from the origin $\norm{\mathbf{r_{i}}},i=5$ of S-V is 87.5 mm.
		Distance between S-V and others $\norm{\mathbf{r}_{i,j}},j=6, 7$ is 118.2 mm.
		
		\item Distance from the origin $\norm{\mathbf{r_{i}}},i=6$ of S-VI is 47.1 mm.
		Distance between S-VI and others $\norm{\mathbf{r}_{i,j}},j=7$ is 122.2 mm.
		
		\item Distance from the origin $\norm{\mathbf{r_{i}}},i=7$ of S-VII is 82.6.
	\end{enumerate}
	
	The white noise of amplitude 40 \% of the maximum measurement amplitude is added to the same.
	
	The CARSS identified only four out of all the sources and reduced the SS to 3980 out of 15012 in the 5K-SS. The CARSS identified four out of seven sources and reduced the SS to 1819 out of 6012 in the 2K-SS. The same sources are identified even under the presence of noise.
	
	The details can be observed from Fig. \ref{tc41}.

	\begin{table}[]
		\centering
		\caption{\textit{Sample test cases}: CARSS report for all test cases. Thr total number of certain sources out of 15,012 (5K) and 6,012 (2K).}
		\resizebox{.95\textwidth}{!}{
			\begin{tabular}{|c|c|c|c|c|c|c|c|}
				\hline
				\multirow{2}{*}{\textbf{S. No.}} & \multirow{2}{*}{\textbf{\begin{tabular}[c]{@{}c@{}}Test\\ Cases\end{tabular}}} & \multicolumn{3}{c|}{\textbf{5K}}                                                                                                                                     & \multicolumn{3}{c|}{\textbf{2K}}                                                                                                                                     \\ \cline{3-8} 
				&                                                                                & \textit{NN}                                            & \textit{NC-1}                                        & \textit{NC-2}                                        & \textit{}                                              & \textit{NC-1}                                        & \textit{NC-2}                                        \\ \hline
				1.                               & TC-1                                                                           & \begin{tabular}[c]{@{}c@{}}Y\\ 1256\end{tabular}       & \begin{tabular}[c]{@{}c@{}}Y\\ 1295\end{tabular}     & \begin{tabular}[c]{@{}c@{}}Y\\ 1401\end{tabular}     & \begin{tabular}[c]{@{}c@{}}Y\\ 875\end{tabular}        & \begin{tabular}[c]{@{}c@{}}Y\\ 933\end{tabular}      & \begin{tabular}[c]{@{}c@{}}Y\\ 997\end{tabular}      \\ \hline
				2.                               & TC-2                                                                           & \begin{tabular}[c]{@{}c@{}}YYY\\ 4258\end{tabular}     & \begin{tabular}[c]{@{}c@{}}YYY\\ 4311\end{tabular}   & \begin{tabular}[c]{@{}c@{}}YYY\\ 4456\end{tabular}   & \begin{tabular}[c]{@{}c@{}}YYY\\ 1313\end{tabular}     & \begin{tabular}[c]{@{}c@{}}YYY\\ 1396\end{tabular}   & \begin{tabular}[c]{@{}c@{}}YYY\\ 1455\end{tabular}   \\ \hline
				3.                               & TC-3                                                                           & \begin{tabular}[c]{@{}c@{}}YYYYY\\ 6930\end{tabular}   & \begin{tabular}[c]{@{}c@{}}YYYYY\\ 7009\end{tabular} & \begin{tabular}[c]{@{}c@{}}YYYYY\\ 7256\end{tabular} & \begin{tabular}[c]{@{}c@{}}YYYYY\\ 2843\end{tabular}   & \begin{tabular}[c]{@{}c@{}}YYYYY\\ 2925\end{tabular} & \begin{tabular}[c]{@{}c@{}}YYYYY\\ 3351\end{tabular} \\ \hline
				4.                               & TC-4                                                                           & \begin{tabular}[c]{@{}c@{}}NNYYYYN\\ 3980\end{tabular} & \multicolumn{2}{c|}{\begin{tabular}[c]{@{}c@{}}NNYYYYN\\ 4042\end{tabular}}                                 & \begin{tabular}[c]{@{}c@{}}NNYYYYN\\ 1819\end{tabular} & \multicolumn{2}{c|}{\begin{tabular}[c]{@{}c@{}}NNYYYYN\\ 1925\end{tabular}}                                 \\ \hline
		\end{tabular}   }
		\label{ttc1}
	\end{table}
	
	\section{Discussions}
	
	\subsection{Monte-carlo simulations}
	
	\subsubsection{General observations}
	
	\begin{enumerate}
		\item The source localization algorithms performed significantly better with CARSS. The results show the same.
		\item Although all the methods localized the active sources correct for one, three, and five sources, they are challenged by the fourth test case where the active sources are seven. The results are evident from the success rates of the active sources in Table. \ref{sr1} and \ref{sr2}. 
		\item For the test case-iv alone, the prominence of CARSS in the improvement of the results is not much evident.  In the rest of the test cases, CARSS had significantly improved the results.
		\item The results in Table. \ref{mts3} show that not alone deep source localization is more challenging. The neighboring source localization is much more strenuous. The results from Table. \ref{mts4}, \ref{sr1} and \ref{sr2} show the same.  
		
		\item The sparse bayesian strategy by (\cite{zhang2011sparse}) has better estimated the sources than the same by (\cite{wipf2011latent}) under the presence of noise.
		\item The advantage of $\ell_{21}$ norm-based method MxNE is visible over the $\ell_2$ norm-based methods in estimating the sources. 
		\item The impotence with ITER's under the presence of noise is very clearly visible in the results. 
		%
	\end{enumerate}
	
	\subsubsection{Case by case observations}
	
	\textit{Test case-I} The \textit{A' metric} values (Table. \ref{tc11} ) show consistency for all the solvers. The ideal value of the \textit{A' metric} is 1. The number of sources to which CARSS reduced to is presented in Table.\ref{mts1}. The number of sources in the SS is reduced by 85 \%. The important factor that affects CARSS is the detection of the peak. After detecting the peak, the certainty is calculated. CARSS substantially improved the results.
	
	\textit{Test case-II} The \textit{A' metric} values (Table. \ref{tc21} ) show consistency for all the solvers. However, slight variations over the first test case are observed (Table. \ref{mts2v}).  The shallow sources emit prominent scalp distribution that makes their peaks visible at the same location. This is evident in Table. \ref{mts2c}. CARSS detected all the sources with the reduction of SS to 25 \%. 
	
	\textit{Test case-III} The \textit{A' metric} values are shown in Table. \ref{tc31}. The test case has two deep sources, an intermittent source, and two shallow sources. All are not near to each other. The spatial extent of the scalp distribution by the shallow sources is dominant but is limited on the scalp. Hence, when two distant shallow sources are active, the interference of peaks does not occur much. The deep sources with a prominent peak on the scalp can be detected by CARSS. This is the reason for CARSS detecting many deep sources in the test case. Also, the presence of shallow sources will attenuate the deep sources. This is a challenge for any solver employed including CARSS. The results are demonstrated in Table. \ref{sr1}. Three observations for CARSS are - (i) many shallow sources are detected by CARSS. (ii) Many intermediate sources too are detected.  (iii) Some deep sources are missed. Also, some sources are missed due to the interference of scalp distribution. A wrong peak led to detecting the wrong cluster of sources.
	
	\textit{Test case-IV}: This test case is a challenge for any solver. Seven sources are chosen of which four are near ($Ds(\mathbf{r}_i, \mathbf{r}_j) \le 55$) deep sources, one intermittent source, and the rest are shallow sources. Enough evidence of source signatures on the scalp is important for any solver to detect. The deep sources are attenuated heavily by the shallow sources. Due to the attenuation, sometimes it may become extremely difficult for any solver to identify. This is what happened in the test case. All the solvers including CARSS are challenged by this test case. The biggest obstacle for CARSS is not the deep sources. It is the near sources that will interfere with each other. 
	
	\subsubsection{Detailed observations}
	
	\textit{1. FOCUSS:}
	\noindent
	
	\begin{enumerate}
		\item The general challenges faced by any ITER are: (i) converging to local minima instead of global, (ii) instability under the presence of noise, (iii) Initial estimate dependent and scattering of the solution. They are evident in the results.
		\item The instability under the presence of noise can be observed in TC-III (Tables. \ref{mts3} and \ref{mts3v}) and TC-IV (Tables \ref{mts4}  and \ref{mts3v}). The localization ability of FOCUSS is challenged a lot under noise although it is better without noise (Table. \ref{sr2} and \ref{sr1}).  
		\item FOCUSS had localized the deep sources that are far apart but failed to localize the deep sources that are near. The convergence to the wrong location is observed instead of true in the fourth test case (Tables. \ref{mts4},\ref{mts4v}, \ref{sr2} and \ref{sr1}).
	\end{enumerate}
	
	\textit{2. MxNE:}
	\noindent
	
	\begin{enumerate}
		\item The trade-off provided by $\ell_{21}$ norm is evident in the results. The results by MxNE show robustness towards the noise. The success rate of MxNE in localizing sources is higher for MxNE compared to $\ell_2$ regularization method (Tables. \ref{sr1} ad \ref{sr2}). 
		\item The reconstructed sources by MxNE are also sparse. 
		\item The MxNE couldn't localize many of the four neighboring deep sources clearly (Tables. \ref{sr1}, \ref{sr2} ).  
	\end{enumerate}
	
	\textit{3. sLORETA:}
	\noindent	
	
	\begin{enumerate}
		\item sLORETA is a well-known $\ell_2$ norm regularizer that is simple to implement. 
		\item The results clearly depict the characteristics  of the $\ell_2$ norm regularization. 
		\item Oversmoothing of the solution is observed in many cases. This can be observed in Fig. \ref{tc31}.
		\item The $\ell_{21}$ norm penalization proved better in many aspects compared to $\ell_2$ norm regularization.
	\end{enumerate}

	\textit{3. Learning methods:}
	\noindent	
	
	\begin{enumerate}
		\item SBL (\cite{wipf2011latent}) is accurate but SBL (\cite{zhang2013extension}) is when no noise is present. However, it is complimentary when noise is present. The SBL (\cite{zhang2013extension}) proved to be robust against noise. The results show the same.
		\item The SBLs performed much better than FOCUSS. They are fast and efficient than FOCUSS. 
		\item Both of them failed to localize the three near deep sources in TC-IV.
	\end{enumerate}
	
	\subsection{CARSS evaluation}
	
	\begin{enumerate}
		\item CARSS can detect deep sources that are far apart. Table. \ref{tcc1}. The only issue with the deep sources is the increase in the number of certain sources. 
		\item The presence of noise shown less impact on the performance of CARSS. The spatial filtering of the measurement vector as suggested in (\cite{michel2019eeg}) is found useful under the presence of noise. 
		\item The efficacy of CARSS is challenged under multiple neighboring and deep sources. The results are depicted in Table. \ref{tcc2}.
		\item The effectiveness of CARSS has been limited in TC-IV (Table. \ref{mts4}). The presence of neighboring and deep sources is responsible for the estimation of the wrong cluster as certain sources. Due to this, CARSS had not provided the right SS.   
	\end{enumerate}
	
	\subsection{Extended source algorithms}
	
	\begin{enumerate}
		\item The extended source localization algorithms (ESls) improved significantly with CARSS. The localization accuracy (DLe), and spatial dispersion improved after providing the reduced SS.
		\item The ESls performed exceptionally well under one source activation. There is not much use of CARSS in the test case alone. 
		\item The presence of deep sources and the presence of noise (in Table. \ref{ext21} and \ref{ext22}) had affected the performance of ESls.
	\end{enumerate}
	
	\subsection{Real data analysis}
	
	The solution space reduced by CARSS is presented in Fig. \ref{rdc}. The CSP profile of the sources can be observed. CARSS reduced the solution space to one-thirds at the maximum. MxNE is utilized thereafter. After estimating the source amplitudes, the label powers in the brain are estimated. The atlas followed is the Destrieux atlas (\cite{destrieux2010automatic}). There are a total of 148 labels. Then, three labels with maximum power are found out and are presented in Table. \ref{rdl}.  
	
	\begin{table}[]
		\caption{\textbf{REAL DATA ANALYSIS}: The maximum label powers observed in different experimental conditions. 'cingul' - cingulum; 'ant' - anterior; 'sup' - superior; 'temp' - temporal; 'sup' - superior; 'trans' = transverse; 'G' - gyrus; 'S' - sulcus; 'L' - left; 'R' - right}
		\label{rdl}
		\resizebox{.95\textwidth}{!}{
			\begin{tabular}{|l|l|l|l|}
				\hline
				\multicolumn{1}{|c|}{\textbf{S.No.}} & \multicolumn{1}{c|}{\textbf{Condition}} & \multicolumn{1}{c|}{\textbf{Alcoholic}}                                                                               & \multicolumn{1}{c|}{\textbf{Control}}                                                                             \\ \hline
				1.                                   & \textit{Single stimuli}                 & \begin{tabular}[c]{@{}l@{}}G and S cingul Ant R  \\  S circular insula ant R  \\  S circular insula sup R\end{tabular} & \begin{tabular}[c]{@{}l@{}}S circular insula ant L  \\  S circular insula ant R  \\  G subcallosal R\end{tabular} \\ \hline
				2.                                   & \textit{Matched}                        & \begin{tabular}[c]{@{}l@{}}G cingul Post ventral L  \\  G precuneus L  \\  S occipital ant R\end{tabular}              & \begin{tabular}[c]{@{}l@{}}G temp sup G T transv R  \\ S occipital ant R  \\ G precuneus L\end{tabular}           \\ \hline
				3.                                   & \textit{Non-matched}                    & \begin{tabular}[c]{@{}l@{}}S occipital ant R  \\  S circular insula ant R  \\  S suborbital L\end{tabular}             & \begin{tabular}[c]{@{}l@{}}S suborbital L  \\  S circular insula sup R  \\  G cingul Post ventral L\end{tabular}  \\ \hline
		\end{tabular}   }
	\end{table}
	
	\subsection{Sample tests}
	
	After a detailed study of the results, as a summary, some key interpretations eluded regarding CARSS are:
	
	\begin{enumerate}
		\item The presence of a deep active source increases the number of certain sources. The third test case (Fig. \ref{tc31}) has two deep sources that are identified but the number of certain sources is increased. 
		\item The presence of many deep sources will also affect the detection of shallow sources. The fourth test case (Fig. \ref{tc41}) has a shallow source unidentified due to the presence of four deep sources.
		\item The presence of tangential sources has a more prominent effect than deep sources to be deficient in being picked as a possible candidate. The fourth test case (Fig. \ref{tc41}) has three tangential sources. Tangentially oriented sources may show two approximately equal peaks in the topography that may result in a change of peak locations.
		\item The presence of near active sources may mislead and may create a fake cluster of sources. The presence of two near active sources in the fourth test case (Fig. \ref{tc41}) is responsible for the detection of very wrong sources in the CARSS solution space.
		\item The spatial filtering of the measurement vector as suggested in (\cite{michel2019eeg}) is found useful under the presence of noise. No much difference in picking the active source and in the number of certain sources under the presence of noise is observed. 
	\end{enumerate}
	
	The EEG source localization is a challenging problem due to its ill-posedness. The paper presents the current state-of-the-art sparse techniques. The merits and demerits of the solvers are studied. A vast comparative study is performed for eight solvers using a sixty-four channel EEG setup. The MAP solvers, an $\ell_2$ norm-based solver, and $\ell_{21}$ norm-based solver are chosen to compare. CARSS is also tested. some important observations are drawn examining the study.

	\bibliography{references}

\begin{thebibliography}{70}
\providecommand{\natexlab}[1]{#1}
\providecommand{\url}[1]{\texttt{#1}}
\expandafter\ifx\csname urlstyle\endcsname\relax
  \providecommand{\doi}[1]{doi: #1}\else
  \providecommand{\doi}{doi: \begingroup \urlstyle{rm}\Url}\fi

\bibitem[Baillet and Garnero(1997)]{baillet1997bayesian}
Sylvain Baillet and Line Garnero.
\newblock A bayesian approach to introducing anatomo-functional priors in the
  eeg/meg inverse problem.
\newblock \emph{IEEE transactions on Biomedical Engineering}, 44\penalty0
  (5):\penalty0 374--385, 1997.

\bibitem[Baillet et~al.(2001)Baillet, Mosher, and
  Leahy]{baillet2001electromagnetic}
Sylvain Baillet, John~C Mosher, and Richard~M Leahy.
\newblock Electromagnetic brain mapping.
\newblock \emph{IEEE Signal processing magazine}, 18\penalty0 (6):\penalty0
  14--30, 2001.

\bibitem[Becker et~al.(2012)Becker, Comon, Albera, Haardt, and
  Merlet]{becker2012multi}
Hanna Becker, Pierre Comon, Laurent Albera, Martin Haardt, and Isabelle Merlet.
\newblock Multi-way space--time--wave-vector analysis for eeg source
  separation.
\newblock \emph{Signal Processing}, 92\penalty0 (4):\penalty0 1021--1031, 2012.

\bibitem[Becker et~al.(2014)Becker, Albera, Comon, Haardt, Birot, Wendling,
  Gavaret, B{\'e}nar, and Merlet]{becker2014eeg}
Hanna Becker, Laurent Albera, Pierre Comon, Martin Haardt, Gw{\'e}na{\"e}l
  Birot, Fabrice Wendling, Martine Gavaret, Christian-George B{\'e}nar, and
  Isabelle Merlet.
\newblock Eeg extended source localization: tensor-based vs. conventional
  methods.
\newblock \emph{NeuroImage}, 96:\penalty0 143--157, 2014.

\bibitem[Becker et~al.(2017)Becker, Albera, Comon, Nunes, Gribonval, Fleureau,
  Guillotel, and Merlet]{becker2017sissy}
Hanna Becker, Laurent Albera, Pierre Comon, J-C Nunes, R{\'e}mi Gribonval,
  Julien Fleureau, Philippe Guillotel, and Isabelle Merlet.
\newblock Sissy: An efficient and automatic algorithm for the analysis of eeg
  sources based on structured sparsity.
\newblock \emph{NeuroImage}, 157:\penalty0 157--172, 2017.

\bibitem[Birot et~al.(2011)Birot, Albera, Wendling, and
  Merlet]{birot2011localization}
Gw{\'e}na{\"e}l Birot, Laurent Albera, Fabrice Wendling, and Isabelle Merlet.
\newblock Localization of extended brain sources from eeg/meg: the exso-music
  approach.
\newblock \emph{NeuroImage}, 56\penalty0 (1):\penalty0 102--113, 2011.

\bibitem[Cotter et~al.(2005)Cotter, Rao, Engan, and
  Kreutz-Delgado]{cotter2005sparse}
Shane~F Cotter, Bhaskar~D Rao, Kjersti Engan, and Kenneth Kreutz-Delgado.
\newblock Sparse solutions to linear inverse problems with multiple measurement
  vectors.
\newblock \emph{IEEE Transactions on Signal Processing}, 53\penalty0
  (7):\penalty0 2477--2488, 2005.

\bibitem[Dale and Sereno(1993)]{dale1993improved}
Anders~M Dale and Martin~I Sereno.
\newblock Improved localizadon of cortical activity by combining eeg and meg
  with mri cortical surface reconstruction: a linear approach.
\newblock \emph{Journal of cognitive neuroscience}, 5\penalty0 (2):\penalty0
  162--176, 1993.

\bibitem[Darvas et~al.(2004)Darvas, Pantazis, Kucukaltun-Yildirim, and
  Leahy]{darvas2004mapping}
Felix Darvas, D~Pantazis, E~Kucukaltun-Yildirim, and RM~Leahy.
\newblock Mapping human brain function with meg and eeg: methods and
  validation.
\newblock \emph{NeuroImage}, 23:\penalty0 S289--S299, 2004.

\bibitem[Destrieux et~al.(2010)Destrieux, Fischl, Dale, and
  Halgren]{destrieux2010automatic}
Christophe Destrieux, Bruce Fischl, Anders Dale, and Eric Halgren.
\newblock Automatic parcellation of human cortical gyri and sulci using
  standard anatomical nomenclature.
\newblock \emph{Neuroimage}, 53\penalty0 (1):\penalty0 1--15, 2010.

\bibitem[Ding(2009)]{ding2009reconstructing}
Lei Ding.
\newblock Reconstructing cortical current density by exploring sparseness in
  the transform domain.
\newblock \emph{Physics in Medicine \& Biology}, 54\penalty0 (9):\penalty0
  2683, 2009.

\bibitem[Friston(2008)]{friston2008hierarchical}
Karl Friston.
\newblock Hierarchical models in the brain.
\newblock \emph{PLoS computational biology}, 4\penalty0 (11):\penalty0
  e1000211, 2008.

\bibitem[Friston et~al.(2008)Friston, Harrison, Daunizeau, Kiebel, Phillips,
  Trujillo-Barreto, Henson, Flandin, and Mattout]{friston2008multiple}
Karl Friston, Lee Harrison, Jean Daunizeau, Stefan Kiebel, Christophe Phillips,
  Nelson Trujillo-Barreto, Richard Henson, Guillaume Flandin, and
  J{\'e}r{\'e}mie Mattout.
\newblock Multiple sparse priors for the m/eeg inverse problem.
\newblock \emph{NeuroImage}, 39\penalty0 (3):\penalty0 1104--1120, 2008.

\bibitem[Friston et~al.(2002)Friston, Penny, Phillips, Kiebel, Hinton, and
  Ashburner]{friston2002classical}
Karl~J Friston, William Penny, Christophe Phillips, S~Kiebel, G~Hinton, and
  John Ashburner.
\newblock Classical and bayesian inference in neuroimaging: theory.
\newblock \emph{NeuroImage}, 16\penalty0 (2):\penalty0 465--483, 2002.

\bibitem[Gon{\c{c}}alves et~al.(2003)Gon{\c{c}}alves, de~Munck, Verbunt, Bijma,
  Heethaar, and da~Silva]{gonccalves2003vivo}
S{\'o}nia~I Gon{\c{c}}alves, Jan~C de~Munck, Jeroen~PA Verbunt, Fetsje Bijma,
  Rob~M Heethaar, and F~Lopes da~Silva.
\newblock In vivo measurement of the brain and skull resistivities using an
  eit-based method and realistic models for the head.
\newblock \emph{IEEE Transactions on Biomedical Engineering}, 50\penalty0
  (6):\penalty0 754--767, 2003.

\bibitem[Gorodnitsky and Rao(1997)]{gorodnitsky1997sparse}
Irina~F Gorodnitsky and Bhaskar~D Rao.
\newblock Sparse signal reconstruction from limited data using focuss: A
  re-weighted minimum norm algorithm.
\newblock \emph{IEEE Transactions on signal processing}, 45\penalty0
  (3):\penalty0 600--616, 1997.

\bibitem[Gorodnitsky et~al.(1995)Gorodnitsky, George, and
  Rao]{gorodnitsky1995neuromagnetic}
Irina~F Gorodnitsky, John~S George, and Bhaskar~D Rao.
\newblock Neuromagnetic source imaging with focuss: a recursive weighted
  minimum norm algorithm.
\newblock \emph{Electroencephalography and clinical Neurophysiology},
  95\penalty0 (4):\penalty0 231--251, 1995.

\bibitem[Gramfort(2009)]{gramfort2009mapping}
Alexandre Gramfort.
\newblock \emph{Mapping, timing and tracking cortical activations with MEG and
  EEG: Methods and application to human vision}.
\newblock PhD thesis, 2009.

\bibitem[Gramfort et~al.(2012)Gramfort, Kowalski, and
  H{\"a}m{\"a}l{\"a}inen]{gramfort2012mixed}
Alexandre Gramfort, Matthieu Kowalski, and Matti H{\"a}m{\"a}l{\"a}inen.
\newblock Mixed-norm estimates for the m/eeg inverse problem using accelerated
  gradient methods.
\newblock \emph{Physics in medicine and biology}, 57\penalty0 (7):\penalty0
  1937, 2012.

\bibitem[Gramfort et~al.(2013)Gramfort, Strohmeier, Haueisen,
  H{\"a}m{\"a}l{\"a}inen, and Kowalski]{gramfort2013time}
Alexandre Gramfort, Daniel Strohmeier, Jens Haueisen, Matti~S
  H{\"a}m{\"a}l{\"a}inen, and Matthieu Kowalski.
\newblock Time-frequency mixed-norm estimates: Sparse m/eeg imaging with
  non-stationary source activations.
\newblock \emph{NeuroImage}, 70:\penalty0 410--422, 2013.

\bibitem[Gramfort et~al.(2014)Gramfort, Luessi, Larson, Engemann, Strohmeier,
  Brodbeck, Parkkonen, and H{\"a}m{\"a}l{\"a}inen]{gramfort2014mne}
Alexandre Gramfort, Martin Luessi, Eric Larson, Denis~A Engemann, Daniel
  Strohmeier, Christian Brodbeck, Lauri Parkkonen, and Matti~S
  H{\"a}m{\"a}l{\"a}inen.
\newblock Mne software for processing meg and eeg data.
\newblock \emph{Neuroimage}, 86:\penalty0 446--460, 2014.

\bibitem[Grech et~al.(2008)Grech, Cassar, Muscat, Camilleri, Fabri, Zervakis,
  Xanthopoulos, Sakkalis, and Vanrumste]{grech2008review}
Roberta Grech, Tracey Cassar, Joseph Muscat, Kenneth~P Camilleri, Simon~G
  Fabri, Michalis Zervakis, Petros Xanthopoulos, Vangelis Sakkalis, and Bart
  Vanrumste.
\newblock Review on solving the inverse problem in eeg source analysis.
\newblock \emph{Journal of neuroengineering and rehabilitation}, 5\penalty0
  (1):\penalty0 25, 2008.

\bibitem[Hallez et~al.(2007)Hallez, Vanrumste, Grech, Muscat, De~Clercq,
  Vergult, D'Asseler, Camilleri, Fabri, Van~Huffel, et~al.]{hallez2007review}
Hans Hallez, Bart Vanrumste, Roberta Grech, Joseph Muscat, Wim De~Clercq,
  Anneleen Vergult, Yves D'Asseler, Kenneth~P Camilleri, Simon~G Fabri, Sabine
  Van~Huffel, et~al.
\newblock Review on solving the forward problem in eeg source analysis.
\newblock \emph{Journal of neuroengineering and rehabilitation}, 4\penalty0
  (1):\penalty0 46, 2007.

\bibitem[H{\"a}m{\"a}l{\"a}inen and
  Ilmoniemi(1984)]{hamalainen1984interpreting}
Matti~S H{\"a}m{\"a}l{\"a}inen and Risto~J Ilmoniemi.
\newblock \emph{Interpreting measured magnetic fields of the brain: estimates
  of current distributions}.
\newblock Helsinki University of Technology, Department of Technical Physics,
  1984.

\bibitem[Hansen and O’Leary(1993)]{hansen1993use}
Per~Christian Hansen and Dianne~Prost O’Leary.
\newblock The use of the l-curve in the regularization of discrete ill-posed
  problems.
\newblock \emph{SIAM Journal on Scientific Computing}, 14\penalty0
  (6):\penalty0 1487--1503, 1993.

\bibitem[Haufe and Ewald(2019)]{haufe2019simulation}
Stefan Haufe and Arne Ewald.
\newblock A simulation framework for benchmarking eeg-based brain connectivity
  estimation methodologies.
\newblock \emph{Brain topography}, 32\penalty0 (4):\penalty0 625--642, 2019.

\bibitem[Haufe et~al.(2013)Haufe, Nikulin, M{\"u}ller, and
  Nolte]{haufe2013critical}
Stefan Haufe, Vadim~V Nikulin, Klaus-Robert M{\"u}ller, and Guido Nolte.
\newblock A critical assessment of connectivity measures for eeg data: a
  simulation study.
\newblock \emph{Neuroimage}, 64:\penalty0 120--133, 2013.

\bibitem[Huang et~al.(2006)Huang, Dale, Song, Halgren, Harrington, Podgorny,
  Canive, Lewis, and Lee]{huang2006vector}
Ming-Xiong Huang, Anders~M Dale, Tao Song, Eric Halgren, Deborah~L Harrington,
  Igor Podgorny, Jose~M Canive, Stephen Lewis, and Roland~R Lee.
\newblock Vector-based spatial--temporal minimum l1-norm solution for meg.
\newblock \emph{NeuroImage}, 31\penalty0 (3):\penalty0 1025--1037, 2006.

\bibitem[Huang et~al.(2016)Huang, Parra, and Haufe]{huang2016new}
Yu~Huang, Lucas~C Parra, and Stefan Haufe.
\newblock The new york head—a precise standardized volume conductor model for
  eeg source localization and tes targeting.
\newblock \emph{NeuroImage}, 140:\penalty0 150--162, 2016.

\bibitem[Ingber(1997)]{ingber1997statistical}
Lester Ingber.
\newblock Statistical mechanics of neocortical interactions: Canonical momenta
  indicatorsof electroencephalography.
\newblock \emph{Physical Review E}, 55\penalty0 (4):\penalty0 4578, 1997.

\bibitem[Ingber(1998)]{ingber1998statistical}
Lester Ingber.
\newblock Statistical mechanics of neocortical interactions: Training and
  testing canonical momenta indicators of eeg.
\newblock \emph{Mathematical and computer modelling}, 27\penalty0 (3):\penalty0
  33--64, 1998.

\bibitem[Kolda and Bader(2009)]{kolda2009tensor}
Tamara~G Kolda and Brett~W Bader.
\newblock Tensor decompositions and applications.
\newblock \emph{SIAM review}, 51\penalty0 (3):\penalty0 455--500, 2009.

\bibitem[Krol et~al.(2018)Krol, Pawlitzki, Lotte, Gramann, and
  Zander]{krol2018sereega}
Laurens~R Krol, Juliane Pawlitzki, Fabien Lotte, Klaus Gramann, and Thorsten~O
  Zander.
\newblock Sereega: Simulating event-related eeg activity.
\newblock \emph{Journal of neuroscience methods}, 309:\penalty0 13--24, 2018.

\bibitem[Limpiti et~al.(2006)Limpiti, Van~Veen, and Wakai]{limpiti2006cortical}
Tulaya Limpiti, Barry~D Van~Veen, and Ronald~T Wakai.
\newblock Cortical patch basis model for spatially extended neural activity.
\newblock \emph{IEEE Transactions on Biomedical Engineering}, 53\penalty0
  (9):\penalty0 1740--1754, 2006.

\bibitem[Liu and Schimpf(2006)]{liu2006efficient}
Hesheng Liu and Paul~H Schimpf.
\newblock Efficient localization of synchronous eeg source activities using a
  modified rap-music algorithm.
\newblock \emph{IEEE transactions on biomedical engineering}, 53\penalty0
  (4):\penalty0 652--661, 2006.

\bibitem[M{\"a}kel{\"a} et~al.(2018)M{\"a}kel{\"a}, Stenroos, Sarvas, and
  Ilmoniemi]{makela2018truncated}
Niko M{\"a}kel{\"a}, Matti Stenroos, Jukka Sarvas, and Risto~J Ilmoniemi.
\newblock Truncated rap-music (trap-music) for meg and eeg source localization.
\newblock \emph{NeuroImage}, 167:\penalty0 73--83, 2018.

\bibitem[Mannepalli and Routray(2019)]{mannepalli2019certainty}
Teja Mannepalli and Aurobinda Routray.
\newblock Certainty-based reduced sparse solution for dense array eeg source
  localization.
\newblock \emph{IEEE Transactions on Neural Systems and Rehabilitation
  Engineering}, 27\penalty0 (2):\penalty0 172--178, 2019.

\bibitem[Matsuura and Okabe(1995)]{matsuura1995selective}
Kanta Matsuura and Yoichi Okabe.
\newblock Selective minimum-norm solution of the biomagnetic inverse problem.
\newblock \emph{IEEE Transactions on Biomedical Engineering}, 42\penalty0
  (6):\penalty0 608--615, 1995.

\bibitem[Michel and Brunet(2019)]{michel2019eeg}
Christoph~M Michel and Denis Brunet.
\newblock Eeg source imaging: a practical review of the analysis steps.
\newblock \emph{Frontiers in neurology}, 10:\penalty0 325, 2019.

\bibitem[Michel and Murray(2012)]{michel2012towards}
Christoph~M Michel and Micah~M Murray.
\newblock Towards the utilization of eeg as a brain imaging tool.
\newblock \emph{Neuroimage}, 61\penalty0 (2):\penalty0 371--385, 2012.

\bibitem[Michel et~al.(2004)Michel, Murray, Lantz, Gonzalez, Spinelli, and
  de~Peralta]{michel2004eeg}
Christoph~M Michel, Micah~M Murray, G{\"o}ran Lantz, Sara Gonzalez, Laurent
  Spinelli, and Rolando~Grave de~Peralta.
\newblock Eeg source imaging.
\newblock \emph{Clinical neurophysiology}, 115\penalty0 (10):\penalty0
  2195--2222, 2004.

\bibitem[Michel et~al.(2009)Michel, Koenig, Brandeis, Wackermann, and
  Gianotti]{michel2009electrical}
Christoph~M Michel, Thomas Koenig, Daniel Brandeis, Ji{\v{r}}{\'\i} Wackermann,
  and Lorena~RR Gianotti.
\newblock \emph{Electrical neuroimaging}.
\newblock Cambridge University Press, 2009.

\bibitem[Molins et~al.(2008)Molins, Stufflebeam, Brown, and
  H{\"a}m{\"a}l{\"a}inen]{molins2008quantification}
A~Molins, Steven~M Stufflebeam, Emery~N Brown, and Matti~S
  H{\"a}m{\"a}l{\"a}inen.
\newblock Quantification of the benefit from integrating meg and eeg data in
  minimum l2-norm estimation.
\newblock \emph{Neuroimage}, 42\penalty0 (3):\penalty0 1069--1077, 2008.

\bibitem[Mosher and Leahy(1999)]{mosher1999source}
John~C Mosher and Richard~M Leahy.
\newblock Source localization using recursively applied and projected (rap)
  music.
\newblock \emph{IEEE Transactions on signal processing}, 47\penalty0
  (2):\penalty0 332--340, 1999.

\bibitem[Mosher et~al.(2005)Mosher, Baillet, Darvas, Pantazis, Yildirim, and
  Leahy]{mosher2005brainstorm}
John~C Mosher, Sylvain Baillet, Felix Darvas, Dimitrios Pantazis, E~Yildirim,
  and R~Leahy.
\newblock Brainstorm electromagnetic imaging software.
\newblock In \emph{5th International Symposium on Noninvasive Functional Source
  Imaging within the Human Brain and Heart (NFSI 2005)}, 2005.

\bibitem[Murray et~al.(2008)Murray, Brunet, and Michel]{murray2008topographic}
Micah~M Murray, Denis Brunet, and Christoph~M Michel.
\newblock Topographic erp analyses: a step-by-step tutorial review.
\newblock \emph{Brain topography}, 20\penalty0 (4):\penalty0 249--264, 2008.

\bibitem[Nolte and Sundsten(2009)]{nolte2009human}
J.~Nolte and J.W. Sundsten.
\newblock \emph{The Human Brain: An Introduction to Its Functional Anatomy}.
\newblock Mosby/Elsevier, 2009.

\bibitem[Nunez et~al.(2006)Nunez, Srinivasan, et~al.]{nunez2006electric}
Paul~L Nunez, Ramesh Srinivasan, et~al.
\newblock \emph{Electric fields of the brain: the neurophysics of EEG}.
\newblock Oxford University Press, USA, 2006.

\bibitem[Oikonomou and Kompatsiaris(2020)]{oikonomou2020novel}
Vangelis~P Oikonomou and Ioannis Kompatsiaris.
\newblock A novel bayesian approach for eeg source localization.
\newblock \emph{Computational Intelligence and Neuroscience}, 2020, 2020.

\bibitem[Oostenveld et~al.(2011)Oostenveld, Fries, Maris, and
  Schoffelen]{oostenveld2011fieldtrip}
Robert Oostenveld, Pascal Fries, Eric Maris, and Jan-Mathijs Schoffelen.
\newblock Fieldtrip: open source software for advanced analysis of meg, eeg,
  and invasive electrophysiological data.
\newblock \emph{Computational intelligence and neuroscience}, 2011:\penalty0 1,
  2011.

\bibitem[Ou et~al.(2009)Ou, H{\"a}m{\"a}l{\"a}inen, and
  Golland]{ou2009distributed}
Wanmei Ou, Matti~S H{\"a}m{\"a}l{\"a}inen, and Polina Golland.
\newblock A distributed spatio-temporal eeg/meg inverse solver.
\newblock \emph{NeuroImage}, 44\penalty0 (3):\penalty0 932--946, 2009.

\bibitem[Pascual-Marqui et~al.(1994)Pascual-Marqui, Michel, and
  Lehmann]{pascual1994low}
Roberto~D Pascual-Marqui, Christoph~M Michel, and Dietrich Lehmann.
\newblock Low resolution electromagnetic tomography: a new method for
  localizing electrical activity in the brain.
\newblock \emph{International Journal of psychophysiology}, 18\penalty0
  (1):\penalty0 49--65, 1994.

\bibitem[Pascual-Marqui et~al.(2002)]{pascual2002standardized}
Roberto~Domingo Pascual-Marqui et~al.
\newblock Standardized low-resolution brain electromagnetic tomography
  (sloreta): technical details.
\newblock \emph{Methods Find Exp Clin Pharmacol}, 24\penalty0 (Suppl
  D):\penalty0 5--12, 2002.

\bibitem[Sanei and Chambers(2013)]{sanei2013eeg}
Saeid Sanei and Jonathon~A Chambers.
\newblock \emph{EEG signal processing}.
\newblock John Wiley \& Sons, 2013.

\bibitem[Schmidt(1986)]{schmidt1986multiple}
Ralph Schmidt.
\newblock Multiple emitter location and signal parameter estimation.
\newblock \emph{IEEE transactions on antennas and propagation}, 34\penalty0
  (3):\penalty0 276--280, 1986.

\bibitem[Snodgrass and Vanderwart(1980)]{snodgrass1980standardized}
Joan~G Snodgrass and Mary Vanderwart.
\newblock A standardized set of 260 pictures: norms for name agreement, image
  agreement, familiarity, and visual complexity.
\newblock \emph{Journal of experimental psychology: Human learning and memory},
  6\penalty0 (2):\penalty0 174, 1980.

\bibitem[Strohmeier et~al.(2016)Strohmeier, Bekhti, Haueisen, and
  Gramfort]{strohmeier2016iterative}
Daniel Strohmeier, Yousra Bekhti, Jens Haueisen, and Alexandre Gramfort.
\newblock The iterative reweighted mixed-norm estimate for spatio-temporal
  meg/eeg source reconstruction.
\newblock \emph{IEEE transactions on medical imaging}, 35\penalty0
  (10):\penalty0 2218--2228, 2016.

\bibitem[Tibshirani et~al.(2005)Tibshirani, Saunders, Rosset, Zhu, and
  Knight]{tibshirani2005sparsity}
Robert Tibshirani, Michael Saunders, Saharon Rosset, Ji~Zhu, and Keith Knight.
\newblock Sparsity and smoothness via the fused lasso.
\newblock \emph{Journal of the Royal Statistical Society: Series B (Statistical
  Methodology)}, 67\penalty0 (1):\penalty0 91--108, 2005.

\bibitem[Vincent and Soille(1991)]{vincent1991watersheds}
Luc Vincent and Pierre Soille.
\newblock Watersheds in digital spaces: an efficient algorithm based on
  immersion simulations.
\newblock \emph{IEEE Transactions on Pattern Analysis \& Machine Intelligence},
  13\penalty0 (06):\penalty0 583--598, 1991.

\bibitem[Wipf and Nagarajan(2009)]{wipf2009unified}
David Wipf and Srikantan Nagarajan.
\newblock A unified bayesian framework for meg/eeg source imaging.
\newblock \emph{NeuroImage}, 44\penalty0 (3):\penalty0 947--966, 2009.

\bibitem[Wipf and Rao(2007)]{wipf2007empirical}
David~P Wipf and Bhaskar~D Rao.
\newblock An empirical bayesian strategy for solving the simultaneous sparse
  approximation problem.
\newblock \emph{IEEE Transactions on Signal Processing}, 55\penalty0
  (7):\penalty0 3704--3716, 2007.

\bibitem[Wipf et~al.(2010)Wipf, Owen, Attias, Sekihara, and
  Nagarajan]{wipf2010robust}
David~P Wipf, Julia~P Owen, Hagai~T Attias, Kensuke Sekihara, and Srikantan~S
  Nagarajan.
\newblock Robust bayesian estimation of the location, orientation, and time
  course of multiple correlated neural sources using meg.
\newblock \emph{NeuroImage}, 49\penalty0 (1):\penalty0 641--655, 2010.

\bibitem[Wipf et~al.(2011)Wipf, Rao, and Nagarajan]{wipf2011latent}
David~P Wipf, Bhaskar~D Rao, and Srikantan Nagarajan.
\newblock Latent variable bayesian models for promoting sparsity.
\newblock \emph{IEEE Transactions on Information Theory}, 57\penalty0
  (9):\penalty0 6236--6255, 2011.

\bibitem[Wipf(2006)]{wipf2006bayesian}
David~Paul Wipf.
\newblock \emph{Bayesian methods for finding sparse representations}.
\newblock PhD thesis, UC San Diego, 2006.

\bibitem[Xu et~al.(2007)Xu, Tian, Chen, and Yao]{xu2007lp}
Peng Xu, Yin Tian, Huafu Chen, and Dezhong Yao.
\newblock Lp norm iterative sparse solution for eeg source localization.
\newblock \emph{IEEE transactions on biomedical engineering}, 54\penalty0
  (3):\penalty0 400--409, 2007.

\bibitem[Yao and Dewald(2005)]{yao2005evaluation}
Jun Yao and Julius~PA Dewald.
\newblock Evaluation of different cortical source localization methods using
  simulated and experimental eeg data.
\newblock \emph{Neuroimage}, 25\penalty0 (2):\penalty0 369--382, 2005.

\bibitem[Zhang et~al.(2008)Zhang, Wu, and Guo]{zhang2008eeg}
De-xiang Zhang, Xiao-pei Wu, and Xiao-jing Guo.
\newblock The eeg signal preprocessing based on empirical mode decomposition.
\newblock In \emph{Bioinformatics and Biomedical Engineering, 2008. ICBBE 2008.
  The 2nd International Conference on}, pages 2131--2134. IEEE, 2008.

\bibitem[Zhang and Rao(2011)]{zhang2011sparse}
Zhilin Zhang and Bhaskar~D Rao.
\newblock Sparse signal recovery with temporally correlated source vectors
  using sparse bayesian learning.
\newblock \emph{IEEE Journal of Selected Topics in Signal Processing},
  5\penalty0 (5):\penalty0 912--926, 2011.

\bibitem[Zhang and Rao(2013)]{zhang2013extension}
Zhilin Zhang and Bhaskar~D Rao.
\newblock Extension of sbl algorithms for the recovery of block sparse signals
  with intra-block correlation.
\newblock \emph{IEEE Transactions on Signal Processing}, 61\penalty0
  (8):\penalty0 2009--2015, 2013.

\bibitem[Zhu et~al.(2014)Zhu, Zhang, Dickens, and Ding]{zhu2014reconstructing}
Min Zhu, Wenbo Zhang, Deanna~L Dickens, and Lei Ding.
\newblock Reconstructing spatially extended brain sources via enforcing
  multiple transform sparseness.
\newblock \emph{NeuroImage}, 86:\penalty0 280--293, 2014.

\end{thebibliography}

	\begin{wrapfigure}{l}{25mm} 
		\includegraphics[width=1in,height=1.25in,clip,keepaspectratio]{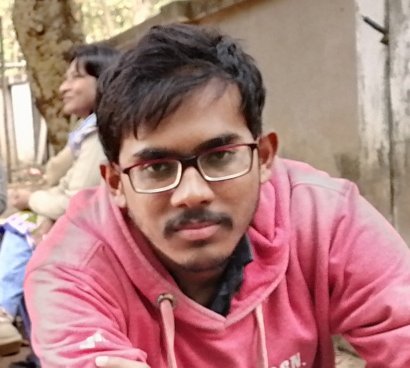}
	\end{wrapfigure}\par
	\textbf{Teja Mannepalli}
	is born in Chirala, Andhra Pradesh, India, on $13^{th} April \text{ 1994}$. He received his B.Tech. degree in Electrical engineering, in 2015 and M.Tech. in Instrumentation and signal processing in 2016 from Indian Institute of Technology, Kharagpur. He is currently a research scholar from IIT Kharagpur working towards Ph.D. His known areas are signal processing, biomedical instrumentation, and linear algebra. \par
	
	\begin{wrapfigure}{l}{25mm} 
		\includegraphics[width=1in,height=1.25in,clip,keepaspectratio]{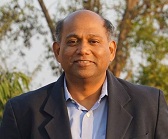}
	\end{wrapfigure}\par
	\textbf{Aurobinda Routray}
	is born on March 31, 1965. He received the B.Sc. Eng. from National Institute of Technology, Rourkela, India, M. Tech from, Indian Institute of Technology, Kanpur, India, and Ph. D from Sambalpur University, Odisha, India in 1989, 1991, and 1999 respectively. 
	
	He is currently the Associate Dean, Infrastructure (Electrical), in IIT Kharagpur, India. He is associated with the Department of Electrical Engineering as a Professor. He won Samsung GRO Award in 2012 and State Young Scientist Award in 1999. His research interests include biomedical, statistical signal processing, and embedded systems. He is a member of the IEEE. \par
	
\end{document}